\documentclass{article}

\usepackage{graphicx} % Required for inserting images
\usepackage{booktabs}
\usepackage{float}
\usepackage{setspace}
\usepackage{adjustbox}
\usepackage{threeparttable}
\usepackage{multirow, makecell}
\usepackage{PRIMEarxiv}
\usepackage{wrapfig,lipsum,booktabs}
\usepackage{titletoc}

\usepackage[utf8]{inputenc} % allow utf-8 input
\usepackage[T1]{fontenc}    % use 8-bit T1 fonts
\usepackage{hyperref}       % hyperlinks
\usepackage{url}            % simple URL typesetting
\usepackage{booktabs}       % professional-quality tables
\usepackage{amsfonts}       % blackboard math symbols
\usepackage{amsmath} 
\usepackage{multirow}
\usepackage{array}
\usepackage{nicefrac}       % compact symbols for 1/2, etc.
\usepackage{microtype}      % microtypography
\usepackage{lipsum}
\usepackage{tcolorbox}
\usepackage{lmodern}
\usepackage{fancyhdr}       % header
\usepackage{graphicx}       % graphics
\usepackage{listings}       % For verbatim writing (discomat schema)
\usepackage{chngcntr}       % Change counter for appendix
\usepackage{soul}
\usepackage{url}
\usepackage{amsmath, amssymb, amsfonts}
\usepackage{hyperref}
\usepackage{algorithm}
\usepackage{amsmath}
\usepackage[commandnameprefix=always]{changes}
\usepackage{amsmath,amsfonts,bm}

\def\eqref#1{equation~\ref{#1}}
\def\1{\bm{1}}

\DeclareMathAlphabet{\mathsfit}{\encodingdefault}{\sfdefault}{m}{sl}
\SetMathAlphabet{\mathsfit}{bold}{\encodingdefault}{\sfdefault}{bx}{n}

\newcommand{\mace}{\textsc{MACE}{}}
\newcommand{\chgnet}{\textsc{CHGNet}{}}
\newcommand{\mgnet}{\textsc{M3GNet}{}}
\newcommand{\sevennet}{\textsc{SevenNet}{}}
\newcommand{\orb}{\textsc{Orb}{}}
\newcommand{\mattersim}{\textsc{MatterSim}{}}
\newcommand{\maceomat}{\textsc{Mace-OMat}{}}
\newcommand{\uma}{\textsc{UMA}{}}
\newcommand{\uniffbench}{\textsc{UniFFBench}}
\newcommand{\minx}{\textsc{MinX}}
\newcommand{\minxeq}{\textsc{MinX-EQ}}
\newcommand{\minxhtp}{\textsc{MinX-HTP}}
\newcommand{\minxpocc}{\textsc{MinX-POcc}}
\newcommand{\minxem}{\textsc{MinX-EM}}

\usepackage{graphicx}%
\usepackage{multirow}%
\usepackage{amsmath,amssymb,amsfonts}%
\usepackage{amsthm}%
\usepackage{mathrsfs}%
\usepackage[title]{appendix}%
\usepackage{textcomp}%
\usepackage{manyfoot}%
\usepackage{adjustbox}
\usepackage{setspace}
\usepackage{booktabs}%
\usepackage{algorithm}%
\usepackage{algorithmicx}%
\usepackage{algpseudocode}%
\usepackage{listings}%
\usepackage{longtable}%
\usepackage{array}%
\usepackage[justification=centering]{caption} % Add in preamble
\usepackage{hyperref}
\usepackage{cleveref}
\usepackage{longtable}
\usepackage{titletoc}
\usepackage[numbers]{natbib}
\usepackage{tabularx}
\title{UniFFBench: Evaluating Universal Machine Learning Force Fields Against Experimental Measurements}

\author{
    Sajid Mannan\textsuperscript{1}, Vaibhav Bihani\textsuperscript{2}, Carmelo Gonzales\textsuperscript{3*}, Kin Long Kelvin Lee\textsuperscript{3*}, \\
    \textbf{Nitya Nand Gosvami}\textsuperscript{4}, \textbf{Sayan Ranu}\textsuperscript{2,5}, \textbf{Santiago Miret}\textsuperscript{3\#}, \textbf{N M Anoop Krishnan}\textsuperscript{1,2\#}\\
    \\
    \textsuperscript{1}Department of Civil Engineering, Indian Institute of Technology Delhi\\
    \textsuperscript{2}Yardi School of Artificial Intelligence, Indian Institute of Technology Delhi\\
    \textsuperscript{3}Intel Labs, California, USA\\
    \textsuperscript{4}Department of Materials Science and Engineering, Indian Institute of Technology Delhi\\
    \textsuperscript{5}Department of Computer Science and Engineering, Indian Institute of Technology Delhi\\
    \textsuperscript{*}Current affiliation: NVIDIA Corporation\\
    \textsuperscript{\#}Corresponding authors: \texttt{santiago.miret@gmail.com, krishnan@iitd.ac.in}
}

\begin{document}
\maketitle
\begin{abstract}
\noindent Universal machine learning force fields (UMLFFs) promise to revolutionize materials science by enabling rapid atomistic simulations across the periodic table. However, their evaluation has been limited to computational benchmarks that may not reflect real-world performance. We introduce \uniffbench{}, a comprehensive evaluation framework featuring the \minx{} dataset---a diverse collection of 1,500+ mineral systems spanning 85 elements, extreme thermodynamic conditions (0--5000 K, 0--1000 GPa), and structural complexity, including partial occupancy and disorder. This diversity, combined with experimental reference values for validation, enables assessment of UMLFF generalization across chemical space and conditions substantially beyond typical training scenarios. Our systematic evaluation of six state-of-the-art UMLFFs reveals a substantial ``reality gap'': models achieving impressive performance on computational benchmarks often fail when confronted with experimental complexity. Even the best-performing models exhibit higher density prediction error than the threshold required for practical applications. We observe disconnects between simulation stability and mechanical property accuracy, with prediction errors correlating with training data representation rather than the modeling method. 

\end{abstract}
% \keywords{Molecular Dynamics, UMLIPs, Minearls}
%TC:endignore
\section{Main}

Translating the computational promise of universal machine learning force fields (UMLFFs), that is, atomistic simulations with quantum accuracy~\citep{duval2023hitchhiker,bihani2024egraffbench,batatia2025foundation,musaelian2023learning,fuforces,miret2023the,saal2013materials,hautier2012computer,merchant_scaling_2023,axelrod2022learning,yuan2025foundation}, to real-world impact requires them to accurately predict material behavior under physically relevant conditions\citep{schmidt2024improving,lee2023matsciml,deringer_machine_2019,friederich2021machine}, where prediction failures can lead to costly experimental dead ends in modern materials discovery pipelines~\citep{miret2023the,zeni2025generative,merchant_scaling_2023,yuan2025foundation}. While current evaluation practices have been instrumental for rapid screening and model development, they often lack experimental grounding. This creates a growing disconnect between benchmark success and real-world applicability, highlighting the need for complementary validation against experimental data. State-of-the-art models, including \chgnet{}~\citep{deng2023chgnet}, \mgnet{}~\citep{chen2022universal}, \mace{}~\citep{batatia2025foundation}, \mattersim{}~\citep{yang2024mattersim}, \sevennet{}~\citep{park_scalable_2024}, and \orb{}~\citep{neumann2024orbfastscalableneural}, are exclusively trained on DFT datasets. Note that the crystal structures on many of the DFT datasets are sourced from experimental databases like ICSD and are hence indirectly grounded on experimental crystal structures, but target properties (energies, forces, stresses) are computed via DFT~\citep{deng2023chgnet,barrosoluque2024openmaterials2024omat24,levine2025open}. The models trained on these datasets are then predominantly evaluated against computational benchmarks that validate performance on similar DFT-derived properties~\citep{chanussot2021open,matbench,miret2025energy,wood2025umafamilyuniversalmodels,chiang2025mliparenaadvancingfairness,fu2023forcesenoughbenchmarkcritical}. While this approach has driven rapid progress in energy and force prediction accuracy, it creates a potential training-evaluation circularity where models learn to reproduce DFT predictions and are validated against similar DFT-computed properties. This leaves open the question of whether models can reliably predict diverse material properties under finite-temperature conditions relevant for practical applications.

In spite of over 20 presently available UMLFFs~\cite{matbench} and numerous computational benchmarks, systematic validation of these force fields against experimental measurements under realistic conditions remains virtually absent~\cite{miret2025energy,fu2025learning,fu2023forcesenoughbenchmarkcritical}, with no studies covering extensive chemical spaces and environmental conditions. This paucity contrasts sharply with other machine learning domains, such as large language models, where real-world testing is considered essential for deployment~\cite{zaki_mascqa_2024,miret2025enabling,alampara2025probing}. Existing evaluation protocols focus predominantly on energy and force prediction errors for static configurations~\citep{matbench,gonzales2024benchmarking}, neglecting essential aspects required for practical applications. While computational datasets provide controlled comparison conditions, they cannot capture experimental complexity including thermal and pressure effects, structural disorder, and dynamic phenomena such as thermal expansion and mechanical response that ultimately determine material performance~\citep{fuforces,deringer_machine_2019}. Moreover, compositional biases in training data may lead to models ``over-fitted'' to specific chemical environments rather than being truly universal~\cite{miret2025energy,yang2024mattersim,levine2025open,schmidt2024improving}. However, a systematic framework for evaluating these critical limitations has been lacking.

Here, we present \uniffbench{}, a comprehensive benchmarking framework for evaluating UMLFFs against experimental measurements. The framework integrates \minx{}, a hand-curated dataset comprising $\sim$1,500 experimentally determined mineral structures organized into four complementary subsets that systematically probe distinct aspects of materials behavior: ambient conditions, extreme thermodynamic environments, compositional disorder through partial occupancies, and mechanical properties via experimentally measured elastic tensors. Our evaluation extends beyond conventional energy and force metrics to assess MD simulation stability, structural fidelity at finite temperatures, bond length accuracy, and elastic property prediction capabilities. Our systematic analysis reveals that prediction errors correlate directly with training data representation, demonstrating systematic biases rather than universal predictive capability. Furthermore, we uncover a striking disconnect between structural stability and mechanical property accuracy, suggesting that current training protocols require modification to incorporate higher-order derivative information beyond energies and forces~\citep{miret2025energy}. By providing standardized protocols and reference datasets grounded in experimental reality, \uniffbench{} establishes essential benchmarks for advancing reliable UMLFF deployment in practical materials discovery and design.

\section{Results}

\subsection{\uniffbench{} Framework}

\begin{figure}[!htbp]
    \centering
    % \vspace{0.1in}
    \includegraphics[width=0.95\textwidth]{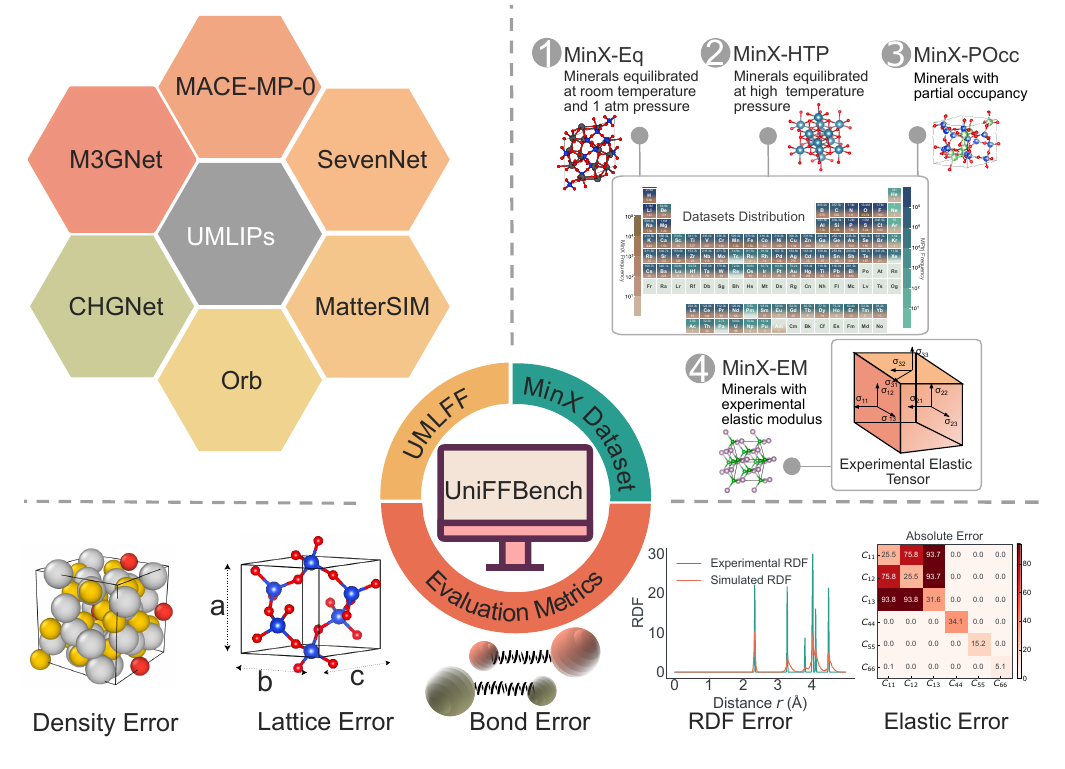}
    % \vspace{-0.3in}
    \caption{\textbf{\uniffbench{} framework for systematic evaluation of UMLFFs.} The framework integrates three core components: six state-of-the-art UMLFF models (\textsc{CHGNet}, \textsc{M3GNet}, \textsc{MACE}, \textsc{MatterSim}, \textsc{SevenNet}, and \textsc{Orb}) evaluated under standardized protocols; the \minx{} dataset comprising four experimental mineral subsets (\minxeq{} for ambient conditions, \minxhtp{} for extreme thermodynamic conditions, \minxpocc{} for partial occupancy structures, and \minxem{} for elastic property validation); and comprehensive evaluation metrics spanning structural accuracy (lattice and density errors), atomic-scale organization (radial distribution functions and bond length analysis), dynamic stability (MD simulations), and mechanical properties (elastic tensor prediction). This multi-dimensional approach enables systematic assessment of model performance across the diverse chemical and structural landscape of real-world minerals.}
    \label{fig:summary_plot}
    \vspace{-0.1in}
\end{figure}

\uniffbench{} establishes a systematic evaluation framework to address the gap between computational model development and real-world materials applications through three integrated components (\Cref{fig:summary_plot}, see Supplementary Section 1 for the design principles of \uniffbench{}). First, we evaluate six state-of-the-art UMLFF models---\chgnet{}, \mgnet{}, \mace{}, \mattersim{}, \sevennet{}, and \orb{} (see Supplementary Section 2 for brief model description)---under standardized computational protocols (see \Cref{meth:MD,meth:error}) to ensure fair performance comparisons across different architectural approaches. Second, the \minx{} dataset provides experimental grounding through approximately 1,500 carefully curated mineral structures organized into four complementary subsets: \minxeq{} for standard ambient conditions representative of typical laboratory environments; \minxhtp{} for extreme thermodynamic regimes that test model robustness; \minxpocc{} for minerals with partial atomic site occupancies that challenge compositional disorder handling; and \minxem{} for direct validation of mechanical property predictions using experimentally measured elastic moduli. Third, our evaluation methodology extends beyond conventional energy and force metrics to encompass structural fidelity through lattice parameters and density accuracy, atomic-scale organization via radial distribution functions and bond length analysis, dynamic stability through finite-temperature MD simulations, and mechanical response via elastic tensor prediction.

\begin{figure}[!htbp]
    \centering
    \includegraphics[width=0.95\textwidth]{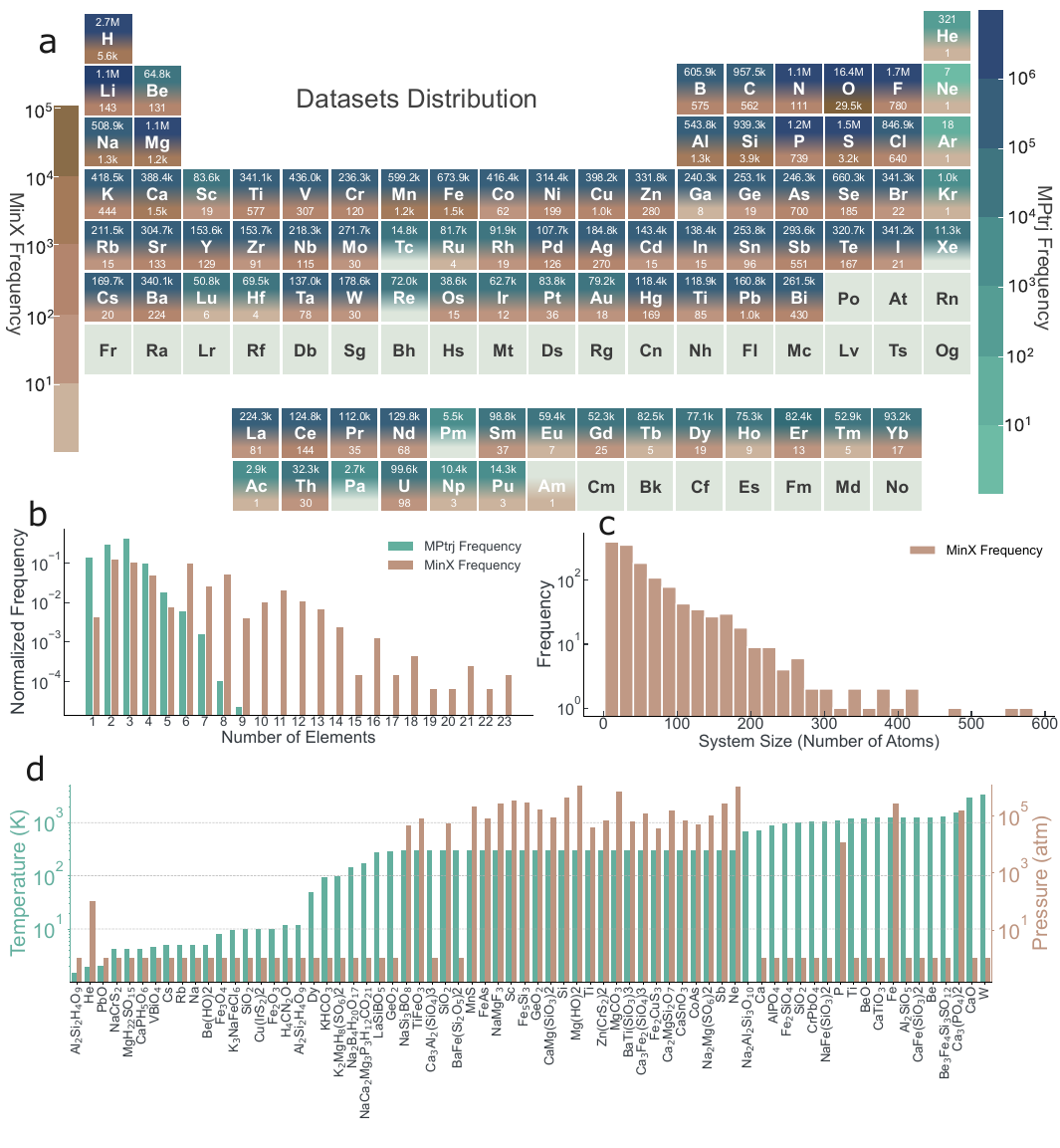}
    % \caption{(a) Distribution of atomic elements and their frequency in the MPtrj (teal-blue gradient) and MinX (brown) datasets respectively. (b) Number of elements present in the unit cells of MPtrj and MinX. (c) Distribution of the number of atoms in the unit cell of MinX. (d) Distribution of temperature (left abscissa) and pressure (right abscissa) across extreme thermodynamic scenarios.}
    \caption{\textbf{\minx{} dataset.} \textbf{a}, Elemental frequency distribution comparison between MPtrj training dataset (teal-blue gradient) and \minx{} evaluation dataset (brown) across the periodic table, revealing significant compositional biases in current training data toward specific element families (H, Li, Mg, O, F, S) while achieving near-complete elemental main group coverage with only Americium absent from both datasets. \textbf{b}, Compositional complexity comparison showing maximum number of unique elements per structure: MPtrj structures contain at most 9 elements while \minx{} minerals exhibit up to 23 distinct elements, reflecting the extraordinary chemical diversity of real-world materials. \textbf{c}, Unit cell size distribution for \minx{} dataset demonstrating structural complexity with many minerals containing hundreds of atoms, substantially exceeding typical computational training configurations. \textbf{d}, Thermodynamic condition distribution across \minxhtp{} subset spanning extreme temperature and pressure regimes (left and right axes respectively), enabling model evaluation under challenging conditions rarely represented in standard training datasets.}
    \label{fig:data-distribution}
\end{figure}

Current UMLFF training relies predominantly on specialized DFT datasets, including MPtrj~\citep{deng2023chgnet}, OC22~\citep{tran2023open}, and Alexandria~\citep{schmidt2024improving}, which may not capture experimental complexities. Comparing \minx{} with the widely-used MPtrj dataset reveals key differences of our evaluation set (see \Cref{fig:data-distribution}, Supplementary Section 7, and Supplementary Figures 5 and 9). While both achieve near-complete periodic table coverage, 
% MPtrj exhibits severe compositional biases toward specific element families. Elements such as H, Li, Mg, O, F, and S are substantially overrepresented compared to their natural abundance in mineral systems, creating potential blind spots for real-world applications. More critically, structural complexity analysis demonstrates that 
MPtrj structures exhibit limited compositional diversity with a maximum of 9 unique elements per structure, whereas \minx{} minerals contain up to 23 distinct elements, reflecting the extraordinary chemical complexity of naturally occurring materials (\Cref{fig:data-distribution}b). Similarly, \minx{} unit cells contain substantially larger numbers of atoms---often hundreds compared to typical MPtrj configurations (\Cref{fig:data-distribution}c). The \minxhtp{} subset further challenges model performance under extreme thermodynamic conditions spanning wide temperature and pressure ranges (\Cref{fig:data-distribution}d), providing essential tests of model robustness beyond standard ambient conditions. This multi-dimensional evaluation approach enables systematic identification of model strengths, limitations, and failure modes across the diverse chemical and structural landscape of naturally occurring minerals.

\subsection{MD simulation stability}

\begin{figure}[!ht]
    \centering
    \includegraphics[width=0.95\textwidth]{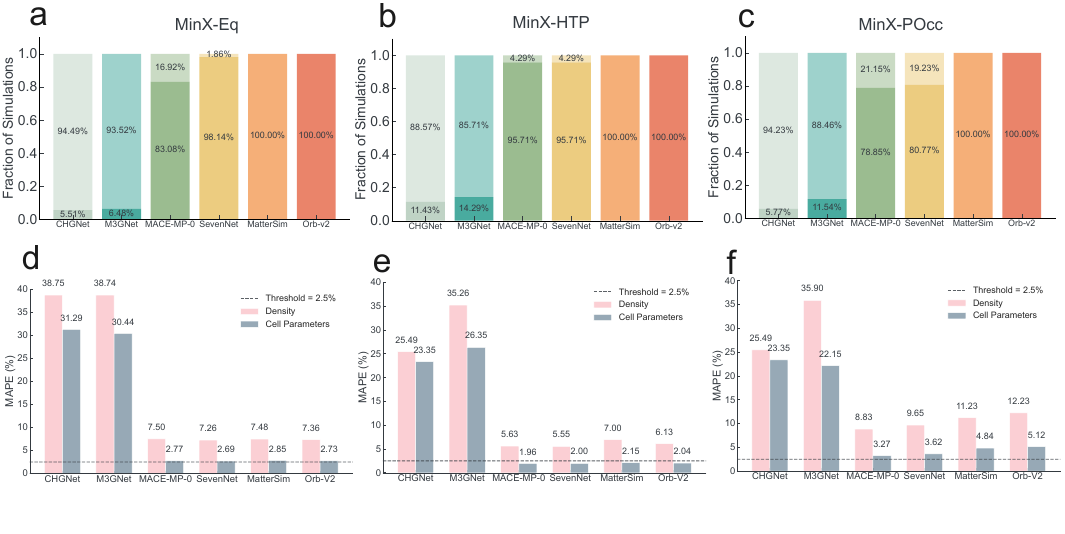}
    \caption{\textbf{Systematic evaluation of state-of-the-art UMLFFs.} \textbf{a--c}, Simulation completion rates across \minx{} datasets showing fraction of successfully completed MD simulations (dark segments) versus failed simulations (light segments) for \minxeq{} (ambient conditions), \minxhtp{} (extreme thermodynamic conditions), and \minxpocc{} (partial occupancy structures). \textbf{d--f}, Structural prediction accuracy quantified through mean absolute percentage error (MAPE) for density (pink) and lattice parameters (grey) relative to experimental values, calculated only for successfully completed simulations. Results demonstrate clear performance hierarchies with some models achieving 100\% completion rates while others fail on over 85\% of realistic mineral structures.}
    \label{fig:fraction-completions}
\end{figure}
First, we systematically evaluate six UMLFFs through MD simulations on three subsets of the \minx{} datasets: \minxeq{}, \minxhtp{}, and \minxpocc{}. Simulation failures in our benchmark are defined by two criteria: (1) inability to complete the full 50 ps MD trajectory due to numerical instabilities (for example, force divergence, bond breaking, or unphysical atomic configurations), or (2) completion with property prediction errors exceeding 100\% relative to experimental reference values. The first criterion assesses intrinsic model stability under finite-temperature dynamics, while the second excludes simulations that, despite completing, produce predictions too inaccurate to provide meaningful physical insight. The 100\% error threshold ensures our evaluation reflects model performance on physically meaningful predictions that could reasonably inform materials science applications. We note that including simulations with errors exceeding 100\% significantly skewed statistical metrics due to extreme outliers, obscuring genuine performance differences between models. However, sensitivity analysis using alternative thresholds (150\%, 250\%, 500\%) demonstrates that our conclusions remain robust across different cutoff values (see Supplementary Figure 4). All analyses of structural and mechanical properties were performed only on simulations satisfying both criteria---numerical stability throughout the trajectory and prediction accuracy within the 100\% threshold.

MD simulation reveal a pronounced performance hierarchy (see (\Cref{fig:fraction-completions}a--c)). \orb{} and \mattersim{} demonstrate strong robustness, achieving 100\% simulation completion rates across all experimental conditions, while \chgnet{} and \mgnet{} suffer failure rates exceeding 85\% across all datasets. \mace{} and \sevennet{} show intermediate performance, with completion rates degrading from $\sim$95\% for \minxhtp{} to $\sim$75\% for \minxpocc{}, suggesting poor generalization to compositional disordered system potentially due to insufficient representation of such systems in training data. These failures may arise from several mechanisms: unphysically large predicted forces (>100 eV/\AA) arising due to interactions that are unseen in the training data that prevent stable integration, structural instabilities that lead to bond breaking or atomic overlap, or insufficient model accuracy for the specific chemical system or thermodynamic conditions.

Failures occur without clear warning indicators that would allow practitioners to identify problematic cases \textit{a priori}. Standard energy and force error metrics during initial equilibration stages show poor correlation with subsequent simulation stability. This disconnect means that low energy and force errors do not guarantee stable long-term simulations, confirming that current evaluation protocols may overestimate real-world reliability of these models.

\subsection{Structural Accuracy}
We next perform the structural analysis of the systems obtained through MD simulations in terms of the density and lattice parameters (see \Cref{fig:fraction-completions}d--f). The four most stable models---\orb{}, \mattersim{}, \sevennet{}, and \mace{}---achieve mean absolute percentage errors (MAPE) below or $\sim 10\%$ for both density and lattice parameters across all datasets. However, even these best-performing models systematically exceed the experimentally acceptable density variation threshold of $\pm$2-3\%. \chgnet{} and \mgnet{} exhibit substantially higher errors exceeding 10\% in their limited successful predictions, consistent with their poor simulation stability. 

All models demonstrate increased prediction errors for the \minxpocc{} subset, with MAPE values typically 2--3 times higher than for ambient conditions. This degradation exemplifies the challenges in modeling compositional disorder and partial occupancy---features commonly encountered in real-world materials but underrepresented in DFT generated training datasets. We further evaluated two state-of-the-art models, \uma{} and \maceomat, trained on significantly larger datasets. As shown in Supplementary Figure 3, while these models achieve excellent stability (100\% completion similar to \orb{} and \mattersim), their density prediction errors (5.34\% and 6.89\% MAPE) still exceed practical application thresholds, demonstrating that increased training data improves stability but does not proportionally enhance experimental accuracy. The consistent pattern across all model architectures suggests that this limitation stems from training data bias rather than specific algorithmic deficiencies. Complete parity plots and additional error metrics are provided in the Supplementary Section 6 and Supplementary Figures 1 and 2).

\subsection{Temporal Evolution of Errors during Simulations}
\begin{figure}[!ht]
    \centering
    \includegraphics[width=0.95\textwidth]{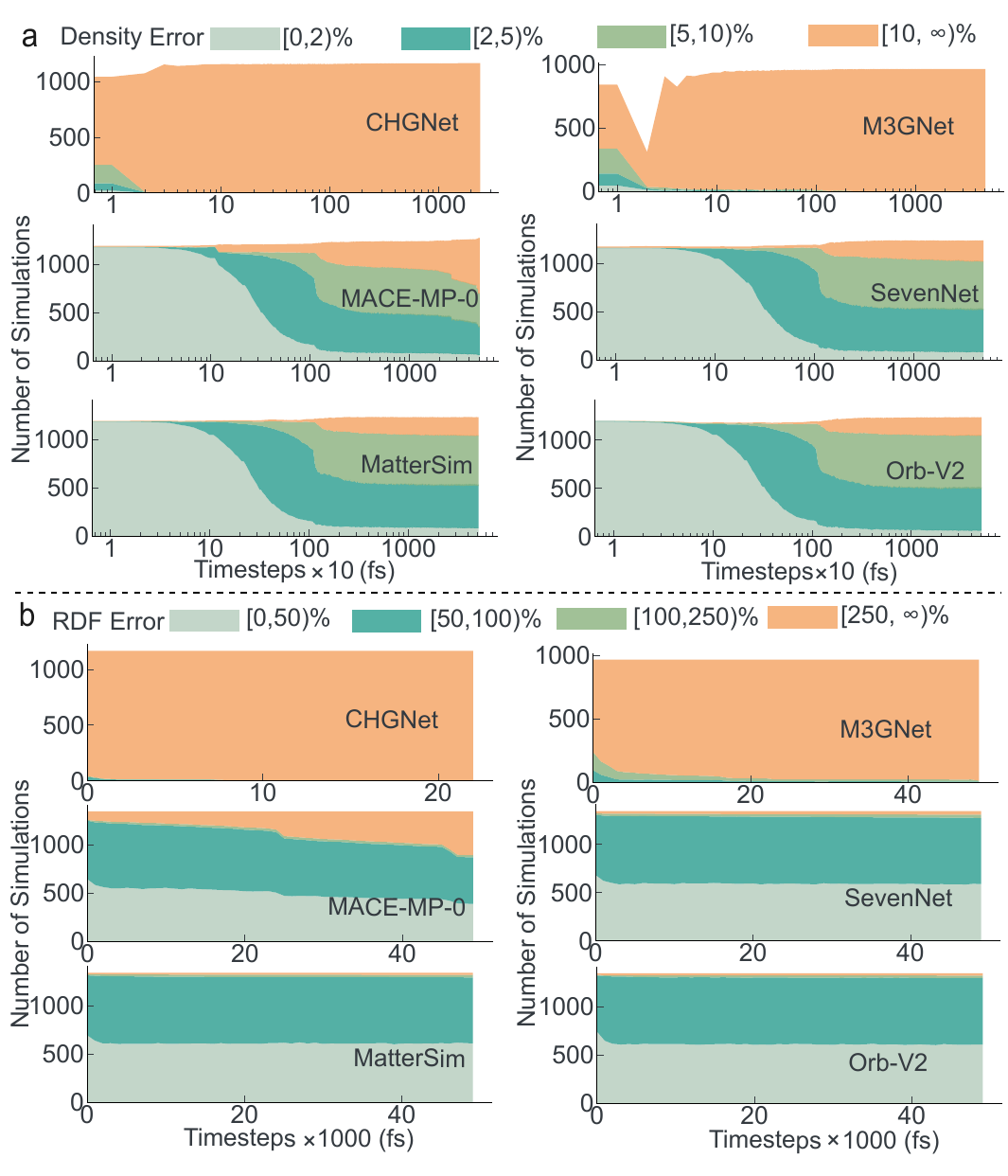}
    \caption{\textbf{Temporal evolution reveals divergent stability patterns among UMLFFs.} \textbf{a}, Density error evolution during MD simulations with stacked areas representing error distributions across four ranges ([0,2)\%, [2,5)\%, [5,10)\%, [10,$\infty$)\%). Simulation timesteps shown on logarithmic scale to capture behavior across multiple time regimes for \minxeq{} (ambient conditions). \textbf{b}, Radial distribution function (RDF) error evolution showing atomic spatial organization accuracy with error ranges ([0,50)\%, [50,100)\%, [100,250)\%, [250,$\infty$)\%). Results demonstrate that stable models converge to consistent error ranges while unstable models exhibit persistent high errors throughout simulation periods for \minxeq{} (ambient conditions).}
    \label{fig:rdf-plots}
\end{figure}

\textbf{Density evolution.} To understand the temporal behavior of these errors, we analyze the structural evolution during MD simulations, focusing on density and radial distribution function (RDF) errors (see Methods Section~\ref{meth:error}). Here, we focus on the \minxeq{} dataset; results on the other subsets are included in the Supplementary Sections 8.2 and 8.3. Density evolution reveals striking differences in simulation stability across models (see \Cref{fig:rdf-plots}a). \chgnet{} and \mgnet{} exhibit instability, with virtually all simulations displaying density errors exceeding 10\% throughout the entire 50 ps simulation window. Even in rare cases where these models complete the simulation, they maintain unacceptably high errors, indicating failures in preserving structural integrity during thermal fluctuations. In stark contrast, \mace{}, \mattersim, \sevennet{}, and \orb{} demonstrate markedly superior performance, with density errors consistently converging below 10\% for the majority of mineral systems. Critically, these stable models demonstrate the expected equilibration behavior: initial timesteps exhibit transient dynamics with increasing errors, while successful models progressively converge to equilibrium states characterized by stable, low error values.

\textbf{RDF evolution.} The spatial arrangement of atoms represents a stringent test of model accuracy, as RDFs capture both short-range chemical bonding and medium-range structural correlations essential for material properties. Atomic-scale structural analysis through RDF evolution (\Cref{fig:rdf-plots}b) corroborates the density findings while providing deeper insights into local atomic organization. \chgnet{} and \mgnet{} again demonstrate poor performance with consistently high RDF errors, confirming their inability to maintain realistic atomic spatial distributions during finite-temperature dynamics. The superior models achieve approximately 50\% of cases within a 50\% error threshold---a relatively lenient criterion adopted due to inherent challenges in comparing crystalline reference structures with finite-temperature simulated systems. This threshold accounts for natural broadening of RDF peaks during thermal motion and incorporates systematic noise introduced through our validation protocol, which perturbs atomic positions by 0.005 \AA{} to enable fair comparison between sharp experimental peaks and thermally broadened simulation profiles (see Methods Sections~\ref{meth:MD}, \ref{meth:post-process}, \ref{meth:error}). Similar behavior is observed for \minxhtp{} and \minxpocc{} subsets (see Supplementary Figures 12 and 13). A detailed description of RDF calculations and a representative comparison for a subset minerals are provided in the Supplementary Section 10 and Supplementary Figure 17). To verify dataset size adequacy, we evaluated model performance on systematically sampled subsets of 100, 500, and 1,000 minerals (Supplementary Figures 14 to 16). The analysis demonstrates that the performance in terms of the errors (density and radial distribution function errors), model rankings, and key conclusions in terms of the typical errors hold across all subsets of 100, 500, and 1,000 minerals. This confirms that \minx{} dataset provide statistically robust evaluation across the chemical and structural diversity of characterized mineral systems.

\subsection{Elastic Tensor Analysis}
\begin{figure}[!ht]
    \centering
    \includegraphics[width=0.95\textwidth]{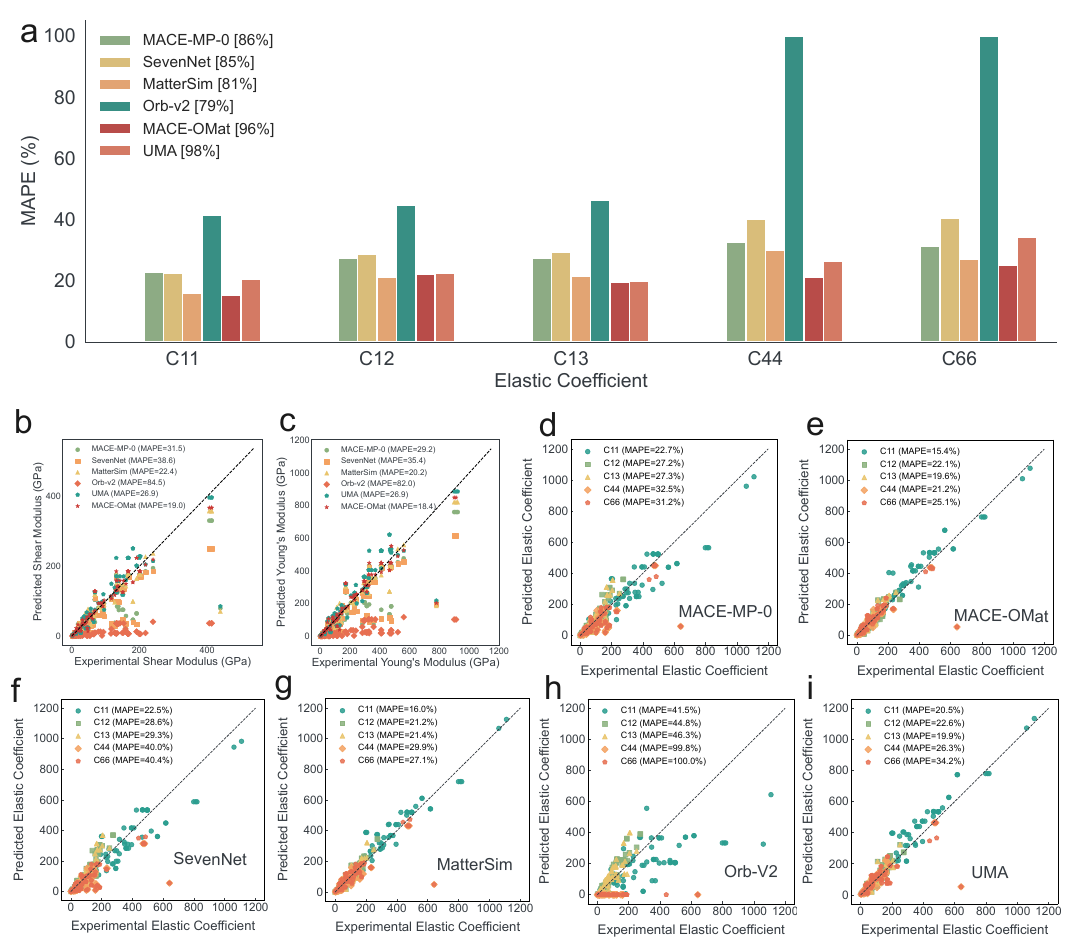}
    \caption{\textbf{Elastic tensor simulations on \minxem{} dataset.} \textbf{a}, Mean absolute percentage error (MAPE) for different elastic coefficients (C$_{11}$, C$_{12}$, C$_{13}$, C$_{44}$, C$_{66}$) across all models. Numbers in parentheses indicate fraction of successful simulations. Parity plots comparing predicted versus experimental \textbf{b}, shear modulus (Voigt average), \textbf{c}, Young's modulus, and {\textbf{d-i}}}, elastic tensors (C$_{11}$, C$_{12}$, C$_{13}$, C$_{44}$, C$_{66}$) for individual models, with MAPE values indicated in legends.
    \label{fig:elastic-plots}
\end{figure}

Accurate prediction of elastic tensor components represents a stringent test of UMLFF capabilities, since they are not explicitly trained on elastic tensors, thereby providing a true test of their generalizability beyond energy and force prediction. We computed elastic coefficients for 100 minerals in the \minxem{} dataset through systematic energy minimization followed by strain application in relevant crystallographic directions, comparing results with experimental measurements (see Methods Section~\ref{meth:elastic}). To assess state-of-the-art models, we additionally evaluated \uma{} and \maceomat{} on this benchmark. Our analysis reveals systematic deterioration in prediction accuracy across different elastic tensor components (\Cref{fig:elastic-plots}a). While C$_{11}$ predictions achieve reasonable mean absolute percentage errors (MAPE) of 20-25\% for most stable models (\mace{}, \sevennet{}, \mattersim{},\uma{}, and \maceomat{}), accuracy degrades progressively for other components, with C$_{44}$ and C$_{66}$ showing particularly poor performance. Notably, both \uma{} and \maceomat{} achieved high completion rates (98\% for \uma{}, 96\% for \maceomat{}), demonstrating robust energy minimization capabilities. However, their elastic tensor predictions show substantial errors comparable to other stable models, with MAPE values ranging from 20--45\% across different tensor components.

\orb---despite exceptional performance in structural stability, density accuracy, and bond length prediction---exhibits failure across all elastic tensor components, with MAPE values consistently exceeding 80\% and reaching 100\% for C$_{66}$ (\Cref{fig:elastic-plots}g). This failure is particularly notable given that \orb{} directly predicts forces rather than computing gradients of energy functions, which may compromise its ability to accurately capture the second-order derivatives of the potential energy surface required for elastic property prediction. \uma{}---trained on one of the largest datasets ($\sim$500M structures from OMat24/OMol25)---exhibits some of the poorest predictions for Young's modulus (MAPE: 79.2\%) and shear modulus (MAPE: 77.3\%), with accuracy better only to \orb{}. The result suggests limitations in current training paradigms: while larger datasets effectively sample configurational space to prevent catastrophic failures, they may not adequately capture the subtle energy-strain relationships and anharmonic effects critical for elastic property prediction. This finding reinforces our central thesis that broader validation on mechanical properties reveals model limitations orthogonal to those assessed by benchmarks focusing on energy and forces, and that achieving reliable mechanical property predictions may require training objectives explicitly incorporating elastic response data or physics-informed constraints beyond energy and force matching.

Parity plot analysis reveals extensive scatter and systematic deviations across all evaluated models. Shear modulus predictions (\Cref{fig:elastic-plots}b) and Young's modulus predictions (\Cref{fig:elastic-plots}c) (see \Cref{meth:elastic} for details) show substantial errors, while individual elastic tensor components (\Cref{fig:elastic-plots}d-g) demonstrate poor accuracy with MAPE values ranging from 22-46\% even for the best-performing models. The addition of \uma{} and \maceomat{} to our evaluation confirms that models trained on vastly larger datasets exhibit similar limitations in elastic property prediction, suggesting that current training paradigms may not adequately capture the subtle energy-strain relationships required for accurate mechanical property prediction. \chgnet{} and \mgnet{} provide virtually no reliable predictions due to poor simulation stability. The complete performance metrics for each model, including the $R^2$ and MAPE (\%), is provided in the Supplementary Table 10).

\subsection{Failure Analysis}

\begin{figure}[!ht]
    \centering
    \includegraphics[width=0.95\textwidth]{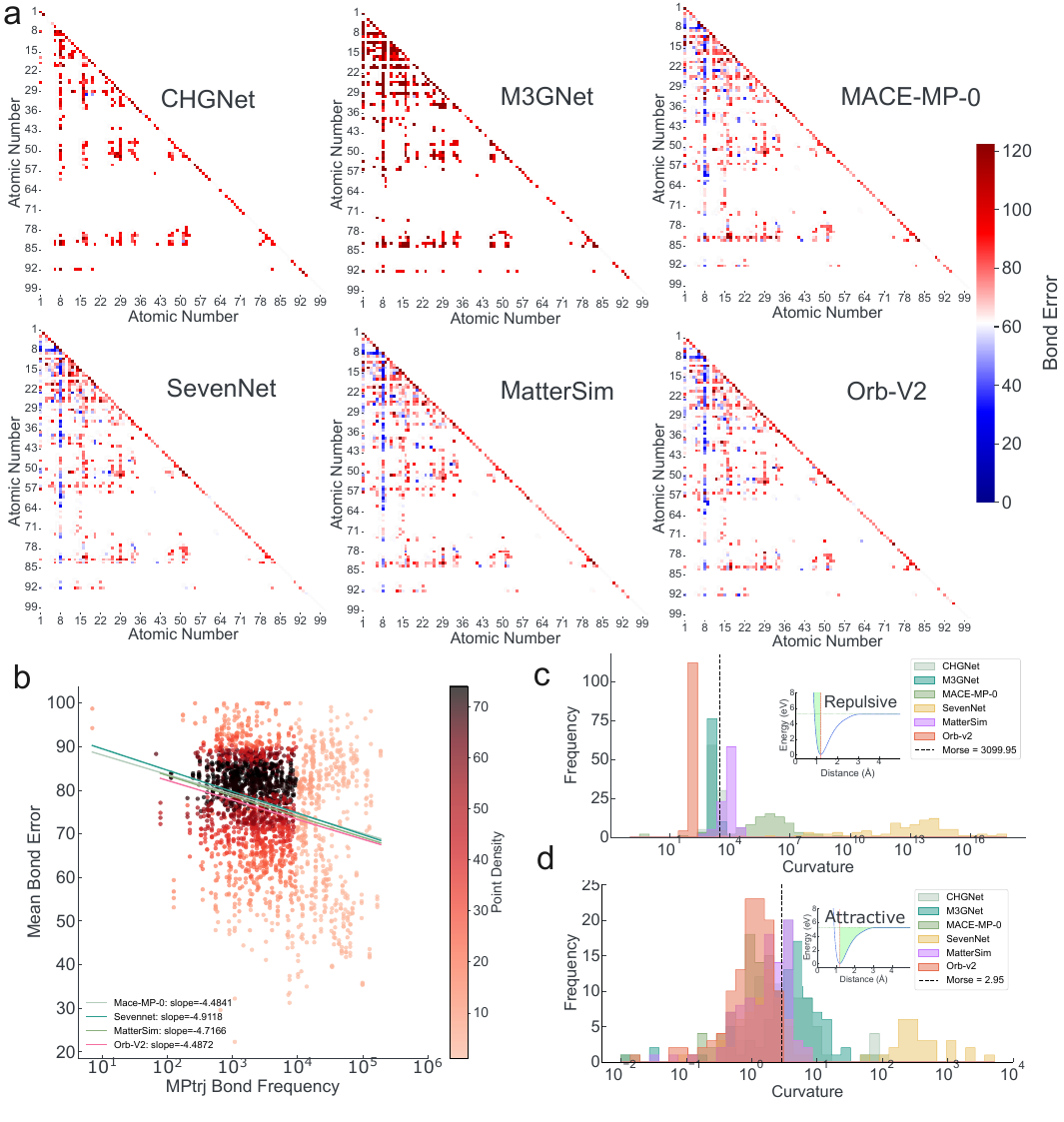}
    \caption{\textbf{Bond length accuracy reveals systematic training data bias in UMLFFs.} \textbf{a}, Comprehensive bond error heatmaps across all atomic pair combinations for each model, revealing systematic patterns in prediction accuracy related to training data composition and bonding type representation. Specifically, low bond error is observed for every element bonded with oxygen confirming the training data bias in the learned interactions. \textbf{b}, Mean bond error versus frequency in MPtrj training dataset, with color density indicating data point concentration. Correlation between MPtrj frequency and mean bond error, albeit noisy, demonstrates that frequently encountered atomic pairs in training data mostly exhibit lower errors, while rare pairs show substantially higher errors. \textbf{c,d}, Frequency versus curvature plot of attractive and repulsive regimes of homo-nuclear (X–X) interaction.}
    \label{fig:bonderror-plots}
\end{figure}

To understand the origins of observed performance limitations, we analyzed training data characteristics, pairwise force-displacement interactions, and their connections to simulation failures and elastic property prediction (full analysis in Supplementary Section 9). Bond length accuracy serves as the primary diagnostic, as interatomic distances govern chemical reactivity, phase stability, and mechanical response.

Atomic pair analysis across all models reveals a systematic oxygen bias: bonds involving oxygen display consistently low errors across all evaluated UMLFFs, manifesting as distinct vertical bands in the error heatmaps (Figure~\ref{fig:bonderror-plots}a; Supplementary Section 9.1). MPtrj and its derivatives are predominantly composed of oxide-based systems, so models trained on these datasets excel at oxygen-containing pairs while struggling with less represented elemental combinations. Quantifying this bias through the correlation between bond error and atomic pair frequency in MPtrj (Figure~\ref{fig:bonderror-plots}b and Supplementary Figure 11) confirms that frequently encountered pairs exhibit lower errors, while underrepresented pairs suffer substantially higher errors. Pair frequency is, however, a necessary condition for accuracy rather than a sufficient one: some high-frequency transition metal pairs still exhibit large errors, attributable to the selective application of Hubbard+$U$ corrections to metals (V, W, Fe, Ni, Co, Cr, Mo, Mn) only in the presence of O or F~\cite{warford2026better}. This creates two incompatible potential energy surfaces in the reference DFT data---a low-energy PBE surface and a high-energy PBE+$U$ surface---forcing MLIPs to interpolate between disjoint surfaces when O or F atoms enter metal coordination spheres. The resulting non-physical repulsive walls propagate into energy, force, and derived property predictions. Comparing UMLFF predictions against DFT values from the Materials Project for a subset of 10 minerals makes this concrete: UMLFFs exhibit average density errors of 5.2\%, roughly 2--3$\times$ larger than the 2.5\% DFT errors on the same structures, and shear modulus errors for Calcite (188.6\%) and Lead ($-$537.2\%) dwarf the corresponding DFT errors of 2.0\% and 14.7\%, respectively. Models trained to near-DFT energy accuracy thus fail to reproduce even their training data's accuracy for derived properties, revealing a fundamental gap between interpolating energy-force landscapes and capturing higher-order physical responses.

Curvature analysis of learned pairwise interactions exposes the mechanistic pathway from these data limitations to simulation failure (Figure~\ref{fig:bonderror-plots}c,d; Supplementary Section 9.2). Interaction smoothness, quantified by the mean absolute curvature of the force-displacement curve relative to the analytically smooth Morse potential, correlates directly with MD stability. \chgnet{} and \mgnet{} exhibit curvature values 10$^2$--10$^3$ times the Morse baseline in the repulsive regime, producing noisy force-displacement curves that require prohibitively small integration timesteps, explaining their $>$85\% failure rates. \orb{}, by contrast, approaches Morse-level smoothness throughout, consistent with its perfect trajectory completion. In the attractive regime, most models behave comparably to classical potentials, indicating that short-range repulsive interactions are the primary instability source.

Smooth force-displacement curves do not, however, guarantee accurate force gradients. Elastic tensor prediction requires second derivatives of the potential energy surface ($\partial^2 E / \partial \varepsilon^2$), and models can interpolate forces accurately while misrepresenting their gradients entirely. This explains why C$_{11}$ components, sensitive to bulk compressibility, are predicted more accurately than shear components C$_{44}$ and C$_{66}$, and why \orb{}'s structural stability does not extend to elastic accuracy. Training data bias toward oxide systems compounds this: models have encountered little elastic deformation physics in diverse chemical environments, so current ``universal'' force fields function largely as interpolation schemes within familiar chemical spaces rather than as physical models with genuine transferability.

\section{Discussion}

Our systematic experimental validation reveals a gap between performance on conventional computational benchmarks and real-world applicability of UMLFFs. Cross-referencing formation energy $R^2$ scores from MatBench Discovery against our experimentally measured $R^2$ values for Young's modulus, shear modulus, density, and lattice parameters reveals poor correlation across all models (Supplementary Figure 10), demonstrating that strong performance on training objectives does not transfer to experimentally measured properties. This disconnect reflects a training-evaluation circularity inherent to current practice: models learn to reproduce DFT predictions and are validated against DFT-derived properties, with limited assessment of whether those predictions align with measurable material behavior (see Supplementary Section 14).

Three architectural fault lines underlie the observed failure modes (see Supplementary Section 13). Invariant models such as \orb{} and equivariant models such as \mace{} and \sevennet{} both achieve structural stability, indicating that the bottleneck lies in training objectives rather than symmetry treatment. \orb{}'s direct force prediction, enabled by denoising diffusion pretraining, yields smooth potential energy surfaces and perfect trajectory completion, yet fails for elastic properties because elastic constants require second energy derivatives ($\partial^2 E / \partial \varepsilon^2$) that cannot be recovered when forces do not derive from a consistent energy function. Scaling training data from $\sim$1.5M structures (MPtrj) to $\sim$500M (OMat24/OMol25) improves simulation stability markedly, as the MACE versus \maceomat{} comparison illustrates, yet density errors decrease by only 4--5$\times$ and remain above the $\pm$2.5\% threshold required for practical application. Current training paradigms, focused on energy and force accuracy at 0~K, do not capture the physics needed for property prediction at finite temperature, including thermal expansion, anharmonic effects, and the energy-strain relationships governing elastic response.

\uniffbench{} breaks this circularity by grounding evaluation in 1,500+ experimental mineral structures assessed under extreme thermodynamic conditions, partial occupancy, and finite-temperature equilibrium, conditions that are either computationally prohibitive at scale or systematically excluded from existing training and evaluation datasets. The benchmark's findings point toward four concrete directions for next-generation UMLFFs (see Supplementary Section 15). First, training protocols should simultaneously optimize energy, forces, and stresses alongside experimental properties, explicitly constraining higher-order derivatives through targets such as elastic tensor components, phonon dispersions~\cite{gangan2025force,thaler2021learning}, and Born stability criteria. Second, \orb{}'s smooth pairwise interactions demonstrate the value of direct force prediction for simulation stability, but recovering elastic accuracy will require either explicit higher-order derivative supervision or property-specific fine-tuning. Third, achieving universal capability requires environmental diversity in training data, not only chemical diversity: atomic pairs must be encountered across varied coordination environments, bonding configurations, and thermodynamic conditions, with balanced inclusion of partial occupancies and compositions exceeding ten elements. Fourth, experimental validation should sit alongside computational benchmarks as standard practice, with community-wide adoption of standardized simulation completion reporting, application-specific accuracy thresholds, and computational resource transparency. While UniFFBench demonstrates a step toward experimentally grounded evaluation of UMLFFs, several limitations remain. First, the benchmark is currently limited to naturally occurring minerals, which may limit its generalizability to other technologically important material classes, such as alloys, glasses, amorphous systems, and two-dimensional materials. Second, the evaluation of mechanical properties focuses primarily on the elastic tensor and could potentially extend to other experimentally measured properties, such as thermal expansion coefficients, viscosity, thermal conductivity, and electrical conductivity, to enable a more comprehensive assessment of model performance. Finally, although this study shows the gap between DFT and experimentally measured properties, a systematic benchmark is needed to quantify gap and understand the differences across diverse material systems, which remains an important direction for future research.

\section{Methods}\label{sec_methods}

\subsection{Data collection and pre-processing}
\label{meth:Data}
To obtain a broad range of minerals covering the periodic table, we curated experimentally determined crystal structures from the literature and primarily from the American Mineralogist Crystal Structure Database (AMCSD) ~\citep{downs2003american}. Although the database offers a wide range of CIF files, several inconsistencies were observed, including incomplete metadata, inconsistent element naming conventions, and variations in space group notation, particularly between the Hermann--Mauguin and Hall notations ~\citep{hahn1983international, burzlaff2016hermann, aroyo2021international}. To address this, we manually standardized the space group representations across all structures to ensure successful parsing using ASE \citep{larsen2017atomic} internal space group parser. After this correction, we explicitly screened all CIF files to identify and exclude those with partial occupancies or structural defects, which are unsuitable for atomistic simulations. This filtering resulted in a curated set of 1,343 stoichiometric and structurally consistent mineral structures (see Supplementary Data 1). Further, we selected a subset of 75 systems, designated as \minxhtp. This subset encompasses a range of configurations, including both high-temperature/high-pressure as well as low-temperature/low-pressure, enabling a systematic assessment of the UMLFFs' robustness and transferability under extreme thermodynamic conditions. The specific details with temperature and pressure conditions for a selected few systems are provided in (Figure \ref{fig:data-distribution}d).

In addition to the \minxeq{} and \minxhtp{}, we selected additional 50 mineral CIF files containing partial occupancies named as \minxpocc{}. To enable realistic atomistic simulations of partially occupied structures, we computed the least common multiple (LCM) of the denominators of all occupancy fractions to determine the necessary supercell size. We then identified three integer factors \((n_x, n_y, n_z)\) such that \(n_x \cdot n_y \cdot n_z = \text{LCM}\), with the aim of keeping the supercell dimensions as isotropic as possible. Atomic species were assigned to sites by random sampling derived from the site occupancy probabilities, ensuring that each site was occupied by only one atom type, thereby generating a fully ordered structure consistent with the original occupancy statistics. The resulting supercell structures were saved in XYZ format and used as input for MD simulations of partially occupied systems.

Furthermore, since elastic tensor data were not available for the majority of the minerals curated from AMCSD, we developed a dataset by extracting both structural and elastic information from the Materials Property Open Database (MPOD) ~\citep{mpod_web}. Specifically, we curated 100 CIF files along with their corresponding elastic tensor data using a Python-based API interface. Metadata including the mineral name, chemical formula, and literature reference for all four datasets: \minxeq{}, \minxhtp{}, \minxpocc{}, and \minxem{} are provided in the GitHub repository (\cref{sec:github}).The complete datasets corresponding to each category are publicly available via Zenodo (\cref{sec:data}).

% \vspace{0.2in}
\subsection{MD Simulation}
\label{meth:MD}
\minx{} crystallographic information files (CIFs) were manually curated and validated during extraction to ensure appropriate experimental conditions (including temperature and pressure), and physicochemical accuracy. These structures were subsequently preprocessed for compatibility with the ASE package API \citep{larsen2017atomic}. The simulation protocol implemented systematic standardization of the simulation supercell: unit cells were replicated to achieve system sizes of 100-200 atoms, with exceptions for structures inherently exceeding this threshold. In the latter case, the original size was maintained. Spatial replication proceeded sequentially along ascending lattice vectors to optimize toward cubic supercell (similar size in all three directions) while preserving crystallographic integrity and minimizing anisotropic effects. 

The computational workflow incorporated a dual-phase equilibration strategy. Initial structural minimization  utilized the  FIRE algorithm \citep{bitzek2006structural} for 1000 steps, followed by a 50 ps NPT equilibration phase. Phase-space sampling using MD simulations was initiated with Maxwell-Boltzmann velocity distributions at experimentally determined temperatures from CIF metadata, with a canonical temperature of 298 K applied for unspecified cases based on the reference literature. The NPT equilibration implemented the Berendsen thermostat and barostat \citep{berendsen1984molecular}, maintaining experimentally reported pressures or a standard state pressure of 1 atm. Production MD runs were executed for 50 ps with an integration timestep of 1 fs, capturing trajectory and thermodynamic data at 10-step intervals. Detailed information regarding the hardware specifications and computational costs is provided in Supplementary Sections 4 and 5.

We employ 50 ps NPT molecular dynamics simulations, balancing statistical convergence with computational feasibility for a benchmark comprising ~8,000 total simulations across 1,500+ systems and 6--8 models. Equilibration analysis demonstrates thermal equilibrium is achieved within 10--15 ps (see Supplementary Figure 6). For structural properties reflecting fast thermal fluctuations, averaging over the remaining snapshots (10 fs sampling) provides reasonable statistics for comparative evaluation. We acknowledge that longer simulations (500--5000 ps) would improve phase-space sampling but would increase computational cost by 10--100$\times$ (365,000--3.65 million CPU days total). Supplementary Section 3 shows detail computational analysis and efficiency metrics for each model (see Supplementary Table 1).

\subsection{Post-processing and analysis of simulation}
\label{meth:post-process}
To analyze the MD simulations, both the trajectory and log files from each simulation were post-processed. All post-processing analyses were carried out exclusively on the equilibrated portions of the MD trajectories, with the initial energy-minimization steps excluded from results. In particular, the density evolution during the simulation was evaluated by extracting atomic configurations at regular intervals, corresponding to a dump frequency of every 10 MD steps. At each extracted frame, the instantaneous density was computed using the mass-to-volume ratio of the simulation cell, expressed in physical units (g/cm$^3$). The total atomic mass was calculated by summing the atomic masses of all atoms in the simulation cell and simultaneously, the volume of the simulation cell was obtained. The density at each frame was then computed as the ratio of mass (in grams) to volume (in cm$^3$). This computation was performed for each dumped frame, and the final reported density corresponds to the time-averaged value over the entire 50 ps MD trajectory. Similarly, the lattice parameters were computed at each dumped frame. Specifically, the simulation cell was extracted for each frame using ASE’s built-in cell analysis tools. The lengths of the lattice vectors parameters (a, b, c) were recorded at each dumped frame and subsequently averaged over the entire MD trajectory to obtain the equilibrium lattice constants.

To evaluate the structural accuracy of the simulated configurations, the radial distribution function (RDF) and bond length distribution were calculated. RDFs were computed using a maximum cutoff radius of $r_{\text{max}} = 6$~\AA{} and a bin width of $\Delta r = 0.01$~\AA{}. The RDF was averaged over the last 100 dumped frames of the trajectory. Given the dump frequency of 10 steps per frame, this corresponds to the last 1000 MD steps or approximately 1~ps of the simulation. For bond length analysis, a similar protocol to RDF calculation was employed. However, instead of the total RDF, partial RDFs for each atomic pair were computed. The bond length for each pair was identified as the position of the first peak in the corresponding partial RDF. This was achieved by locating the index $i_{\text{peak}}$ where the RDF $g(r_i)$ reaches its maximum within the cutoff radius:$i_{\text{peak}} = \arg\max\limits_{i < n_{\text{cut}}} g(r_i),$ .Here, $g(r_i)$ is the RDF value corresponding to $(r_i)^{\text{th}}$ distance value. The corresponding bond length is then given by $r_{\text{peak}} = r[i_{\text{peak}}]$. Furthermore, to ensure smooth and physically meaningful experimental RDF and partial RDF profiles, the initial atomic configuration was subjected to small random perturbations. Specifically, Gaussian noise with a standard deviation $\sigma = 0.005$~\AA{} was added to the atomic positions. This perturbation procedure was repeated 1000 times, and the RDFs and partial RDFs were calculated for each perturbed structure and averaged over all 1000 configuration. Notably, it is worth highlighting that the trajectory and log files generated from MD simulations are often large in size, posing significant challenges in terms of both data storage and computational cost for post processing. In particular, the evaluation of RDF and partial RDF across multiple frames substantially increases the computational time and cost for post-processing. To address this, we developed a parallelized post-processing script which loads the MD trajectory and log files for analysis and return  key structural metrics, including density, lattice, RDF, and bond length, and stores them in structured CSV files. Further, these CSV files can be loaded into a notebook for better understanding of the metrics and plotting. Overall, the use of parallel computing substantially accelerates the analysis pipeline, thereby making the workflow scalable and tractable for extensive simulations.

\subsection{Elastic Tensor Computation}
\label{meth:elastic}
To determine the elastic constants (\(C_{ij}\)), we employed a stress–strain approach based on finite deformations. A small maximum strain value of \(\varepsilon = 1 \times 10^{-4}\) was selected to ensure linear elastic behavior. For each of the six independent strain components in Voigt notation (\(\varepsilon_{xx}\), \(\varepsilon_{yy}\), \(\varepsilon_{zz}\), \(\varepsilon_{yz}\), \(\varepsilon_{xz}\), \(\varepsilon_{xy}\)), twenty linearly spaced strain values were generated in the range \(-\varepsilon\) to \(+\varepsilon\). For each strain value, a corresponding deformation matrix was constructed and applied to the atomic configuration by modifying the unit cell. After applying the deformation, the atomic positions were relaxed using the Fast Inertial Relaxation Engine (FIRE) algorithm  until the maximum atomic force were less than \(0.05\,\text{eV}/\text{\AA}\) or a maximum of 1000 optimization steps were completed, whichever reached first. The stress tensor for each relaxed, strained configuration was then computed in Voigt notation. A reference stress was also calculated for the initial, unstrained configuration. Finally, we performed a linear regression of the computed stress components against the applied strain values. The slope of each stress–strain curve corresponds to the associated elastic stiffness constant, resulting in a full \(6 \times 6\) elastic stiffness tensor \(C_{ij}\).

\subsection{Error metrics}
\label{meth:error}
To evaluate the accuracy of a force field in MD simulations, we use different quantitative error metrics. These metrics provide insights into how well a force field can reproduce various physical properties such as equilibrium density, lattice structure, atomic arrangements, and bond lengths. Below are the definitions and physical interpretations of each metric and corresponding error. If any error value exceeds 100\%, the simulation is classified as failed, even if it completes the full 50 ps MD run.

\textbf{Density Error.} This error measures the deviation in average density during the MD simulation and is an indicator of how accurately the potential captures equilibrium properties.
\begin{equation}
\text{Density Error} = \left( \frac{\rho_f - \rho_i}{\rho_i} \right) \times 100
\label{eq:density_error_eq}
\end{equation}
where $\rho_f$ and $\rho_i$ are the final and initial densities, respectively. 

\textbf{Lattice Error} This error quantifies the change in the lattice parameters and reflects the structural stability maintained by the potential during simulation.
\begin{equation}
\text{Lattice Error} = \left( \frac{a_f - a_i}{a_i} \right) \times 100
\label{eq:lattice_error_eq}
\end{equation}
where $a_f$ and $a_i$ are the final and initial lattice constants.

\textbf{RDF Error} This error evaluates the discrepancy in radial distribution functions (RDF) between the reference (ground truth) and the simulated structure, thus characterizing short-range atomic order.
\begin{equation}
\text{RDF Error} = \frac{ \sum_{i=1}^{n} \left( g(r_i) - g_{\text{ref}}(r_i) \right)^2 }{ \sum_{i=1}^{n} \left( g_{\text{ref}}(r_i) \right)^2 }
\label{eq:rdf_error_eq}
\end{equation}
where $g(r)$ is the RDF from simulation and $g_{\text{ref}}(r)$ is the reference RDF.

\textbf{Bond Error} This error captures the relative change in bond lengths, providing insight into the force field's ability to characterize phase change behavior and thermal stability.
\begin{equation}
\text{Bond Error} = \left( \frac{L_{\text{current}} - L_{\text{initial}}}{L_{\text{initial}}} \right) \times 100
\label{eq:bond_error_eq}
\end{equation}
where $L_{\text{initial}}$ and $L_{\text{current}}$ are the initial and current bond lengths.

\textbf{Elastic Error} This metric quantifies the accuracy of mechanical property predictions and captures the anisotropic elastic response of the material. More importantly, it assesses the performance of UMLFFs in reproducing the second derivatives of the potential energy surface, thereby reflecting the fidelity with which the force field captures the underlying energy landscape.
\begin{equation}
\textrm{Elastic Error}_{ij} = \left(\sqrt{ \frac{1}{n} \sum_{k=1}^{n} \left( C_{ij}^{(k)} - \hat{C}_{ij}^{(k)} \right)^2 } \right) 
\label{eq:elastic_error_eq}
\end{equation}
where \( C_{ij}^{(k)} \) and \( \hat{C}_{ij}^{(k)} \) are the ground truth and predicted elastic constants for the \(k^\text{th}\) material, respectively.

\subsection{Voigt Averaging Method}
We employed the Voigt average method to calculate the shear and Young's moduli of minerals. This method assumes a uniform strain distribution across the material, which typically results in an overestimation of the material's stiffness. Despite this limitation, it remains widely acceptable by the community due to its simplistic nature. In the Voigt approach, the shear modulus \( G_V \) and Young's modulus \( E_V \) are computed from the stiffness tensor components \( C_{ij} \) by averaging the elastic contributions from all crystallographic orientations. The bulk modulus is similarly derived using standard elasticity relations based on the stiffness constants. The Voigt average bulk modulus \( K_V \) and shear modulus \( G_V \) are mathematically expressed as:

\begin{equation}
K_V = \frac{1}{9} \left( C_{11} + C_{22} + C_{33} + 2(C_{12} + C_{13} + C_{23}) \right)
\label{eq:bulk_modi_eq}
\end{equation}

\begin{equation}
G_V = \frac{1}{15} \left( C_{11} + C_{22} + C_{33} - C_{12} - C_{13} - C_{23} + 3(C_{44} + C_{55} + C_{66}) \right)
\label{eq:shear_modi_eq}
\end{equation}

Using the bulk and shear moduli obtained, Young's modulus \( E_V \) can be calculated using the standard isotropic elasticity relations:

\begin{equation}
E_V = \frac{9K_V G_V}{3K_V + G_V}
\label{eq:youngs_modi_eq}
\end{equation}

where \( K_V \), \( G_V \), and \( E_V \) are the Bulk Modulus, Shear Modulus and Young's Modulus respectively.

\subsection{Success Rate (\%) Calculation}
The success rate for each model was computed as a weighted average of the fraction of successful completions across three datasets: \minxeq{}, \minxhtp{}, and \minxpocc{}. The success rate (\%) is defined as:
\begin{equation}
\textrm{Success Rate (\%)} = \frac{{\sum_{i=1}^{3} w_i \cdot s_i}}{\sum_{i=1}^{3} w_i}\
\label{eq:success_rate_eq}
\end{equation}

where, \( w_i \) represents the number of mineral in dataset \(i\), \( s_i \) denotes the fraction of successful simulations in  dataset \(i\), with less than 100(\%) error in density and lattice parameter. 

\subsection{Pair-wise forcefield analysis}
To evaluate the pairwise interactions captured by the UMLFFs, we generated element-specific datasets consisting of two-atom configurations. In each configuration, one atom was fixed at the origin while the second atom was displaced along a linear path (x-direction) in increments of 0.1\,\AA, ranging from a near-contact of 0.1\,\AA~to max-distance up to 5\,\AA. For each separation distance, energies and forces were computed using the corresponding UMLFFs via the ASE calculator interface. These pairwise potentials for each element are presented in the Supplementary Sections 11 and 12. Additionally, a similar analysis was performed for heteronuclear pairs involving oxygen to assess element-specific interactions with oxygen across varying distances.

Furthermore, to quantitatively assess the interaction behavior learned by the UMLFFs, we performed a detailed curvature analysis of the pair potential curves. Specifically, we identified the minimum of each pairwise potential and segmented the curve into two distinct regions based on minima: the \textit{repulsive regime} (preceding the minimum) and the \textit{attractive regime} (following the minimum). This segmentation enables a more explicit investigation of the force field’s performance across different interaction regimes. For each region, we computed the absolute mean of the local curvature of the potential curve at each point, providing a measure of the steepness and nature of interaction. The mathematical formulations of the curvature calculation are as follows. For all models except \orb{}, the curvature was computed as the second derivative of the pairwise potential energy curve \( V(r) \) with respect to \(r\)  and in case of \orb{} the curvature was estimated from the first derivative of the force–distance curve. The mean absolute curvature in the repulsive and attractive regions of the pairwise potential is computed as:

\begin{equation}
\bar{\kappa}_{\text{repulsive}} = \frac{1}{N_{\text{before}}} \sum_{r_i < r_{\min}} \left| \frac{d^2 V(r_i)}{dr^2} \right|,
\quad
\bar{\kappa}_{\text{attractive}} = \frac{1}{N_{\text{after}}} \sum_{r_i > r_{\min}} \left| \frac{d^2 V(r_i)}{dr^2} \right|
\label{eq:pair_pot_eq}
\end{equation}

where:

\( r_{\text{min}} \) denote the interatomic distance corresponding to the minimum of the potential:
\[
r_{\text{min}} = \arg\min_{r \in [r_1, r_2]} V(r),
\]
where \( [r_1, r_2] \) is a cutoff range choosen to find minima. Here we kept 1\,\AA \ to 3\,\AA. 

\( \bar{\kappa}_{\text{before}} \) : denotes the mean absolute curvature in the repulsive region,\\ 
\( \bar{\kappa}_{\text{after}} \) : denotes the mean absolute curvature in the attractive region. \\
\( N_{\text{before}} \) and \( N_{\text{after}} \) represent the number of data points before and after 
the minimum distance \( r_{\min} \), respectively. 

Further, to assess the deviation of UMLIFFs from classical interatomic potentials, we performed the same curvature analysis on the oxygen--oxygen (O--O) interaction using the Morse potential ~\citep{matsumoto2002introduction}, which is mathematically defined as:
\begin{equation}
V(r) = D_e \left(1 - e^{-a(r - r_e)}\right)^2
\label{eq:morse_pot_eq}
\end{equation}
where \( D_e \) is the bond dissociation energy, \( r_e \) is the equilibrium bond distance, and \( a \) controls the width of the potential well. Finally, these curvature values were then aggregated across all element pairs, and visualized as histograms of frequency versus curvature for both repulsive and attractive regions. The resulting distributions are shown in Figures~\ref{fig:bonderror-plots}(b--c).

\section{Data availability}\label{sec:data}
The crystal structure files for all mineral species are publicly accessible and can be downloaded from \cite{MinX Dataset}. Source data for Figures 2-6 are available with this manuscript. Source data for supplementary figures is available as Supplementary Data 2.
\section{Code availability}\label{sec:github}
The \uniffbench{} framework code can be accessed from the Github repo: \url{https://github.com/M3RG-IITD/UniFFBench}. A version of the code corresponding to this manuscript is archived on Zenodo  \cite{UniFFBench-v1.0.1}.

% \url{https://doi.org/10.5281/zenodo.20165458}

\section*{Acknowledgements}
N. M. A. Krishnan and S. Miret acknowledge financial support for this research from Intel. The authors thank the High Performance Computing (HPC) facility at IIT Delhi for providing the computational and storage resources used in the post-processing of the simulations. We also thank Ayush Maheshwari for providing GPU computing resources and support with running the simulations.  N. M. A. Krishnan acknowledges the support from Alexander von Humboldt foundation and Anusandhan National Research Foundation through the grant number ANRF/ARG/2025/007405/ENS. S. Mannan acknowledges financial support from the Prime Minister's Research Fellowship (PMRF), Ministry of Education, Government of India.

\section*{Author contributions}

\noindent S. Mannan and N. M. A. Krishnan\ jointly proposed the idea of benchmarking MLIPs against experimental data; S. Mannan, V. Bihani, C. Gonzales, and K. L. Kelvin Lee\ led the implementation of different models and developed the \uniffbench{} framework. N. N. Gosvami, S. Ranu, and S. Miret contributed to the analysis of results and provided scientific insights. Further, N. M. A. Krishnan and S. Miret acquired the funding, administered the project, provided resources, and supervised the work. All authors reviewed, edited, and approved the final version of the manuscript.

\section*{Competing interests}
The authors declare no competing interests.

\clearpage
% \appendix
% \clearpage
% \beginsupplement

% \begin{center}
%     % {\Large {\textbf{Supplementary Information}}}\\[1.5em]

%     {\Large {UniFFBench: Evaluating Universal Machine Learning Force Fields Against Experimental Measurements}}\\[2em]

%     % \normalsize
%     Sajid Mannan\textsuperscript{1}, 
%     Vaibhav Bihani\textsuperscript{2}, 
%     Carmelo Gonzales\textsuperscript{3}, 
%     Kin Long Kelvin Lee\textsuperscript{3},\\
%     Nitya Nand Gosvami\textsuperscript{4}, 
%     Sayan Ranu\textsuperscript{2,5}, 
%     Santiago Miret\textsuperscript{3,*}, 
%     N M Anoop Krishnan\textsuperscript{1,2,*}\\[2em]

%     \textsuperscript{1} Department of Civil Engineering, Indian Institute of Technology Delhi, Hauz Khas, New Delhi, 110016, Delhi, India\\
%     \textsuperscript{2} Yardi School of Artificial Intelligence, Indian Institute of Technology Delhi, Hauz Khas, New Delhi, 110016, Delhi, India\\
%     \textsuperscript{3} Intel Labs, California, USA\\
%     \textsuperscript{4} Department of Materials Science and Engineering, Indian Institute of Technology Delhi, Hauz Khas, New Delhi, 110016, Delhi, India\\
%     \textsuperscript{5} Department of Computer Science and Engineering, Indian Institute of Technology Delhi, Hauz Khas, New Delhi, 110016, Delhi, India\\[2em]
%     *Corresponding authors: santiago.miret@intel.com, krishnan.iitd@ac.in
% \end{center}

% \vspace{2em}

% \clearpage
\appendix

\setcounter{section}{0}
\renewcommand{\thesection}{\arabic{section}}
\renewcommand{\thesubsection}{\thesection.\arabic{subsection}}

\makeatletter
\renewcommand\@seccntformat[1]{%
  \ifnum\pdfstrcmp{#1}{section}=0
    Supplementary Section~\csname the#1\endcsname:\quad
  \else
    \csname the#1\endcsname\quad
  \fi
}
\makeatother

% Section numbering

% \renewcommand{\thesection}{\arabic{section}}
% \setcounter{section}{0}

\renewcommand{\figurename}{Supplementary Figure}
\renewcommand{\tablename}{Supplementary Table}

\setcounter{figure}{0}
\setcounter{table}{0}

% \renewcommand{\thefigure}{\arabic{figure}}
% \renewcommand{\thetable}{\arabic{table}}
% \renewcommand{\thetable}{Supplementary\arabic{table}}
% \setcounter{table}{0}  % Optional: reset table counter to 0
% \renewcommand{\thefigure}{Supplementary\arabic{figure}}
% \setcounter{figure}{0}

% \vspace{1em}

% \section*{Table of Contents}
% \vspace{1em}
% \noindent
% S1. Hardware Specifications for Molecular Dynamics Simulations \dotfill 33 \\
% S2. Hardware Specifications for Elastic Tensor Calculations \dotfill 33 \\
% S3. Dataset Distribution and Composition Details \dotfill 39 \\
% S4. Parity Plot for Predicted vs. Reference Densities \dotfill 44 \\
% S5. Parity Plot for Predicted vs. Reference Lattice Parameters \dotfill 45 \\
% S6. Bond Length Prediction Error as a Function of Frequency (MPtraj) \dotfill 51 \\
% S7. Pairwise Energy and Force Comparison: Homonuclear Systems \dotfill 52 \\
% S8. Pairwise Energy and Force Comparison: Heteronuclear Systems \dotfill 52 \\

% \clearpage

\setcounter{page}{1}

\startcontents[appendices]

\begin{center}
{\LARGE\bfseries Supplementary Information\par}
\end{center}
% \section*{Supplementary Information}

\printcontents[appendices]{}{1}{\setcounter{tocdepth}{2}}

% \section{Appendix}
% \label{secA1}

\section{Design Principles for Benchmarking UMLFFs}\label{app_design_principles}
% \label{app:uniffbench_design}
The development of \uniffbench{} was based on several critical design principles that can further guide future benchmarking efforts for UMLFFs. These principles emerged from systematic analysis of the gaps between current evaluation practices and the requirements for reliable experimental deployment.

\begin{enumerate}

\item \textbf{Experimental Grounding}: Traditional UMLFF evaluation relies heavily on DFT-generated test sets that share similar computational origins with training data, creating fundamental circularity where models are evaluated against idealized representations rather than experimental reality. Effective benchmarks should prioritize experimentally determined structures and properties, even when such data contains uncertainties or is more limited than computational datasets.

\item \textbf{Multi-Scale Property Assessment}: Energy and force prediction errors provide insufficient insight into model reliability for practical applications. Comprehensive benchmarks should assess multiple property scales including structural accuracy, dynamic stability, and mechanical properties.

\item \textbf{Realistic Simulation Conditions}: Static, zero-temperature configurations fail to capture the dynamic complexity of real materials applications. Effective benchmarks should evaluate performance under realistic simulation conditions including thermal motion, finite integration timesteps, and long-term stability assessment.

\item \textbf{Chemical Diversity and Bias Quantification}: Benchmarks should systematically probe underrepresented chemical environments including complex compositions (>10 elements), partial occupancy systems, and extreme thermodynamic conditions.

\item \textbf{Failure Mode Identification and Transparency}: Benchmarks should systematically characterize failure modes since our analysis reveals that simulation failures occur without clear warning indicators. Effective benchmarks should quantify failure rates, identify systematic patterns, and provide diagnostic tools such as the curvature analysis for recognizing potentially unreliable predictions.

\item \textbf{Practical Accuracy Thresholds}: Academic benchmarks often lack connection to practical requirements. Benchmarks should establish and report against application-specific accuracy thresholds rather than relative rankings, providing actionable feedback for model development.

\item \textbf{Temporal Stability Assessment}: Most evaluations assess instantaneous properties but ignore temporal evolution. Benchmarks should assess convergence behavior and long-term stability through temporal evolution analysis rather than just initial accuracy metrics.

\item \textbf{Computational Resource Transparency}: Benchmark results should include computational cost and resource requirements. This enables cost-benefit analysis for practical deployment decisions alongside accuracy assessment.

\end{enumerate}

\subsection{Implementation Guidelines}

Based on these principles, we recommend that future UMLFF benchmarks should incorporate: (1) experimental reference data prioritized over computational references; (2) multi-metric evaluation including simulation completion rates, structural accuracy, and mechanical properties; (3) systematic failure reporting with diagnostic tools; (4) chemical diversity metrics quantified relative to training data; (5) practical accuracy standards based on application requirements; (6) temporal evolution analysis over simulation timescales; and (7) computational cost documentation for deployment decisions.

\section{Description of UMLFFs Studied}\label{app_model_details}

\subsection{CHGNet (Crystal Hamiltonian Graph Neural Network)}

CHGNet is a graph neural network-based universal machine learning interatomic potential~\cite{deng2023chgnet}. The model is pretrained on energies, forces, stresses, and magnetic moments from the Materials Project Trajectory Dataset, consisting of over 1.5 million inorganic structures spanning ~10 years of DFT calculations. CHGNet's key innovation lies in its explicit inclusion of magnetic moments as charge constraints, enabling the model to learn and accurately represent orbital occupancy of electrons and enhancing its capability to describe both atomic and electronic degrees of freedom. The charge-informed approach allows CHGNet to differentiate ionic states and capture charge distribution effects that are crucial for materials with complex electronic structures.

\textbf{Architecture}: Graph neural network with charge-informed features \\
\textbf{Training Data}: Materials Project Trajectory Dataset ($\sim$1.5M structures, 146k compounds) \\
\textbf{Key Features}: Magnetic moment prediction, charge-informed modeling \\
\textbf{Limitations}: High computational cost, poor simulation stability observed in this study

\subsection{M3GNet (Materials Graph Network with 3-body interactions)}

M3GNet is a universal graph deep learning interatomic potential~\cite{chen2022universal}. The model incorporates three-body interactions within a graph neural network framework and was trained on the massive database of structural relaxations from the Materials Project spanning the past decade. M3GNet covers 89 elements of the periodic table and has demonstrated broad applications in structural relaxation, dynamic simulations, and property prediction across diverse chemical spaces. The model has been successfully applied to materials discovery, with screening of 31 million hypothetical crystal structures identifying 1.8 million potentially stable materials.

\textbf{Architecture}: Graph neural network with explicit three-body interactions \\
\textbf{Training Data}: Materials Project Trajectory Dataset ($\sim$1.5M structures, 146k compounds) \\
\textbf{Key Features}: Three-body interaction modeling, broad chemical coverage \\
\textbf{Limitations}: Poor simulation completion rates, high computational requirements

\subsection{MACE (Higher Order Equivariant Message Passing)}

MACE is an equivariant message passing neural network~\cite{batatia2025foundation} that addresses computational limitations of traditional MPNNs through higher-order message passing. The key innovation is the use of four-body messages, which reduces the required number of message passing iterations to just two layers, resulting in a fast and highly parallelizable model. MACE incorporates atomic cluster expansion (ACE) descriptors and maintains SE(3) equivariance while achieving state-of-the-art accuracy on multiple benchmarks. The model demonstrates excellent learning efficiency and has been trained on various datasets including Materials Project data.

\textbf{Architecture}: Equivariant MPNN with higher-order (four-body) messages \\
\textbf{Training Data}: Materials Project Trajectory Dataset ($\sim$1.5M structures, 146k compounds) \\
\textbf{Key Features}: Reduced message passing iterations, high parallelizability, SE(3) equivariance \\
\textbf{Limitations}: Complex architecture, potential scalability issues for very large systems

\subsection{MatterSim}

MatterSim is a deep learning atomistic model~\cite{Yang_MatterSim_A_Deep} specifically designed for materials modeling across elements, temperatures, and pressures. The model serves as a universal MLFF capable of predicting energies, forces, and stresses for structures containing any combination of the first 89 elements under simulation conditions spanning 0-5000 K and 0-1000 GPa. MatterSim is built using M3GNet and Graphormer architectures as backbones and trained through active learning on large-scale first-principles computations. The model achieves up to 10-fold improvement in accuracy compared to universal force fields trained on relaxation trajectories, particularly for high temperature and pressure conditions.

\textbf{Architecture}: Hybrid M3GNet/Graphormer backbone with active learning \\
\textbf{Training Data}: Large-scale first-principles computations across temperature/pressure space \\
\textbf{Key Features}: Temperature/pressure generalization, active learning approach, high accuracy \\
\textbf{Limitations}: Computational complexity, proprietary training methodology

\subsection{SevenNet (Scalable EquiVariance-Enabled Neural NETwork)}

SevenNet is a graph neural network interatomic potential package~\cite{park_scalable_2024} that focuses on scalable parallel molecular dynamics simulations. Built on the NequIP architecture, SevenNet addresses the parallelization challenges of GNN-based interatomic potentials through an efficient parallelization scheme compatible with LAMMPS ~\cite{LAMMPS}. The model achieves over 80\% parallel efficiency in weak-scaling scenarios and exhibits nearly ideal strong-scaling performance with full GPU utilization. SevenNet-0 is the pretrained universal model trained on Materials Project data, demonstrating excellent performance on various material systems including amorphous structures with over 100,000 atoms.

\textbf{Architecture}: NequIP-based equivariant graph neural network with optimized parallelization \\
\textbf{Training Data}: Materials Project Trajectory Dataset ($\sim$1.5M structures, 146k compounds) \\
\textbf{Key Features}: Excellent parallel scalability, LAMMPS integration, efficient multi-GPU support \\
\textbf{Limitations}: Performance degradation with suboptimal GPU utilization, complexity of parallelization

\subsection{Orb}

Orb is a family of universal interatomic potentials~\cite{neumann2024orbfastscalableneural} at Orbital Materials, designed for fast and scalable atomistic modeling. Unlike other UMLFFs, Orb deliberately avoids architectural constraints for SE(3) equivariance, instead learning invariances from data to achieve superior computational efficiency. The model employs a novel two-stage training approach: first training as a denoising diffusion model on ground-state materials, then supervised training as a neural network potential. Orb achieves 3-6 times faster performance than existing universal potentials while maintaining high accuracy, representing a 31\% reduction in error on the Matbench Discovery benchmark upon release.

\textbf{Architecture}: Attention-augmented Graph Network Simulator (non-equivariant) \\
\textbf{Training Data}: Materials Project (MPtrj), Alexandria datasets \\
\textbf{Key Features}: Exceptional speed and scalability, diffusion pretraining, stable molecular dynamics \\
\textbf{Limitations}: Poor elastic property prediction despite excellent structural performance

\subsection{UMA}

UMA is a family of universal foundation models \citep{wood2025umafamilyuniversalmodels} for atomistic simulations designed to push the frontier of accuracy, and generalization across diverse chemical domains. The models are trained on half a billion unique 3D atomic structures, incorporating data from molecules, materials, and catalysts. UMA’s primary innovation is the Mixture of Linear Experts (MoLE) architecture, which allows for a massive increase in model capacity (up to 1.4B parameters) while keeping active parameters low (approximately 50M).

% \rasp{
\textbf{Architecture}: Equivariant GNN (based on eSEN) with Mixture of Linear Experts (MoLE) \\
\textbf{Training Data}: Combined dataset of ~500M structures (OMat24, OMol25, OC20, OMC25, ODAC25) \\
\textbf{Key Features}: Zero-shot generalization across chemical domains, state-of-the-art accuracy on Matbench Discovery \\
\textbf{Limitations}: Scaling laws suggest performance is highly dependent on massive compute for pre-training.

\section{Computational Cost Analysis}\label{compute_cost}

Computational efficiency analysis reveals critical trade-offs between speed and reliability across UMLFFs (Supplementary Table~\ref{table:md-time-per-step}). While raw execution speed varies by only 4$\times$ across models (0.736--2.794 s per MD step), practical efficiency differs by over 53$\times$ when accounting for simulation completion rates. See (\Cref{sec_methods}) for success rate calculation. \orb{} achieves optimal performance with the fastest execution (0.736 s per step) and perfect reliability (100\% completion), followed closely by \mattersim{} (0.780 s per step, 100\% completion). Despite moderate computational speeds, \chgnet{} and \mgnet{} demonstrate practical inefficiency due to failure rates exceeding 85\%, resulting in substantial computational loss. \mace{} and \sevennet{} represent intermediate cases, achieving high reliability (95-97\%) at the cost of increased computational time. To capture these aspects into a single metric, we define efficiency score as the ratio of success rate with time per MD step (see Supplementary Table~\ref{table:md-time-per-step}). An ideal UMLFF should have high success rate with low inference time. We observe that \orb{} exhibits the highest efficiency score followed by \mattersim{} and \mace{}. It is worth noting that the complete \minx{} evaluation required 36,500 CPU days---demonstrating the extensive scale of benchmarking---with efficiency differences translating to order-of-magnitude variations in computational requirements for equivalent scientific outcomes. Details of the model checkpoints used for each UMLFF and the corresponding computation cost for each dataset (\minxeq{}, \minxhtp, and \minxpocc{}) (see Supplementary~\Cref{table:checkpoints,table:minxeq,table:minxhtp,table:minxpocc}). Further, computational time in GPU are provided (see Supplementary Table~\ref{table:md-time-per-step-cpu}).
% Thus, for high-throughput materials discovery, \orb{} and \mattersim{} emerge as optimal choices, combining computational efficiency with reliability.

\begin{table*}[ht]
\begin{spacing}{1.15}
    \centering
    \begin{threeparttable}
    \caption{Computational efficiency metrics for UMLFF models showing execution speed, resource requirements, and practical efficiency accounting for simulation success rates.} 
    \label{table:md-time-per-step}    
    \vspace{0.5mm}
    \begin{tabular}{c|cccccc}
    \hline
    UMLFFs & \chgnet & \mgnet & \mace  & \sevennet & \mattersim & \orb \\ \hline 
    Time (s) per MD step & 1.452 & 2.794 & 1.087 & 2.153 & 0.780 & 0.736 \\ \hline
    Time (hr) per mineral & 6.68 & 4.89 & 36.41 & 38.63 & 11.64 & 12.44 \\ \hline
    Total time (CPU days) & 393.99 & 1295.84 & 10239.48 & 12873.25 & 3908.15 & 4177.96 \\ \hline
    Success rate(\%) & 5.80 & 7.03 & 83.56 & 97.38 & 100.0 & 100.0 \\ \hline
    % Success rate (\%) & 14.6 & 8.3 & 95.2 & 97.1 & 100.0 & 100.0 \\ \hline
    Efficiency score\tnote{1} & 3.99 & 2.52 & 76.87 & 45.22 & 128.2 & 135.9 \\
    \hline
    \end{tabular}%
    \begin{tablenotes}
    % \item[1] Success rate / Time per step $\times$ 10
    \item[1] Success rate / Time per step 
    \end{tablenotes}
    \end{threeparttable}
\end{spacing}
\end{table*}

% \begin{tcolorbox}[colback=gray!5!white,colframe=gray!75!black,title=\uniffbench{} Framework Summary]
% \textbf{Dataset}: 1,500 experimental mineral structures across 4 subsets\\
% \textbf{Models}: 6 state-of-the-art UMLFFs (CHGNet, M3GNet, MACE, MatterSim, SevenNet, Orb)\\
% \textbf{Evaluation}: MD stability, structural accuracy, mechanical properties\\
% \textbf{Key Innovation}: First systematic experimental validation of UMLFFs\\
% \textbf{Code/Data}: Available at \url{https://github.com/sajidmannan/UIP_EVAL}
% \end{tcolorbox}

\section{Hardware Specifications for Molecular Dynamics Simulations}\label{sec_md_details}
% \label{exp-details}
All experiments were run on an internal cluster, using Kubernetes for orchestration. A single job was launched per material and model in an embarrassingly parallel fashion, using 6 CPUs and 12GB of RAM per run. We used total of 300 CPUs for this benchmark. In simulations that run for 50,000 steps, the computation cost of running a benchmark across materials quickly adds up and becomes a limiting factor given finite computational resources. In addition to computational cost, storage requirements for experiment results and metadata also need to be considered. In total, around 850 GB of data was saved including experiment result logs, metadata, experiment tracking logs, and evaluation data.

All MD simulations reported in this work were conducted on CPU resources to ensure consistent evaluation across models and to reflect typical usage scenarios in many research groups. However, we recognize that modern UMLFFs are increasingly optimized for GPU execution. To provide comprehensive computational cost analysis, we evaluated execution speed for a representative subset of 100 minerals on both CPU (Intel Xeon) and GPU (NVIDIA A100) platforms (Supplementary Table~\ref{table:md-time-per-step-cpu}). GPU implementations provide substantial acceleration, with speedup factors ranging from 1.6$\times$ (CHGNet) to 9.3$\times$ (M3GNet) compared to CPU execution. Despite this acceleration, the total computational cost for comprehensive benchmarking remains significant---evaluating our complete MinX dataset on GPU would still require approximately 2,000--6,000 GPU-hours per model, depending on architectural efficiency. The CPU-based timing reported in the main text (36,500 CPU days total) thus reflects the resource investment required for large-scale UMLFF validation, which remains substantial even with GPU acceleration when scaled to datasets of this size and diversity.

\begin{table*}[ht]
\begin{spacing}{1.15}
    \centering
    \caption{ Details of the pretrained model checkpoints integrated in the \uniffbench{} benchmark and their respective references.} 
    \label{table:checkpoints}
    \vspace{0.5mm}
\begin{tabular}{ccc}
\hline
Model & Checkpoint & Repository/Source \\ \hline
\chgnet & CHGNet-MPtrj-2024.2.13-PES-11M & \citep{Ko_Materials_Graph_Library_2021} \\
\mgnet & M3GNet-MP-2021.2.8-PES & \citep{Ko_Materials_Graph_Library_2021} \\
\mace & 2023-12-10-mace-128-L0{\_}energy{\_}epoch-249 & \citep{batatia2025foundation} \\
\mattersim & mattersim-v1.0.0-1M & \citep{Yang_MatterSim_A_Deep}  \\ 
\orb & orb-v2-20241011 & \citep{neumann2024orbfastscalableneural} \\
\sevennet & 7net-0{\_}11July2024 & \citep{park_scalable_2024} \\
\uma{} & uma-s-1p1.pt & \citep{wood2025family} \\
\maceomat{} & medium-omat-0 & \citep{batatia2025foundation} \\
\hline
\end{tabular}
\end{spacing}
\vspace{-1.5mm}
\end{table*}

\begin{table*}[ht]
\begin{spacing}{1.15}
    \centering
    \caption{ Compute Details \minxeq } 
    \label{table:minxeq}
    \vspace{0.5mm}
\begin{tabular}{ccc}
\hline
Model & Average time per minerals in (hr) & CPU Days \\ \hline
\chgnet & 6.68 & 393.99 \\
\mgnet & 4.89 & 1295.84 \\
\mace & 36.41 & 10239.48 \\
\sevennet & 38.63 & 12873.25 \\
\mattersim & 11.64 & 3908.15 \\ 
\orb & 12.44 & 4177.96 \\

\hline
\end{tabular}
\end{spacing}
\vspace{-1.5mm}
\end{table*}

\begin{table*}[ht]
\begin{spacing}{1.15}
    \centering
    \caption{ Compute Details \minxpocc } 
    \label{table:minxpocc}
    \vspace{0.5mm}
\begin{tabular}{ccc}
\hline
Model & Average time per minerals in (hr) & CPU Days \\ \hline
\chgnet & 6.68 & 20.03 \\
\mgnet & 16.26 & 105.69 \\
\mace & 55.50 & 666.05 \\
\sevennet & 68.78 &756.53\\
\mattersim & 20.25 & 263.27 \\ 
\orb & 26.37 & 342.86 \\
\hline
\end{tabular}
\end{spacing}
\vspace{-1.5mm}
\end{table*}

\begin{table*}[ht]
\begin{spacing}{1.15}
    \centering
    \caption{ Compute Details \minxhtp } 
    \label{table:minxhtp}
    \vspace{0.5mm}
\begin{tabular}{ccc}
\hline
Model & Average time per minerals in (hr) & CPU Days \\ \hline
\chgnet & 12.62 & 37.86 \\
\mgnet & 1.81 & 24.43 \\
\mace & 32.04 & 544.75 \\
\sevennet & 28.46 & 483.81 \\
\mattersim & 10.31 & 180.56\\ 
\orb & 9.90 & 173.33 \\
\hline
\end{tabular}
\end{spacing}
\vspace{1.15mm}
\end{table*}

\begin{table*}[ht]
\centering
\begingroup
\color{black}
\setlength{\tabcolsep}{3pt}
\begin{spacing}{1.15}
\caption{Computational efficiency metrics for UMLFF models showing execution speed and resource requirements.}
\label{table:md-time-per-step-cpu}
\vspace{0.5mm}

\resizebox{\textwidth}{!}{%
\begin{tabular}{c|cccccccc}
\hline
UMLFFs & \chgnet & \mgnet & \mace & \sevennet & \mattersim & \orb & \uma & \maceomat \\ \hline
Time (s) per MD step (CPU)
& 1.452 & 2.794 & 1.087 & 2.153 & 0.780 & 0.736 & 3.50 & 2.25 \\ \hline
Time (s) per MD step (CUDA)
& 0.90 & 0.30 & 0.29 & 0.28 & 0.29 & 0.16 & 0.52 & 0.26 \\ \hline
\end{tabular}
}
\end{spacing}
\endgroup
\end{table*}

% \FloatBarrier

\begin{table*}[ht]
\centering
\begingroup
\color{black}
\setlength{\tabcolsep}{10pt}
\caption{Baseline comparison of density predictions for 10 randomly selected minerals. Experimental values are compared against: (1) UMLFFs predictions (results from stable models: \mace{}, \sevennet{}, \mattersim{}, \orb{}) from finite-temperature MD simulations at 298~K, and (2) DFT-computed densities from Materials Project database (0~K, PBE functional, static optimization). Percent errors relative to experimental values are shown in parentheses. The comparison demonstrates that UMLFFs underperform their DFT training data, with errors approximately 2--3$\times$ larger than DFT despite achieving near-DFT accuracy on energy/force benchmarks.}
% \label{table:md-time-per-step-cpu}
\label{table:baseline}

\resizebox{\columnwidth}{!}{%
\begin{tabular}{lcccccccc}
\hline
\textbf{Mineral} &
\textbf{Experimental} &
\textbf{DFT} &
\textbf{\mace{}} &
% \textbf{\chgnet{}} &
% \textbf{\mgnet{}} &

\textbf{\sevennet{}} &
\textbf{\mattersim{}} &
\textbf{\orb{}} \\
&
\textbf{(g\,cm$^{-3}$)} &
\textbf{(g\,cm$^{-3}$)} &
\textbf{(g\,cm$^{-3}$)} &
\textbf{(g\,cm$^{-3}$)} &
\textbf{(g\,cm$^{-3}$)} &
% \textbf{(g\,cm$^{-3}$)} &
% \textbf{(g\,cm$^{-3}$)} &
\textbf{(g\,cm$^{-3}$)} \\
\hline
Benitoite   & 3.68 & 3.66 (-0.5\%) & 3.52 (-4.2\%) & 3.52 (-4.2\%) & 3.53 (-4.2\%)& 3.53 (-4.1\%) \\
Cinnabar    & 8.17 & 7.38 (-9.7\%) & 7.07 (-13.5\%)  & 6.11 (-25.2\%) & 6.43 (-21.3\%)& 6.87 (-15.9\%) \\
Cuprite     & 6.11 & 6.20 (+1.5\%) & 6.02 (-1.4\%)  & 6.08 (-0.4\%) & 6.04 (-1.1\%)& 6.03 (-1.2\%) \\
Dysprosium  & 8.56 & 8.43 (-1.5\%) & 8.50 (-0.7\%)   & 8.43 (-1.5\%) & 8.50 (-0.7\%)& 8.46 (-1.1\%) \\
Graphite    & 2.28 & 2.18 (-4.4\%) & 2.24 (-1.9\%)  & 2.24 (-1.8\%) & 2.24 (-1.9\%)& 2.28 (+0.1\%) \\
Kaolinite   & 2.57 & 2.50 (-2.7\%) & 2.31 (-10.0\%)  & 2.38 (-7.3\%) & 2.32 (-9.8\%) & 2.40 (-6.6\%) \\
Perovskite  & 4.03 & 4.03 (+0.1\%) & 3.90 (-3.1\%) & 3.89 (-3.5\%) & 3.88 (-3.6\%)& 3.88 (-3.6\%) \\
Pyrite      & 5.12 & 4.94 (-3.5\%) & 5.04 (-1.5\%)  & 4.99 (-2.4\%) & 5.00 (-2.2\%) & 4.87 (-4.9\%) \\
Quartz      & 2.65 & 2.63 (-0.6\%) & 2.46 (-7.1\%)  & 2.45 (-7.5\%) & 2.40 (-9.3\%) & 2.42 (-8.6\%) \\
Tungsten    & 19.26 & 19.16 (-0.5\%) & 18.69 (-3.0\%)   & 18.66 (-3.1\%) & 18.68 (-3.0\%) & 18.79 (-2.5\%) \\
\hline
\end{tabular}
}
\endgroup

\end{table*}

\textbf{Note:} DFT values are from Materials Project database computed using PBE functional with standard precision settings. The systematic underperformance of UMLFFs relative to DFT (despite training on DFT data) indicates fundamental limitations in model learning or generalization rather than simply inheriting DFT systematic errors. The large error for Cinnabar (25.2\%) in UMLFFs predictions versus modest DFT error (1.5\%) exemplifies cases where models fail to reproduce even the accuracy of their training data.

\begin{table}[ht]
\centering
\color{black}
% \begingroup
\renewcommand{\arraystretch}{1.2}
\caption{Baseline comparison of shear modulus prediction. Experimental values are compared against: (1) UMLFFs predictions (results from stable models: \mattersim{}, \sevennet{}, \mace{}, \orb{}, \uma{}) and (2) DFT-computed shear modulus from Materials Project database. Percent errors relative to experimental values are shown in parentheses.}
\label{table:baseline_shear}
\resizebox{\textwidth}{!}{
\begin{tabular}{lccccccc}
\hline
\textbf{Mineral} &
\textbf{Experimental} &
\textbf{DFT} &
\textbf{\mattersim{}} &
\textbf{\sevennet{}} &
\textbf{\mace{}} &
\textbf{\orb{}} &
\textbf{\uma{}} \\
&
\textbf{(GPa)} &
\textbf{(GPa)} &
\textbf{(GPa)} &
\textbf{(GPa)} &
\textbf{(GPa)} &
\textbf{(GPa)} &
\textbf{(GPa)} \\
\hline

Corundum  & 166.08 & 136.00 (-18.1\%) & 175.98 (+6.0\%) & 169.56 (+2.1\%) & 185.24 (+11.5\%) & 6.44 (-96.1\%) & 223.62 (+34.6\%) \\
Calcite   & 37.75  & 37.00 (-2.0\%) & 89.89 (+138.1\%) & 88.42 (+134.2\%) & 92.55 (+145.2\%) & 6.43 (-83.0\%) & 108.95 (+188.6\%) \\
Platinum  & 65.10  & 50.00 (-23.2\%) & 56.24 (-13.6\%) & 63.95 (-1.8\%) & 64.56 (-0.8\%) & 4.58 (-93.0\%) & 39.72 (-39.0\%) \\
Lead      & 10.46  & 12.00 (+14.7\%) & 14.36 (+37.3\%) & 6.30 (-39.8\%) & 14.07 (+34.5\%) & -0.11 (-101.1\%) & -45.73 (-537.2\%) \\
Copper    & 54.00  & 57.00 (+5.6\%) & 64.90 (+20.2\%) & 39.88 (-26.1\%) & 58.38 (+8.1\%) & -2.92 (-105.41\%) & 57.69 (+6.8\%) \\
Silver    & 33.54  & 35.00 (+4.4\%) & 39.67 (+18.3\%) & 23.07 (-31.2\%) & 37.64 (+12.2\%) & 2.81 (-91.6\%) & 36.76 (+9.6\%) \\
Cobalt    & 84.31  & 127.00 (+50.6\%) & 81.62 (-3.2\%) & 59.23 (-29.7\%) & 93.95 (+11.4\%) & 12.33 (-85.4\%) & 94.26 (+11.8\%) \\
Nickle    & 93.16  & 99.00 (+6.3\%) & 99.27 (+6.6\%) & 84.25 (-9.6\%) & 96.52 (+3.6\%) & 11.47 (-87.7\%) & 119.61 (+28.4\%) \\
Iron      & 89.18  & 73.00 (-18.1\%) & 78.09 (-12.4\%) & 73.83 (-17.2\%) & 92.75 (+4.0\%) & 1.08 (-98.8\%) & 206.29 (+131.3\%) \\
Tungsten  & 160.10 & 149.00 (-6.9\%) & 142.74 (-10.8\%) & 37.25 (-76.7\%) & 149.64 (-6.5\%) & 12.93 (-91.9\%) & 160.49 (+0.2\%) \\

\hline

\end{tabular}
}
% \endgroup
\end{table}

\begin{table}[ht]
\centering
\color{black}
\begingroup
\renewcommand{\arraystretch}{1.2}
\caption{Baseline comparison of bulk modulus prediction. Experimental values are compared against: (1) UMLFFs predictions (results from stable models: \mattersim{}, \sevennet{}, \mace{}, \orb{}, \uma{}) and (2) DFT-computed bulk modulus from Materials Project database. Percent errors relative to experimental values are shown in parentheses.}
\label{table:baseline_bulk}
\resizebox{\textwidth}{!}{
\begin{tabular}{lccccccc}
\hline
\textbf{Mineral} &
\textbf{Experimental} &
\textbf{DFT} &
\textbf{\mattersim{}} &
\textbf{\sevennet{}} &
\textbf{\mace{}} &
\textbf{\orb{}} &
\textbf{\uma{}} \\
&
\textbf{(GPa)} &
\textbf{(GPa)} &
\textbf{(GPa)} &
\textbf{(GPa)} &
\textbf{(GPa)} &
\textbf{(GPa)} &
\textbf{(GPa)} \\
\hline

Corundum  & 253.97 & 232.00 (-8.7\%) & 259.29 (+2.1\%) & 270.33 (+6.4\%) & 258.77 (+1.9\%) & 199.61 (-21.4\%) & 264.59 (+4.2\%) \\
Calcite   & 69.21  & 78.00 (+12.7\%) & 154.00 (+122.5\%) & 149.73 (+116.3\%) & 154.13 (+122.7\%) & 190.83 (+175.7\%) & 152.64 (+120.5\%) \\
Platinum  & 282.70 & 248.00 (-12.3\%) & 302.90 (+7.1\%) & 297.98 (+5.4\%) & 299.92 (+6.1\%) & 294.97 (+4.3\%) & 324.65 (+14.8\%) \\
Lead      & 44.76  & 40.00 (-10.6\%) & 50.68 (+13.2\%) & 46.34 (+3.5\%) & 51.74 (+15.6\%) & 62.09 (+38.7\%) & 26.45 (-40.9\%) \\
Copper    & 136.27 & 151.00 (+10.8\%) & 154.19 (+13.2\%) & 142.00 (+4.2\%) & 148.92 (+9.3\%) & 273.24 (-74.09\%) & 154.30 (+13.2\%) \\
Silver    & 101.20 & 88.00 (-13.0\%) & 119.04 (+17.6\%) & 118.59 (+17.2\%) & 114.47 (+13.1\%) & 110.28 (+9.0\%) & 118.11 (+16.7\%) \\
Cobalt    & 189.76 & 201.00 (+5.9\%) & 152.27 (-19.8\%) & 102.64 (-45.9\%) & 176.93 (-6.8\%) & 967.31 (+409.8\%) & 229.29 (+20.8\%) \\
Nickle    & 185.97 & 174.00 (-6.4\%) & 192.16 (+3.3\%) & 212.21 (+14.1\%) & 204.11 (+9.8\%) & 648.08 (+248.5\%) & 278.02 (+49.5\%) \\
Iron      & 166.93 & 208.00 (+24.6\%) & 139.56 (-16.4\%) & 105.96 (-36.5\%) & 285.29 (+70.9\%) & 98.09 (-41.2\%) & 338.33 (+102.7\%) \\
Tungsten  & 310.38 & 302.00 (-2.7\%) & 335.09 (+8.0\%) & 332.13 (+7.0\%) & 350.68 (+13.0\%) & 300.98 (-3.0\%) & 333.47 (+7.4\%) \\

\hline
\end{tabular}
}

\endgroup
\end{table}

% \clearpage

\section{Hardware Specifications and Error Metrics for Elastic Tensor Calculations}\label{sec_elastic_details}
All eleastic tensor experiment were done on an AMD EPYC 7282 16-Core Processor @ 2.80\,GHz with 1\,TB of installed RAM. A single job was launched per materials and model using 1 CPUs per run.

% \section{Elastic Tensor Error metrics}\label{sec_error_metric}
\begin{table}[htbp]
\centering
\caption{Summary of $R^2$ and MAPE (\%) metrics for different UMLFFs across elastic coefficients on the \minxem{} dataset. The $R^2$ and MAPE values for \chgnet{} and \mgnet{} are zero, as no simulations yielded prediction errors below the 100\% threshold.}
\renewcommand{\arraystretch}{1.2}
\begin{tabular}{l l*{5}{>{\centering\arraybackslash}p{2.2cm}}}
\toprule
\textbf{Model} & Metric & \textbf{C11} & \textbf{C12} & \textbf{C13} & \textbf{C44} & \textbf{C66} \\
\midrule
\multirow{2}{*}{CHGNet}     & $R^2$         & 0.000 & 0.000 & 0.000 & 0.000 & 0.000 \\
                            & MAPE (\%)  & 0.0   & 0.0   & 0.0   & 0.0   & 0.0   \\
\multirow{2}{*}{M3GNet}     & $R^2$         & 0.000 & 0.000 & 0.000 & 0.000 & 0.000 \\
                            & MAPE (\%)  & 0.0   & 0.0   & 0.0   & 0.0   & 0.0   \\
\multirow{2}{*}{MACE-MP-0}  & $R^2$         & 0.864 & 0.588 & 0.434 & 0.493 & 0.594 \\
                            & MAPE (\%)  & 22.7  & 27.2  & 27.3  & 32.5  & 31.2  \\
\multirow{2}{*}{SevenNet}   & $R^2$         & 0.856 & 0.505 & 0.235 & 0.361 & 0.475 \\
                            & MAPE (\%)  & 22.5  & 28.6  & 29.3  & 40.0  & 40.4  \\
\multirow{2}{*}{MatterSim}  & $R^2$         & 0.941 & 0.819 & 0.733 & 0.579 & 0.670 \\
                            & MAPE (\%)  & 16.0  & 21.2  & 21.4  & 29.9  & 27.1  \\
\multirow{2}{*}{Orb-v2}     & $R^2$         & 0.266 & 0.174 & -0.410 & -0.759 & -0.898 \\
                            & MAPE (\%)  & 41.5  & 44.8  & 46.3  & 99.8  & 100.0 \\
\bottomrule
\end{tabular}
\end{table}

% \begin{table*}[ht]
% \begin{spacing}{1.15}
%     \centering
%     \caption{Time taken per MD simulation step per model.} 
%     \label{apptable:md-time-per-step}
%     \vspace{0.5mm}
% \begin{tabular}{c|cccccc}
% \hline
% Model & \chgnet & \mgnet & \mace  & \mattersim & \orb & \sevennet \\ \hline
% Time(s) per step & 1.452  & 2.794  & 1.087 & 0.780      & 0.736  & 2.153  \\ \hline
% \end{tabular}
% \end{spacing}
% \vspace{-1.5mm}
% \end{table*}

\section{Density and Lattice Parameter Predictions}\label{sec_parity_plot}

\begin{figure}[!ht]
    % \vspace{-0.3in}
    \centering
    \includegraphics[width=0.95\textwidth]{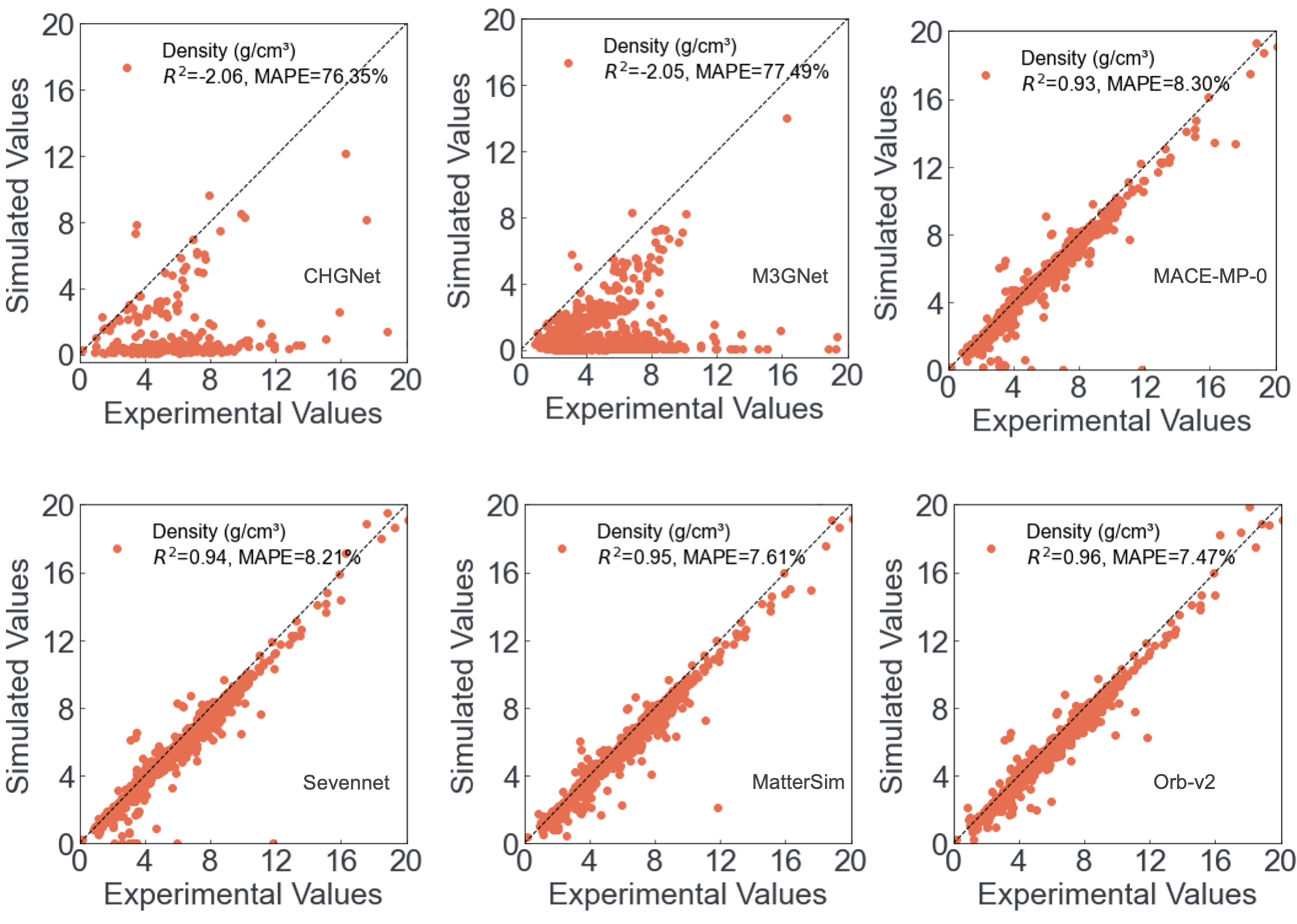}
    \caption{Parity plots comparing the predicted and experimental density (g/cm³) for each UMLFF. The dashed line shows perfect agreement (y = x). Each point corresponds to a distinct minerals, and model accuracy is quantified using the ($R^2$) and (MAPE) values}
    \label{fig:density-plots}
    % \vspace{-0.2in}
\end{figure}

\begin{figure}[!ht]
    % \vspace{-0.3in}
    \centering
    \includegraphics[width=0.95\textwidth]{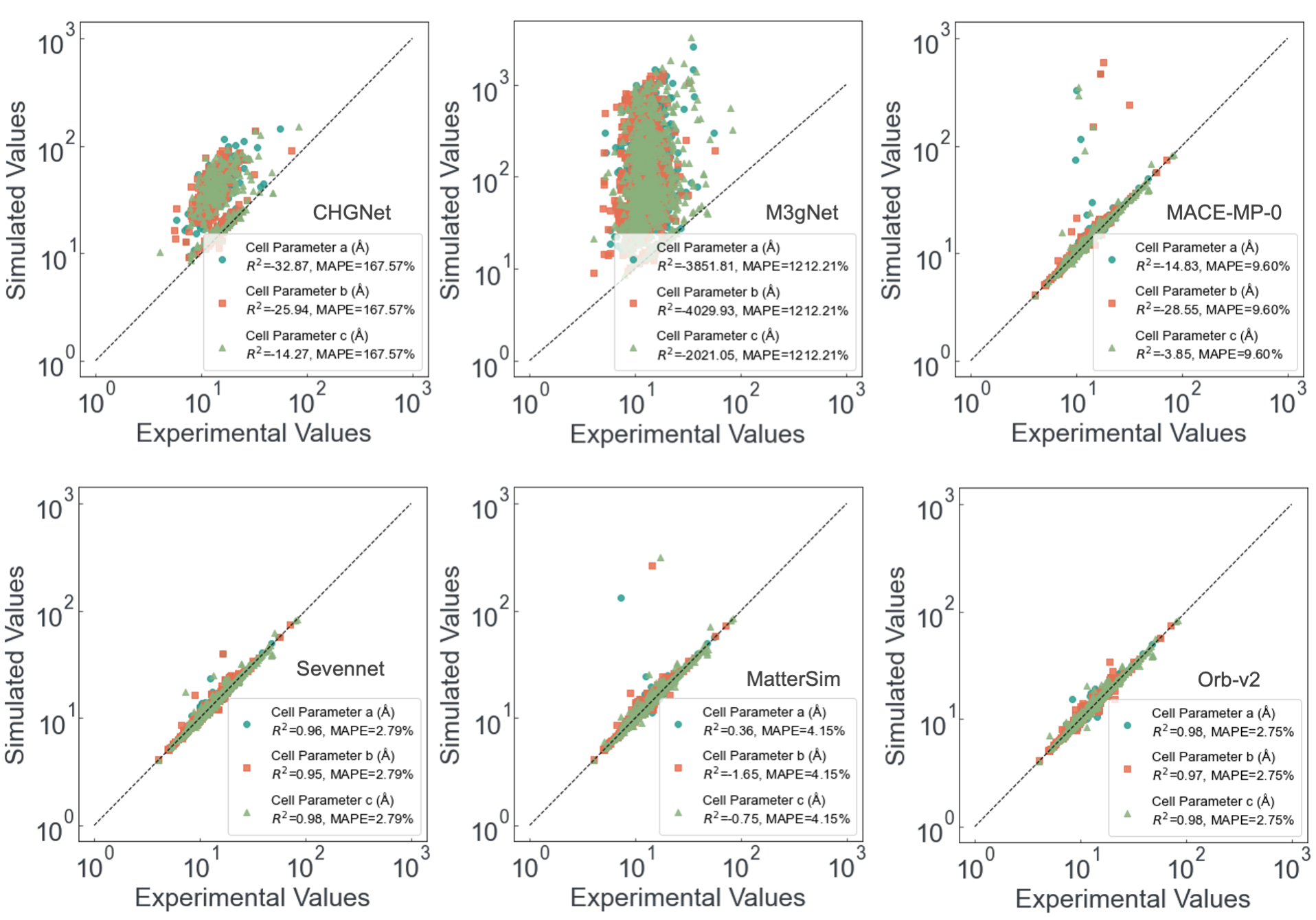}
    \caption{Parity plots comparing the predicted and experimental lattice parameters in (\text{\AA}\ ) for each UMLFF. The dashed line shows perfect agreement (y = x). Each point corresponds to a distinct minerals, and model accuracy is quantified using the ($R^2$) and (MAPE) values}
    \label{fig:lattice-plots_parity}
    \vspace{-0.2in}
\end{figure}

Supplementary \Cref{fig:density-plots,fig:lattice-plots_parity} shows the parity plots for density and lattice parameters predicted by each model for all the structures in \minx (including all the completed simulations and without excluding those that exhibit > 100\%), with corresponding \( R^2 \) and MAPE values indicated in the legend. The UMLFFs achieve \( R^2 \) values greater than 0.9 for density except \chgnet{} and \mgnet, demonstrating strong predictive performance over density. However, it fails in predicting the lattice parameters except the \orb{}  that has more than 0.9 \(R^2\) score. A possible explanation for this discrepancy lies in the inherent difference in how density and lattice parameters contribute to the overall structural representation. Density is inversely proportional to volume, which is a product of the three lattice parameters. Therefore, errors in individual lattice directions may compensate or cancel out the true effect when calculating volume, leading to a lower impact on the predicted density. In contrast, lattice parameters are evaluated directly, and errors in any direction contribute linearly, making them more sensitive to prediction inaccuracies. In other words, density is varying inversely and making even significant change in lattice parameters to zero and hence negligible impact on mean or std value of density. This suggests that while models may approximate the volumetric properties well, capturing the precise geometry remains more challenging.

\begin{figure}[!ht]
    % \vspace{-0.3in}
    \centering
    \includegraphics[width=0.95\textwidth]{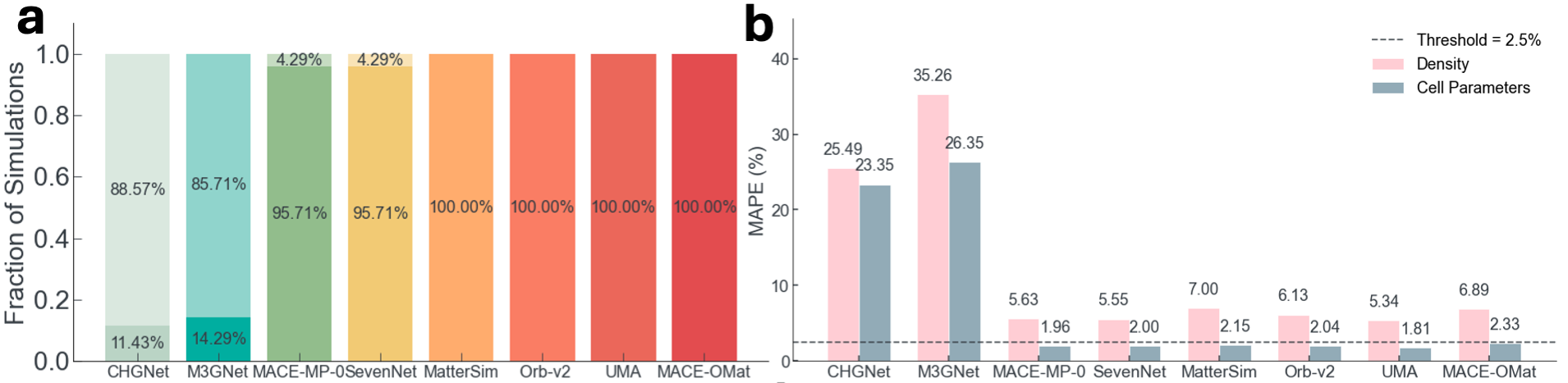}
    \caption{(a) Fraction of simulations completed and (b) MAPE in density and cell parameters on MinX-HTP dataset. Note that two new models, namely, \uma{} and \maceomat{} are included.}
    \label{fig:MAPE-plots}
    \vspace{-0.2in}
\end{figure}

\textbf{Simulation failures.} We note that simulation failures reported in this work are intrinsic to model behavior rather than hardware-dependent. Initial testing suggested possible memory constraints for some failures, but subsequent evaluation with increased computational resources (up to 512 GB RAM, extended walltime limits) produced identical failure patterns. This confirms that the failures arise from fundamental limitations in force field accuracy or numerical stability for these specific structures and thermodynamic conditions, rather than from implementation or resource constraints. Our stability assessment is thus hardware-independent and reflects genuine model limitations in handling the chemical and structural diversity of the \minx{} benchmark.

\begin{figure}[!ht]
    % \vspace{-0.3in}
    \centering
    \includegraphics[width=0.95\textwidth]{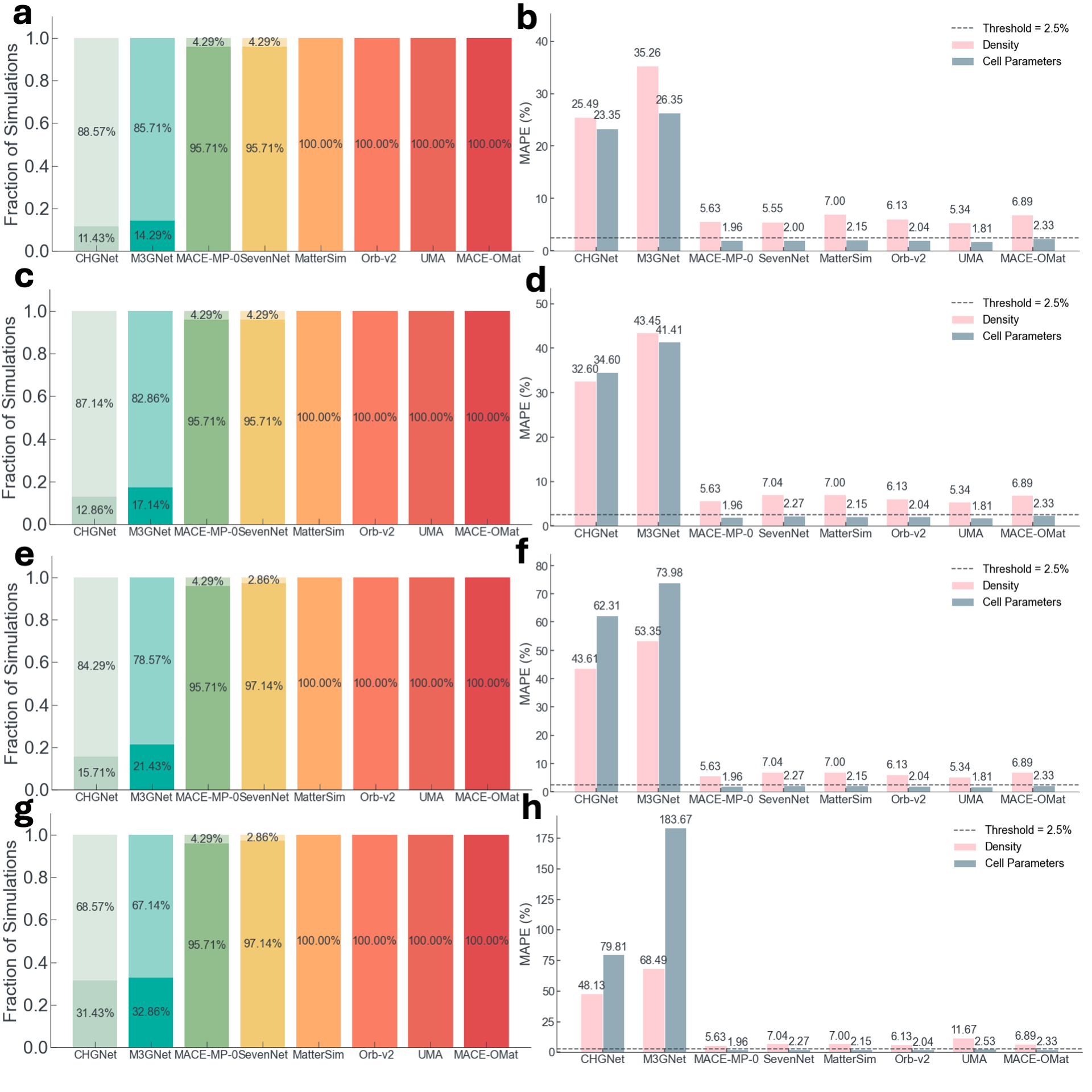}
    \caption{Sensitivity of conclusions to error cutoff thresholds. Panels (a,b) correspond to the 100\% cutoff, (c,d) to 150\%, (e,f) to 250\%, and (g,h) to 500\% cutoff thresholds used to filter unsuccessful simulations. Panels (a,c,e,g) show the fraction of simulations retained for each model under the respective cutoff, while panels (b,d,f,h) report the corresponding mean absolute percentage error (MAPE) for density and cell parameters, with the dashed line indicating the 2.5\%  threshold.}
    \label{fig:lattice-plots-threshold}
    \vspace{-0.2in}
\end{figure}

\textbf{Sensitivity to cutoffs.} Supplementary Figure~\ref{fig:lattice-plots-threshold} shows the sensitivity analysis for error cutoff thresholds (100\%, 150\%, 250\%, 500\%). The figure demonstrates that: (a) the fraction of successful simulations remains relatively stable across thresholds for each model, confirming that most failures are catastrophic rather than marginal; (b) MAPE values for density and lattice parameters show minimal variation across thresholds, indicating our conclusions are robust to the specific cutoff choice; and (c) all models fail to achieve the $\pm$2.5\% accuracy threshold regardless of cutoff, reinforcing our main findings.

\begin{figure}[!ht]
    % \vspace{-0.3in}
    \centering
    \includegraphics[width=0.95\textwidth]{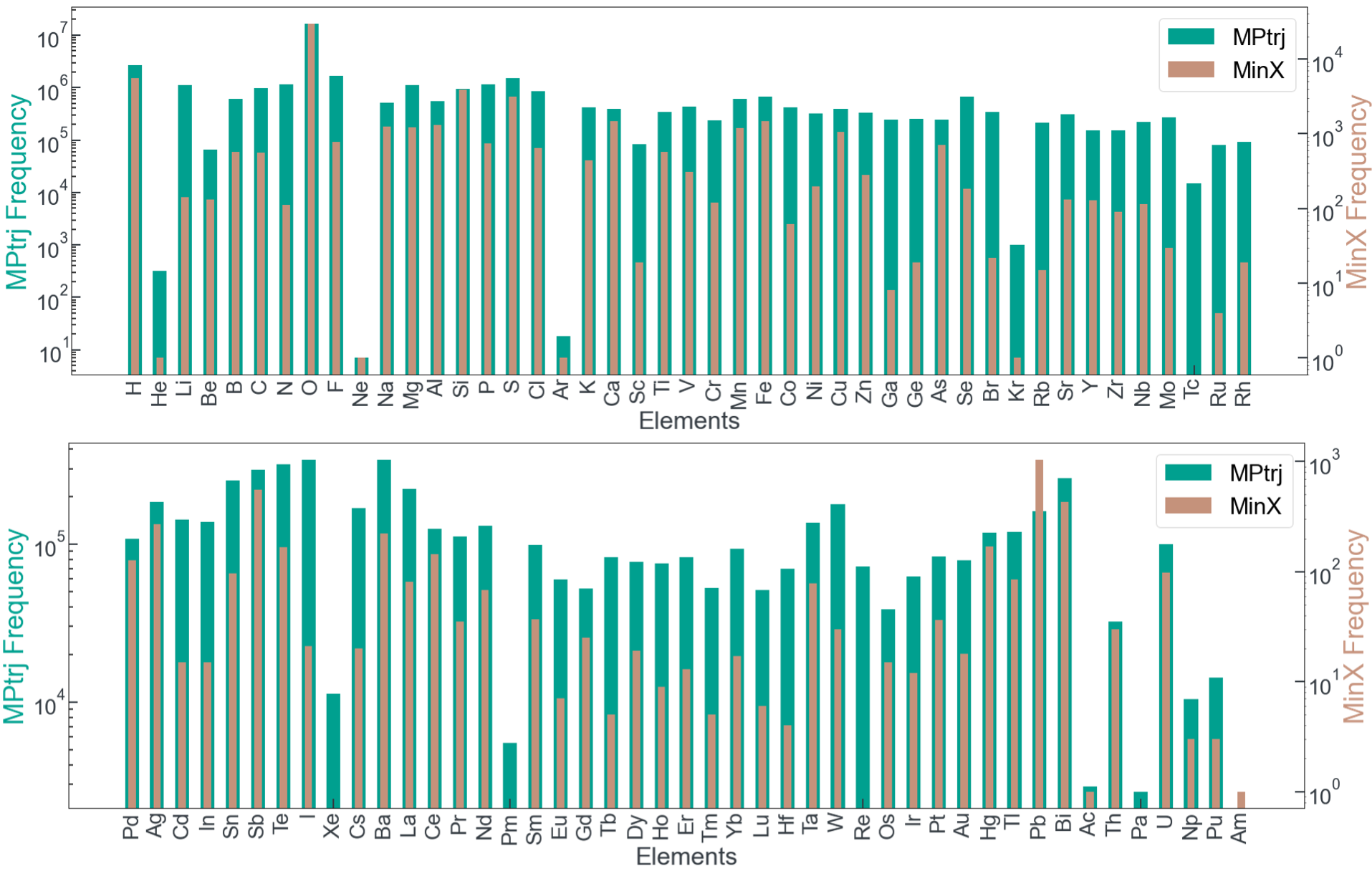}
   \caption{\textbf{\minx{} dataset.} Elemental frequency distribution comparison between MPtrj training dataset (green) and \minx{} evaluation dataset (brown) across the periodic table.}
    % \label{fig:lattice-plots_bar}
    \vspace{-0.2in}
\end{figure}

\textbf{Equilibration}. To analyze whether the systems equilibrate in 50 ps, temporal evolution of temperature and energy of 10 minerals randomly selected from the \minx{} dataset is plotted in Supplementary Figure~\ref{fig:md-equi}. We observe that the minerals equilibrate in 10-15 ps. This was consistently observed across systems and models.

\begin{figure}[!htbp]
    \centering
    \includegraphics[width=0.9\textwidth]{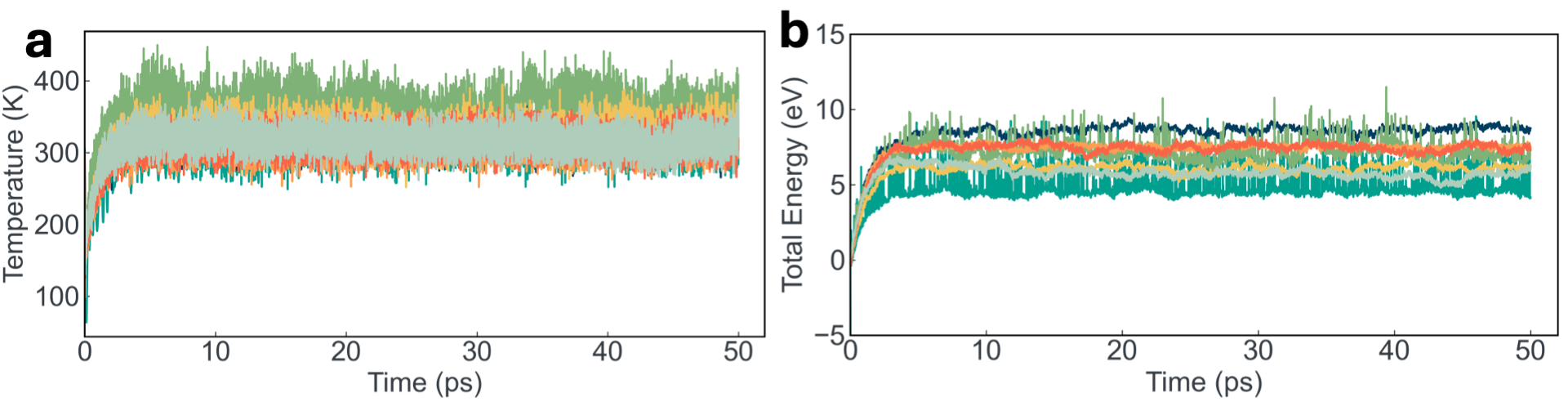}
    \caption{\textbf{Equilibration behavior of MD simulations.} \textbf{a} The temporal evolution of temperature and \textbf{b} total energy of 10 randomly selected minerals from \minxeq{} during MD simulations. The stabilization of both temperature and total energy after an initial transient indicates that the system reaches equilibration within the simulated timescale. All subsequent analyses are performed using the equilibrated portion of the trajectory.}
    \label{fig:md-equi}
\end{figure}
\begin{figure}[!htbp]
    \centering
    \includegraphics[width=0.75\textwidth]{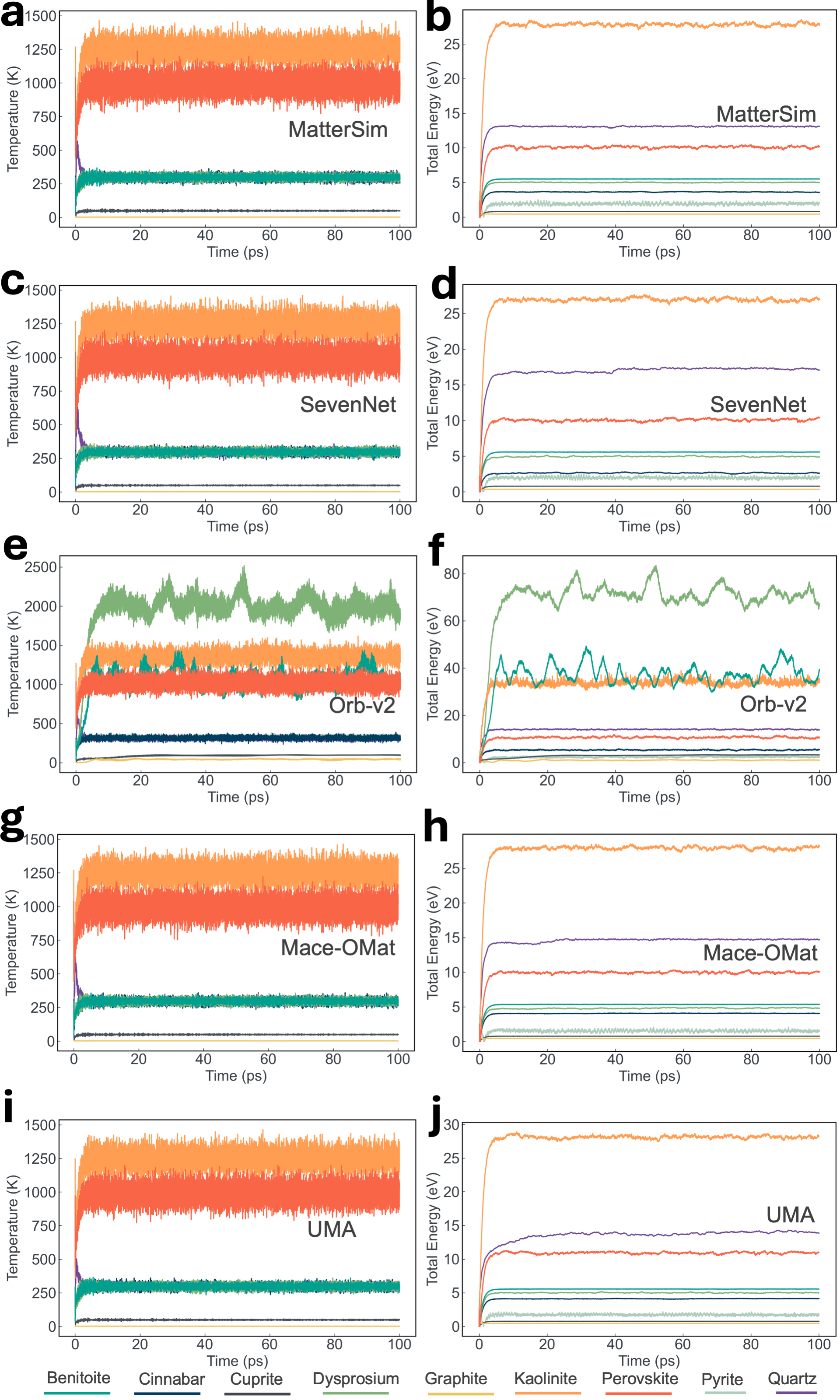}
    % \caption{(a) Distribution of atomic elements and their frequency in the MPtrj (teal-blue gradient) and MinX (brown) datasets respectively. (b) Number of elements present in the unit cells of MPtrj and MinX. (c) Distribution of the number of atoms in the unit cell of MinX. (d) Distribution of temperature (left abscissa) and pressure (right abscissa) across extreme thermodynamic scenarios.}
   
    \caption{\textcolor{black}{\textbf{Equilibration behavior of MD simulations.} \textbf{a} The temporal evolution of temperature and \textbf{b} total energy of 10 randomly selected minerals from \minxhtp during MD simulations. The stabilization of both temperature and total energy after an initial transient indicates that the system reaches equilibration within the simulated timescale.}}
    \label{fig:md-100}
\end{figure}

\begin{figure}[!htbp]
    \centering
    \includegraphics[width=0.75\textwidth]{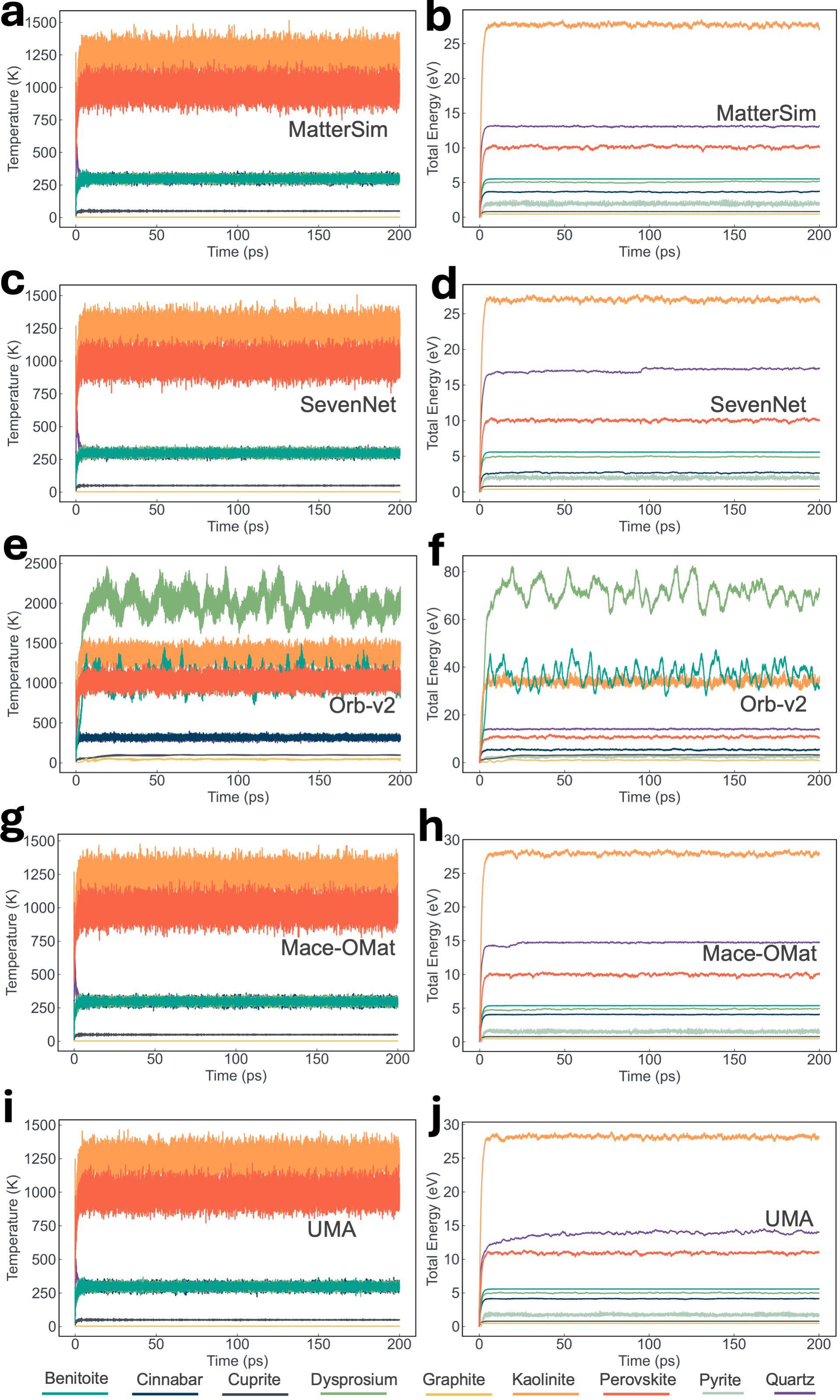}
    % \caption{(a) Distribution of atomic elements and their frequency in the MPtrj (teal-blue gradient) and MinX (brown) datasets respectively. (b) Number of elements present in the unit cells of MPtrj and MinX. (c) Distribution of the number of atoms in the unit cell of MinX. (d) Distribution of temperature (left abscissa) and pressure (right abscissa) across extreme thermodynamic scenarios.}
    \caption{\textcolor{black}{\textbf{Equilibration behavior of MD simulations.} \textbf{a} The temporal evolution of temperature and \textbf{b} total energy of 10 randomly selected minerals from \minxhtp during MD simulations. The stabilization of both temperature and total energy after an initial transient indicates that the system reaches equilibration within the simulated timescale.}}
    \label{fig:md-200}
\end{figure}

% \clearpage

\section{Dataset Details}\label{sec_dataset}

\begin{figure}[!ht]
    % \vspace{-0.3in}
    \centering
    \includegraphics[width=0.9\textwidth]{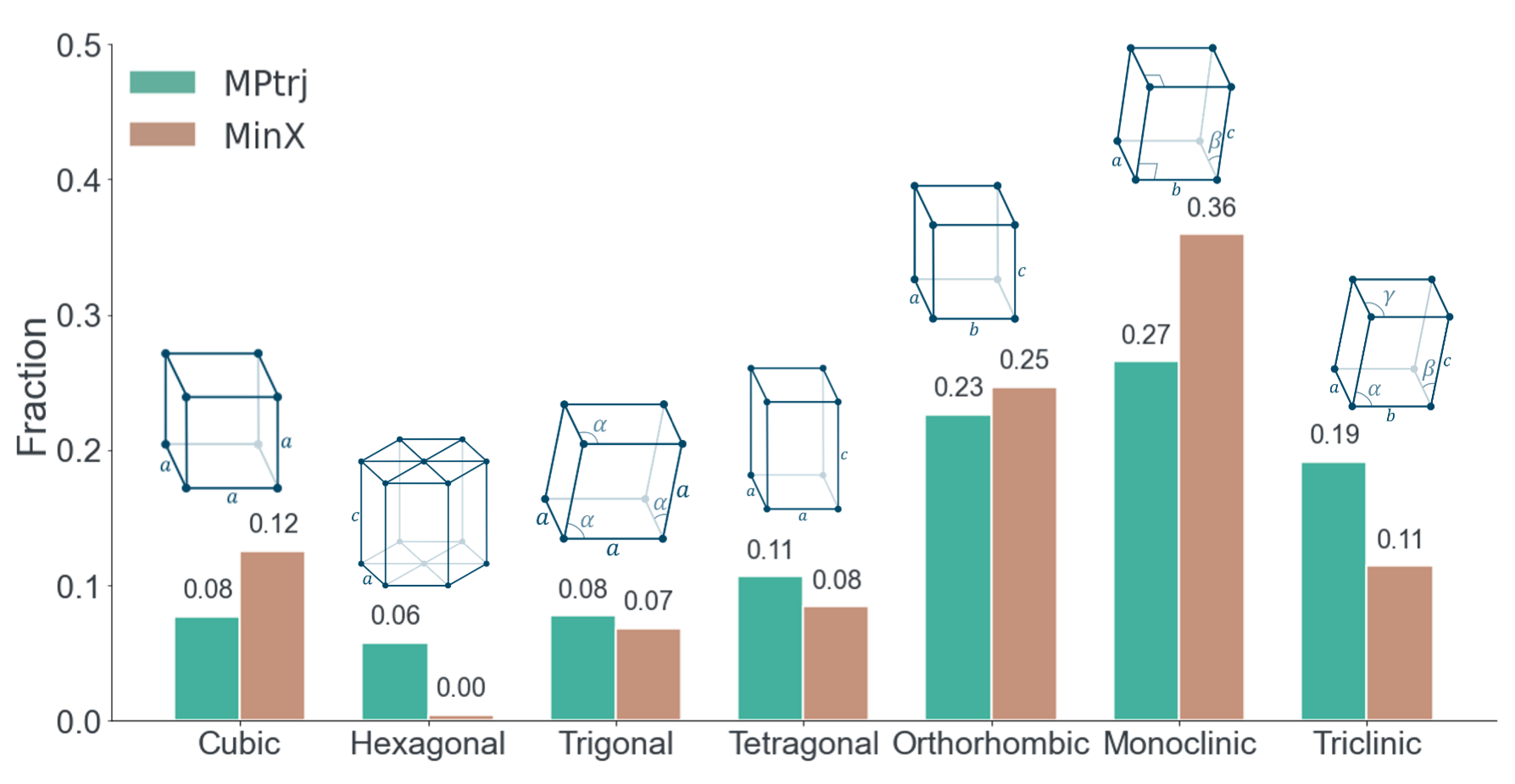}
    \caption{Comparison of the distribution of crystal systems in the MPtrj and \minx datasets. The plot shows the fraction of materials belonging to each crystal system, highlighting differences in structural diversity between the two datasets.}
    \label{fig:system-plots}
    % \vspace{-0.2in}
\end{figure}

 To analyze the distribution of crystal systems across the dataset, we computed the normalized fraction of each crystal system by dividing the number of structures belonging to each system with the total number of crystal structures. Supplementary~\Cref{fig:system-plots} shows the dataset exhibits a disproportionately higher representation of orthorhombic and monoclinic structures in training data compared to other crystal systems. However, this distribution is consistent with the prevalence of orthorhombic and monoclinic structures observed in naturally occurring minerals. 

\begin{figure}[!ht]
    % \vspace{-0.3in}
    \centering
    \includegraphics[width=0.9\textwidth]{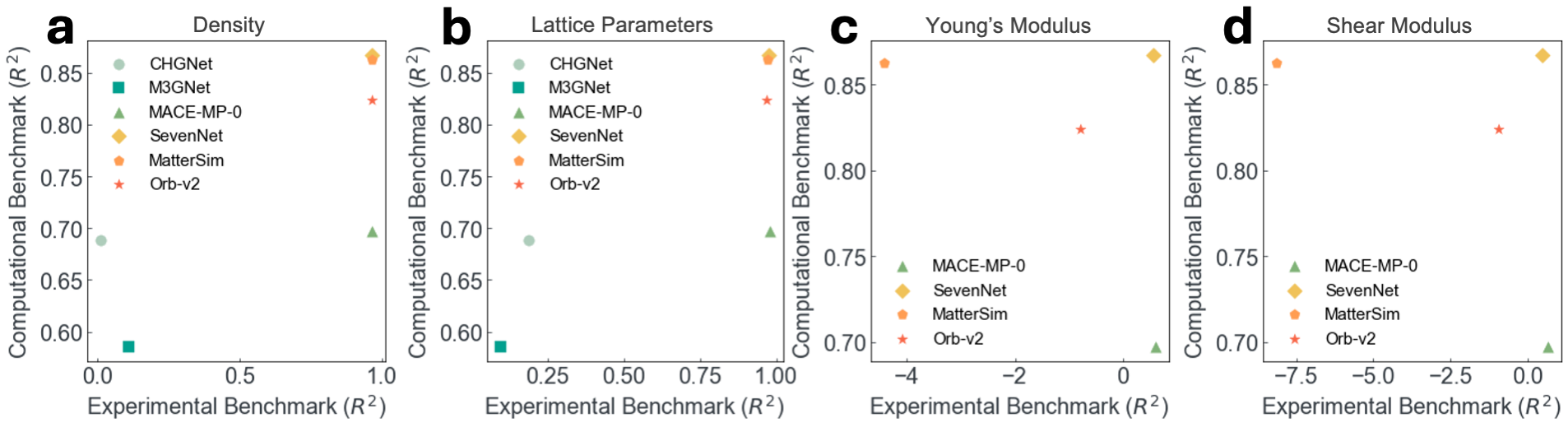}
    \caption{\textbf{Correlation between computational and experimental benchmark performance across structural and mechanical properties.}\textbf{(a)} density, \textbf{(b)} lattice parameters, \textbf{(c)} Young’s modulus, and \textbf{(d)} shear modulus. Computational benchmark $R^2$ scores from MatBench Discovery are plotted against experimentally evaluated R² values for respective models, highlighting differences in predictive}
    \label{fig:comp-exp-comparison}
    % \vspace{-0.2in}
\end{figure}

% \clearpage

\section{Error Analysis}
\subsection{Mean Bond Error}
\begin{figure}[!ht]
    % \vspace{-0.3in}
    \centering
    \includegraphics[width=0.9\textwidth]{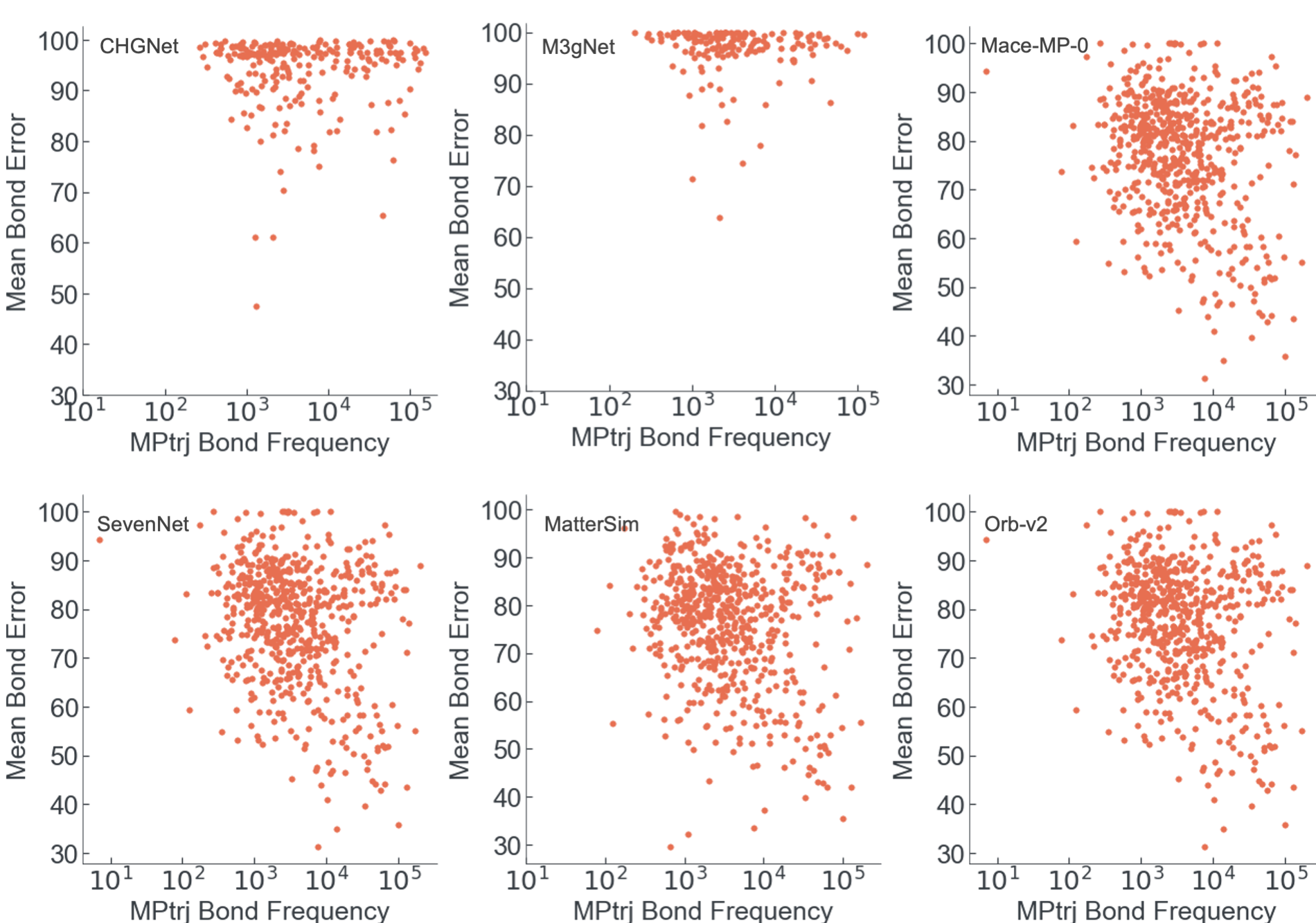}
    \caption{Relationship between mean bond error and bond frequency in the MPtrj dataset. More frequently occurring bonds tend to exhibit lower prediction errors, indicating better model accuracy for frequently occurring bonds}
    \label{fig:bond_freq-plots}
    % \vspace{-0.2in}
\end{figure}

Supplementary~\Cref{fig:bond_freq-plots} presents the mean bond error for all unique bonds present in the MPtrj dataset. A clear negative correlation is observed between bond frequency and bond error, supporting the hypothesis that training data bias affects the model's accuracy, with more frequent bonds being predicted more accurately.

% \clearpage

\subsection{Temporal Evolution: \minxhtp{}}\label{app_temp_minxhtp}
\begin{figure}[!ht]
    % \vspace{-0.3in}
    \centering
    \includegraphics[width=0.95\textwidth]{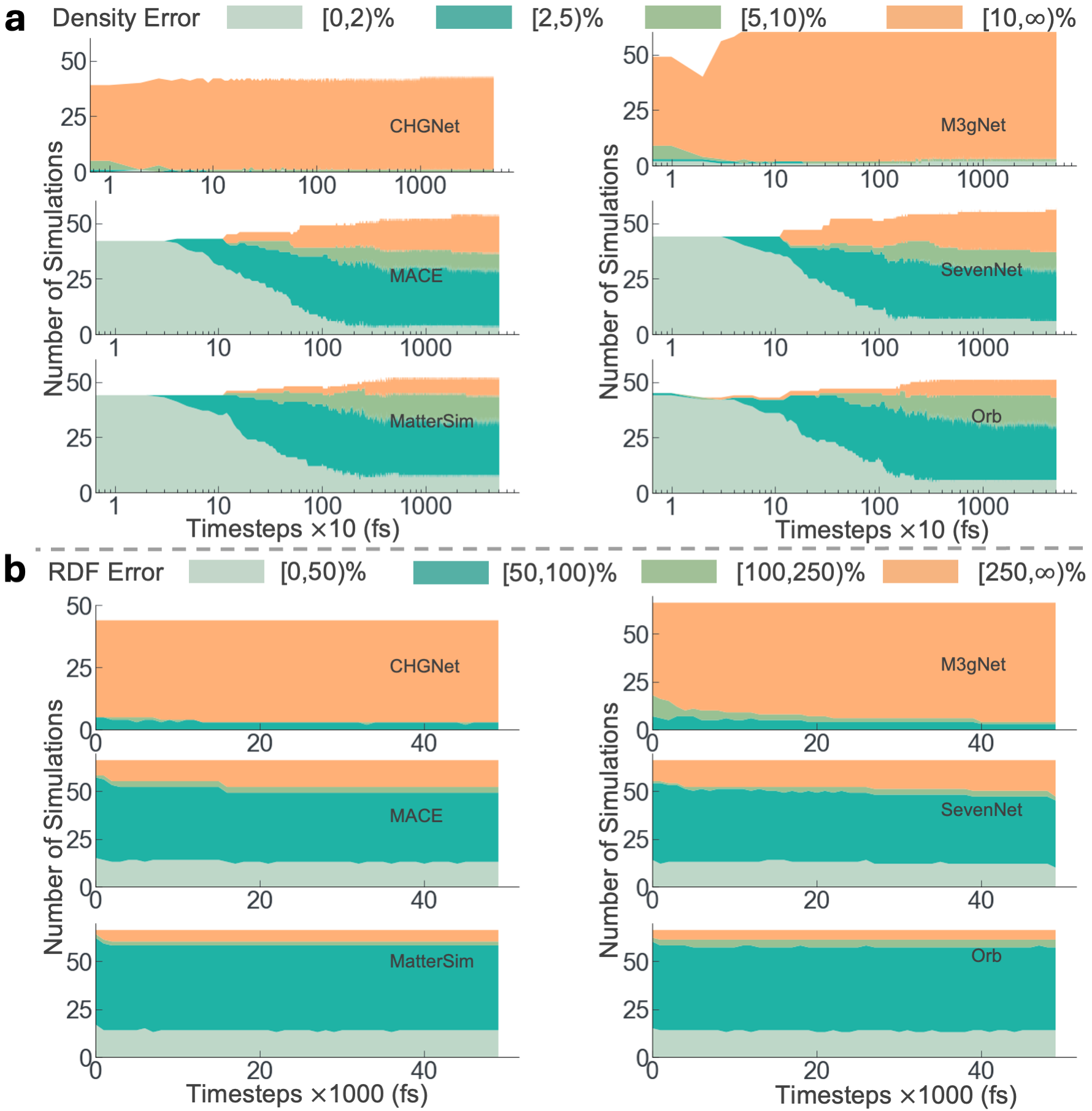}
    \caption{\textbf{Temporal evolution on \minxhtp{} reveals divergent stability patterns among UMLFFs.} \textbf{a}, Density error evolution during MD simulations for \minxhtp{}. \textbf{b}, Radial distribution function (RDF) error evolution showing atomic spatial organization accuracy with error ranges for \minxhtp{}}
    % \label{fig:rdf-plots}
    \label{fig:high_temp-plots}
    \vspace{-0.2in}
\end{figure}

Supplementary~\Cref{fig:high_temp-plots} shows temporal evolution density and RDF for \minxhtp{}. \textbf{a}, Density error evolution during MD simulations with stacked areas representing error distributions across four ranges ([0,2)\%, [2,5)\%, [5,10)\%, [10,$\infty$)\%). Simulation timesteps shown on logarithmic scale to capture behavior across multiple time regimes. \textbf{b}, RDF error evolution showing atomic spatial organization accuracy with error ranges ([0,50)\%, [50,100)\%, [100,250)\%, [250,$\infty$)\%). Results demonstrate that even the stable models converge to consistent error ranges for just 20\% of the structure while unstable models exhibit persistent high errors throughout simulation periods.

% \clearpage

\subsection{Temporal Evolution: \minxpocc{}}\label{app_temp_minxpocc}
\begin{figure}[!ht]
    % \vspace{-0.3in}
    \centering
    \includegraphics[width=0.95\textwidth]{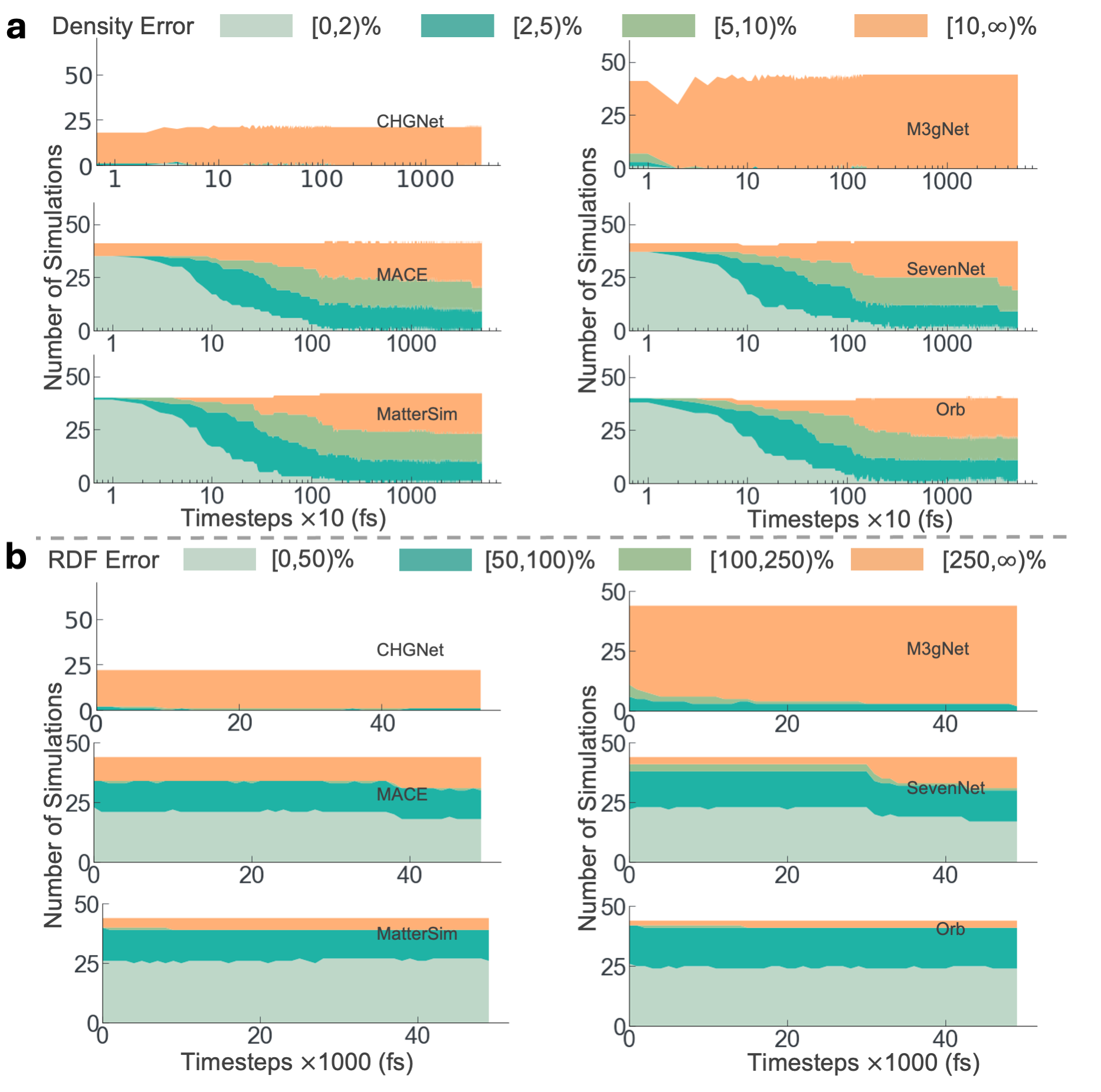}
    \caption{\textbf{Temporal evolution on \minxpocc{} reveals divergent stability patterns among UMLFFs.} \textbf{a}, Density error evolution during MD simulations for \minxpocc{}. \textbf{b}, Radial distribution function (RDF) error evolution showing atomic spatial organization accuracy with error ranges for \minxpocc{}}
    \label{fig:occ}
    \vspace{-0.2in}
\end{figure}

Supplementary~\Cref{fig:occ} shows temporal evolution density and rdf for \minx-POCC. \textbf{a}, Density error evolution during MD simulations with stacked areas representing error distributions across four ranges ([0,2)\%, [2,5)\%, [5,10)\%, [10,$\infty$)\%). Simulation timesteps shown on logarithmic scale to capture behavior across multiple time regimes. \textbf{b}, RDF error evolution showing atomic spatial organization accuracy with error ranges ([0,50)\%, [50,100)\%, [100,250)\%, [250,$\infty$)\%). Results demonstrate that even the stable models unable to converge to consistent error ranges for the disordered structure.

\begin{figure}[!ht]
    % \vspace{-0.3in}
    \centering
    \includegraphics[width=0.95\textwidth]{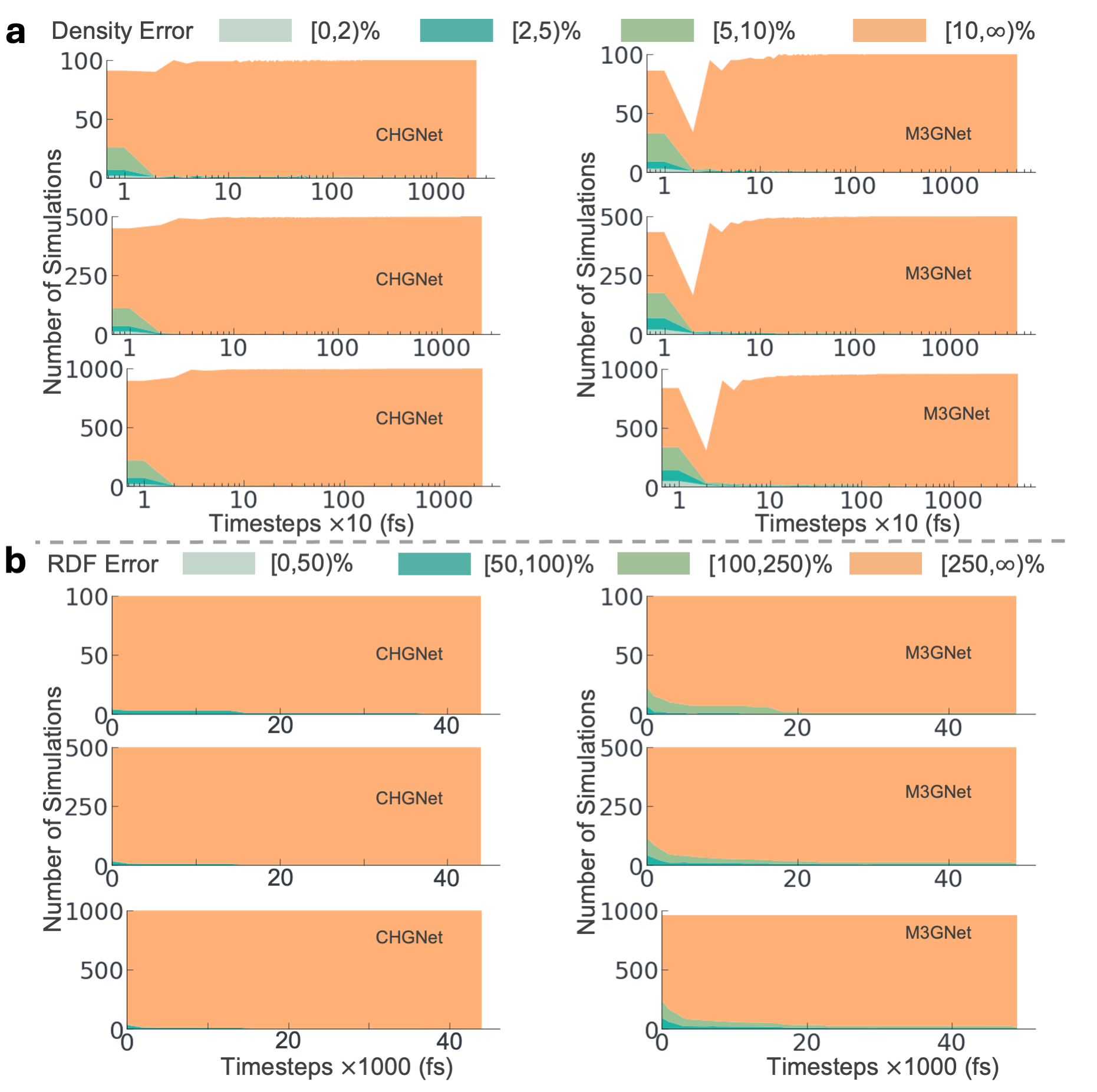}
    \caption{\textbf{(a)} Temporal evolution of the density error using \chgnet{} and \mgnet{}. \textbf{(b)} Corresponding temporal radial distribution function (RDF) error evolution for the same systems. Note, in both panels, the first, second, and third rows correspond to simulations of 100, 500, and 1,000 systems, respectively.}
     \label{fig:temporal-evolution-den-chg-mg}
    \vspace{-0.2in}
\end{figure}

\begin{figure}[!ht]
    % \vspace{-0.3in}
    \centering
    \includegraphics[width=0.95\textwidth]{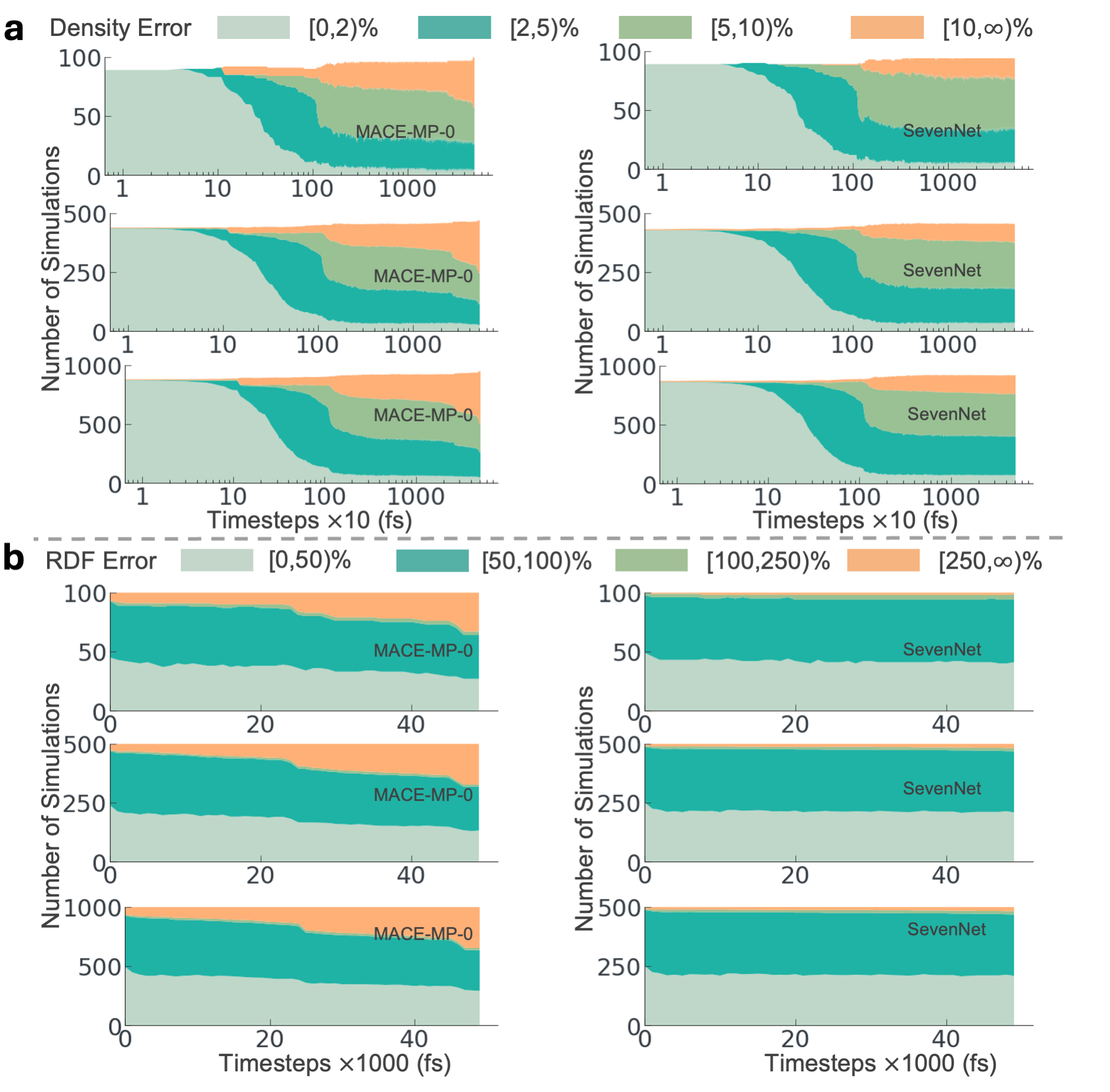}
    \caption{\textbf{(a)} Temporal evolution of the density error using \mace{} and \sevennet{}. \textbf{(b)} Corresponding temporal radial distribution function (RDF) error evolution for the same systems. Note, in both panels, the first, second, and third rows correspond to simulations of 100, 500, and 1,000 systems, respectively.}
    \label{fig:temporal-evolution-den-mace-seven}
    \vspace{-0.2in}
\end{figure}

\begin{figure}[!ht]
    % \vspace{-0.3in}
    \centering
    \includegraphics[width=0.95\textwidth]{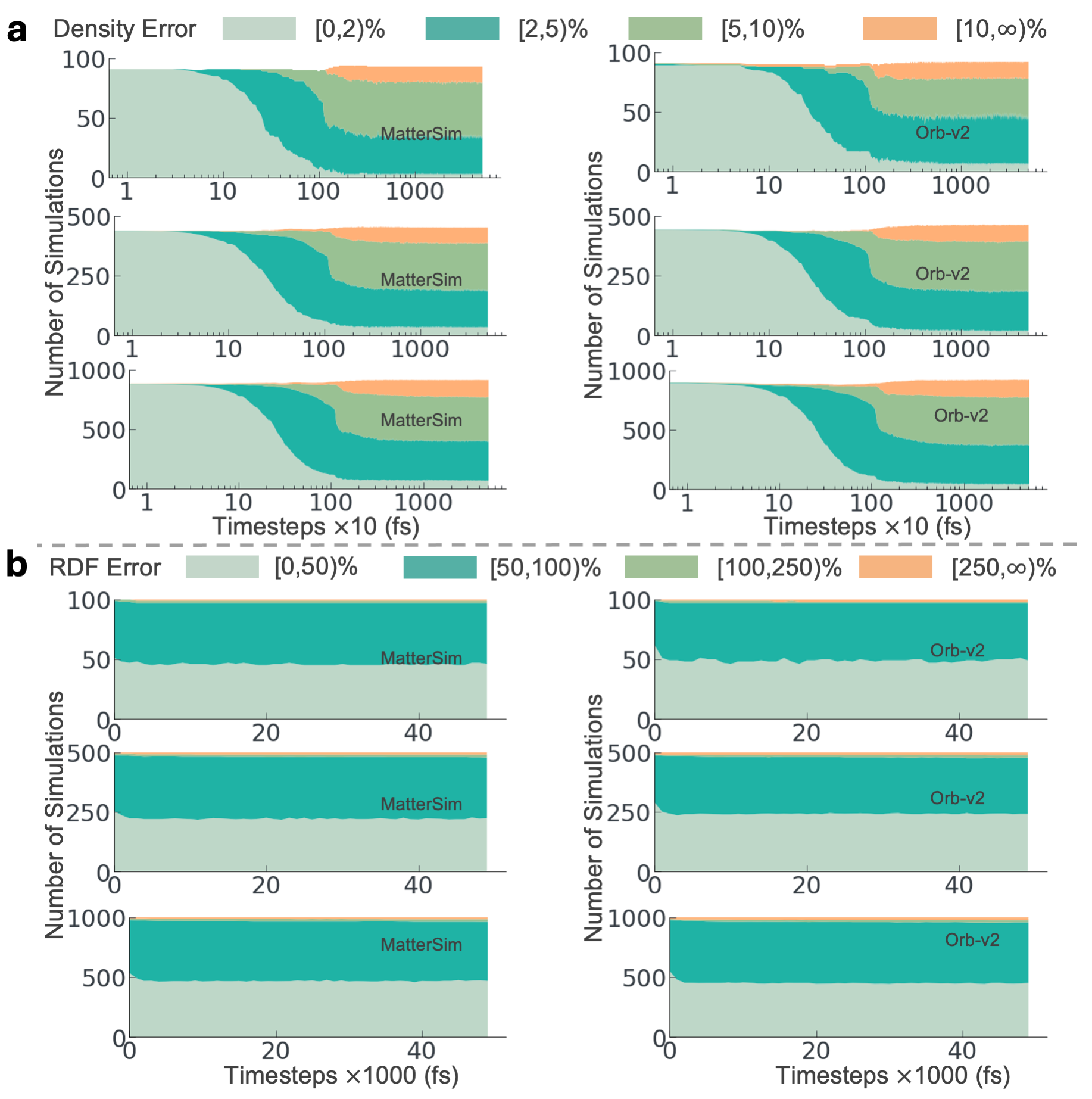}
    \caption{\textbf{(a)} Temporal evolution of the density error using \mattersim{} and \orb{}. \textbf{(b)} Corresponding temporal radial distribution function (RDF) error evolution for the same systems. Note, in both panels, the first, second, and third rows correspond to simulations of 100, 500, and 1,000 systems, respectively.}
    \label{fig:temporal-evolution-den-matter-orb}
    \vspace{-0.2in}
\end{figure}

% \clearpage

\section{Failure Analysis}
\label{app:failure}
To understand the origins of UMLFF performance limitations, we conducted a systematic analysis connecting training data characteristics with observed failure modes, focusing on pairwise element interactions and bond length accuracy. Interatomic bond lengths represent fundamental structural parameters governing chemical reactivity, phase stability, and mechanical properties, making accurate reproduction essential for reliable materials behavior prediction.

\subsection{Training Data Bias} 
\label{app:training_bias}
Comprehensive atomic pair analysis across all models (Figure 6a) reveals systematic patterns in bond length prediction accuracy that transcend individual model architectures. Most strikingly, bonds involving oxygen (atomic number 8) display pronounced accuracy across all evaluated UMLFFs, manifesting as distinct vertical blue regions in the error heatmaps. This universal oxygen bias is particularly evident in \textsc{MatterSim}, \textsc{Orb}, \textsc{MACE}, and \textsc{SevenNet}, indicating systematically superior performance for oxygen-containing pairs compared to other elemental combinations. This consistent pattern suggests a fundamental training data bias. Most UMLFFs in our evaluation are trained on MPtrj or its derivatives, which are predominantly composed of oxide-based systems reflecting the focus on ceramic, mineral, and energy materials in computational databases. The prevalence of oxygen-containing compounds in these training datasets results in models that excel at predicting bonds involving oxygen while struggling with less represented elemental combinations.

To quantify this training data bias, we analyzed the correlation between bond length prediction errors and atomic pair frequencies in the MPtrj training dataset (Figure 6b). By examining first peak positions in partial radial distribution functions---which provide precise measurements of interatomic distances for each atomic pair combination---we compared UMLFF bond errors with the occurrence frequency of these pairs in MPtrj. Despite substantial scatter, we observe that frequently encountered atomic pairs exhibit lower errors, while underrepresented pairs suffer substantially higher prediction errors. Notably, some frequently represented pairwise interactions still exhibit high errors, while none of the low-frequency pairs achieve low errors. \textcolor{black}{This could potentially be attributed to the selective use of Hubbard+U corrections to transition metals (V, W, Fe, Ni, Co, Cr, Mo, Mn) only in the presence of O or F indeed creates two incompatible potential energy surfaces in the reference DFT data. A low-energy PBE surface and a high-energy PBE+U surface. As MLIPs are continuous functions with finite cutoffs, they are forced to interpolate between these disjoint surfaces when O or F atoms enter metal coordination spheres, resulting in non-physical repulsive walls, where the model incorrectly predicts energy, forces and derived quantities \cite{warford2026better}}. This asymmetric relationship indicates that while the same atomic pair may be frequently present in the training data, the local chemical environments can be substantially different between training and evaluation systems. Thus, pairwise frequency represents a necessary but not sufficient condition for low prediction errors. This observation underscores the importance of not only chemical diversity in training datasets but also environmental diversity---ensuring that atomic pairs are encountered across a wide range of local coordination environments, bonding configurations, and chemical contexts to achieve truly universal force field behavior. See Supplementary~\Cref{fig:bond_freq-plots} for mean bond error versus MPTrj bond frequency plots for each model.

% \textcolor{blue}{To contextualize UMLFF performance relative to their training data, we compared experimental densities with DFT-computed values from the Materials Project database (source of training structures for MPtrj and related datasets) for a subset of 10 minerals (Table~\ref{table:baseline}). DFT predictions show typical errors of 2.5\%, consistent with PBE functional accuracy for structural properties. In contrast, UMLFFs exhibit average density errors of 5.2\%---approximately 2--3$\times$ larger than their DFT training data errors. This demonstrates that models \textit{underperform} their training data accuracy: despite achieving energy MAEs of 20--30 meV/atom on computational benchmarks (approaching DFT self-consistency)~\cite{miret2025energy, fu2025learning, wood2025family}, they exhibit substantially larger errors for derived properties. This performance gap reveals fundamental limitations in how models learn and generalize from energy/force training to property predictions, with density and elastic modulus errors substantially exceeding typical DFT prediction errors.}

\textcolor{black}{To contextualize UMLFF performance relative to their training data, we compared experimental densities, shear modulus and bulk modulus with DFT-computed values from the Materials Project database (source of training structures for MPtrj and related datasets) for a subset of 10 minerals (Supplementary Tables~\ref{table:baseline}, \ref{table:baseline_shear}, and \ref{table:baseline_bulk}.
% ,Table~\ref{table:baseline_shear},Table~\ref{table:baseline_bulk}). 
DFT predictions show typical errors of 2.5\%, consistent with PBE functional accuracy for structural properties. In contrast, UMLFFs exhibit average density errors of 5.2\%---approximately 2--3$\times$ larger than their DFT training data errors. Similarly, the experimental values of shear and bulk moduli show systematic under performance of UMLFFs relative to DFT. For instance the large errors in the shear modulus predictions (Table 2) for Calcite (188.6\%) and Lead (-537.2\%) using UMLFFs, compared to the relatively small DFT errors of 2.0\% and 14.7\%, respectively, illustrate cases where the models fail to reproduce even
the accuracy of their underlying training data. A similar trend is observed for the bulk modulus (Table 2), where UMLFF predictions show large deviations for Calcite (175.7\%) and Lead (-40.9\%), whereas the corresponding DFT errors are relatively modest at 12.7\% and 10.6\%, respectively. This demonstrates that models often exhibit varying trends from their underlying training data. Despite achieving energy MAEs of 20--30 meV/atom on computational benchmarks (approaching DFT self-consistency)~\cite{miret2025energy, fu2025learning, wood2025family}, they exhibit substantially larger errors for derived properties. This error could partly be attributed to the selective use of Hubbard+U corrections to transition metals (V, W, Fe, Ni, Co, Cr, Mo, Mn) only in the presence of O or F indeed creates two incompatible potential energy surfaces in the reference DFT data. A low-energy PBE surface and a high-energy PBE+U surface. As MLIPs are continuous functions with finite cutoffs, they are forced to interpolate between these disjoint surfaces when O or F atoms enter metal coordination spheres, resulting in non-physical repulsive walls, where the model incorrectly predicts energy, forces and derived quantities \cite{warford2026better}. However, this large performance gap reveals limitations in how models learn and generalize from energy/force training to property predictions, with density and elastic modulus errors substantially exceeding typical DFT prediction errors}

\subsection{Curvature Analysis of Learned Pair-wise Interactions} 
\label{app:curvature}
To probe the mechanistic origins of simulation failures and poor elastic property prediction, we analyzed the mathematical character of learned pairwise force-displacement interactions through curvature analysis of attractive and repulsive regimes for homonuclear (X–X) interactions (Figure 6c,d; see Methods for details). Complete force displacement plots for the homonuclear (X-X) and heteronuclear (X-O) interactions are included (see Supplementary \Cref{sec_pairwise,app:heteronuclear}). We quantify interaction smoothness through curvature metrics, where higher values indicate mathematical roughness that can lead to numerical instabilities during MD integration and compromise the accurate representation of force-displacement relationships required for elastic property prediction. For reference, we compare all models against the analytically smooth Morse potential, which represents the theoretical baseline for well-behaved interatomic interactions. 

The analysis exposes striking variations in potential smoothness that correlate directly with observed simulation stability patterns. Models with poor MD stability---\chgnet{} and \mgnet{}---exhibit severe mathematical roughness with curvature values exceeding the Morse baseline by 10$^2$–10$^3$ times, particularly in the repulsive region where short-range interactions dominate (see Figure 6c and Supplementary \Cref{sec_pairwise}). This roughness manifests as noisy force-displacement curves (see Supplementary \Cref{sec_pairwise,app:heteronuclear}) with rapid oscillations that necessitate prohibitively small integration timesteps, explaining the computational failures observed during MD simulations. Conversely, \orb{} exhibits remarkably low curvature values approaching the Morse baseline, consistent with its exceptional simulation completion rates and smooth pairwise force representations. In the attractive regime (Figure 6d), where long-range interactions dominate, most UMLFFs demonstrate more uniform behavior comparable to classical potentials, suggesting that short-range repulsive interactions represent the primary challenge for current model architectures.

Accurate elastic tensor calculation requires precise force-displacement relationships under deformation, particularly for capturing the second-order derivatives of the potential energy surface. Models with noisy, poorly resolved short-range interactions cannot reliably represent the subtle force variations required for mechanical property prediction. Furthermore, learning the force-displacement relationships reasonably, and even smoothly, does not guarantee that the gradients are also learned correctly, explaining why even structurally stable models like \orb{} can fail at elastic tensor prediction despite excellent performance in other areas. This correlation between potential smoothness, learned gradients, simulation reliability, and mechanical property accuracy demonstrates that training data limitations manifest through multiple pathways: compositional bias affects bond length accuracy, while inadequate representation of short-range interactions or their gradients compromises both MD stability and elastic property prediction.

\textbf{Elastic Tensor Prediction.} Our analysis reveals three fundamental reasons for poor elastic property prediction. First, elastic tensors require accurate second derivatives of the potential energy surface. Current training protocols focus exclusively on energies (0th derivatives) and forces (1st derivatives), leaving higher-order derivatives poorly constrained. This explains why $C_{11}$ components, which relate to bulk compressibility, show superior accuracy compared to shear components ($C_{44}$, $C_{66}$). Second, the disconnect between \orb{'s} excellent structural stability and elastic property prediction failure reveals that smooth energy landscapes do not guarantee accurate force derivatives. Models can interpolate forces adequately while completely misrepresenting their gradients---a fundamental limitation that has profound implications for mechanical property prediction. Third, training data bias toward oxide systems and limited mechanical property representation means models have never learned the physics of elastic deformation in diverse chemical environments. The systematic correlation between prediction accuracy and atomic pair frequency in training data demonstrates that current ``universal'' force fields mostly represent sophisticated interpolation schemes within familiar chemical spaces rather than truly universal physical models (for instance, in comparison to DFT).

\section{Details of RDF Calculation for Each Minerals}\label{sec_rdf_calc}

\begin{figure}[!ht]
    % \vspace{-0.3in}
    \centering
    \includegraphics[width=0.85\textwidth]{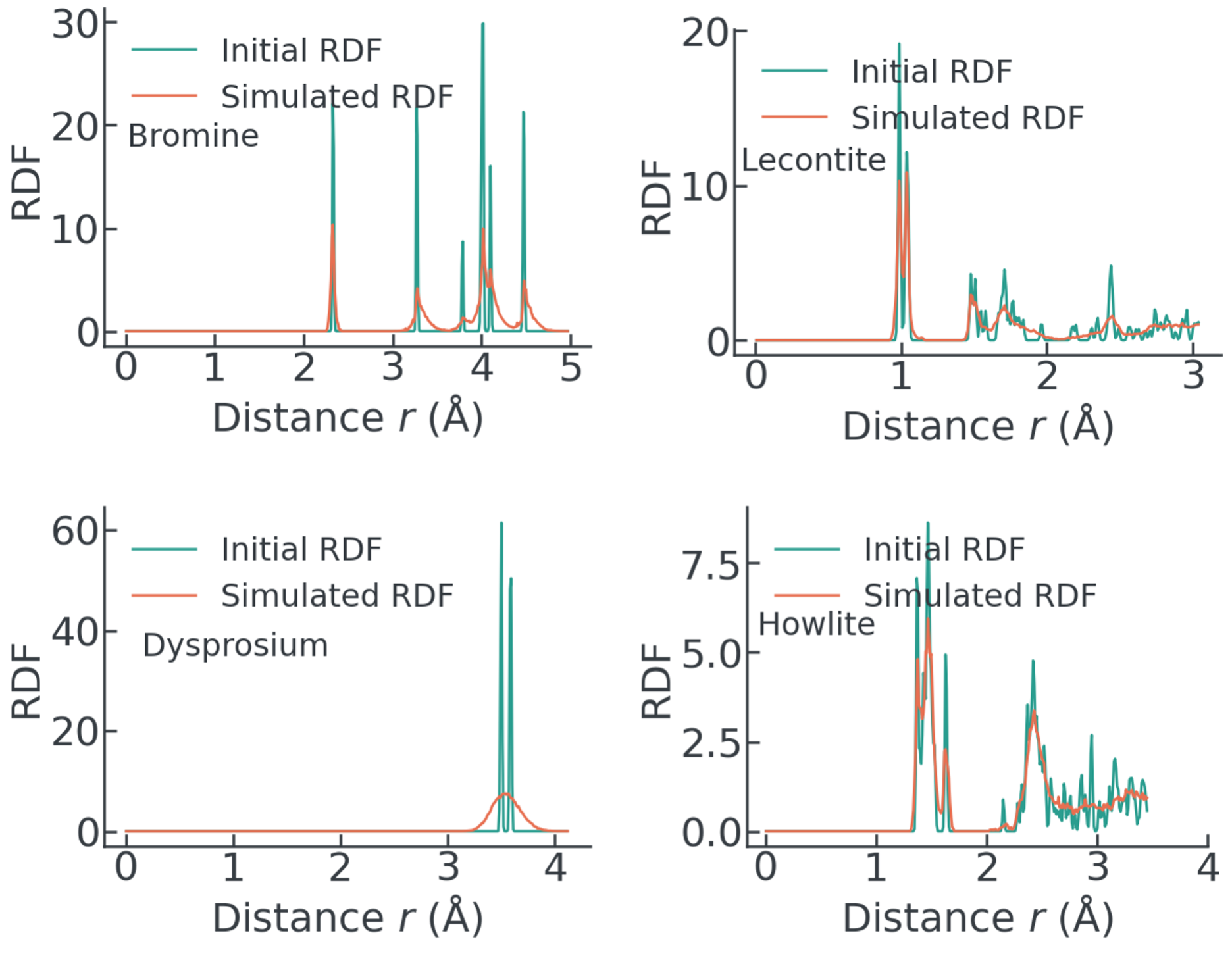}
    \caption{RDFs Comparison: Green shows the initial structure RDF which is averaged over 1000 perturbed configuration with a noise of 0.005 \AA, while brown shows simulated RDF averaged over 1000 frames}
    \label{fig:RDF plot}
    \vspace{-0.2in}
\end{figure}

To evaluate the dynamic structural stability of MD simulations, we compare the simulated RDFs with those of naturally occurring minerals. To do this, we first read all CIFs using ASE and replicate the unit cells to construct supercells containing approximately 100–200 atoms using the same protocol explained in method section.  Further, to obtain a smooth and representative RDF, we introduce random Gaussian noise with a standard deviation of 0.005 \,\AA to the atomic positions of the original (unperturbed) configuration. For each configuration, this noise is added afresh 200 times, followed by RDF calculation after each perturbation. The final RDF is obtained by averaging over all 200 perturbed configurations. It is important to note that the noise is applied independently to the original structure in each iteration, not cumulatively on previously perturbed structures. Supplementary~\Cref{fig:RDF plot} shows the experimental and simulated RDF.

\clearpage

\section{Pairwise Energy and Force: Homo-nuclear Systems}\label{sec_pairwise}

To evaluate the performance of the UMLFFs in capturing pairwise interatomic interactions, we analyzed the energy profiles for each element. As shown in Supplementary ~\Cref{fig:pairwiseplots1,fig:pairwiseplots2,fig:pairwiseplots3,fig:pairwiseplots4}, the interaction curves are smooth after the minima, indicating stable repulsive behavior. However, significant fluctuations and noise are observed in the attractive region (i.e., before the minima), suggesting numerical instabilities. These observations are further corroborated by the corresponding pairwise force plots, Supplementary~\Cref{fig:pairwiseplots5,fig:pairwiseplots6,fig:pairwiseplots7,fig:pairwiseplots8,fig:pairwiseplots9}, obtained by computing the gradient of the energy curves with respect to distance, which highlights the instabilities more evident.

\begin{figure}[!ht]
    % \vspace{-0.3in}
    \centering
    \includegraphics[width=0.95\textwidth]{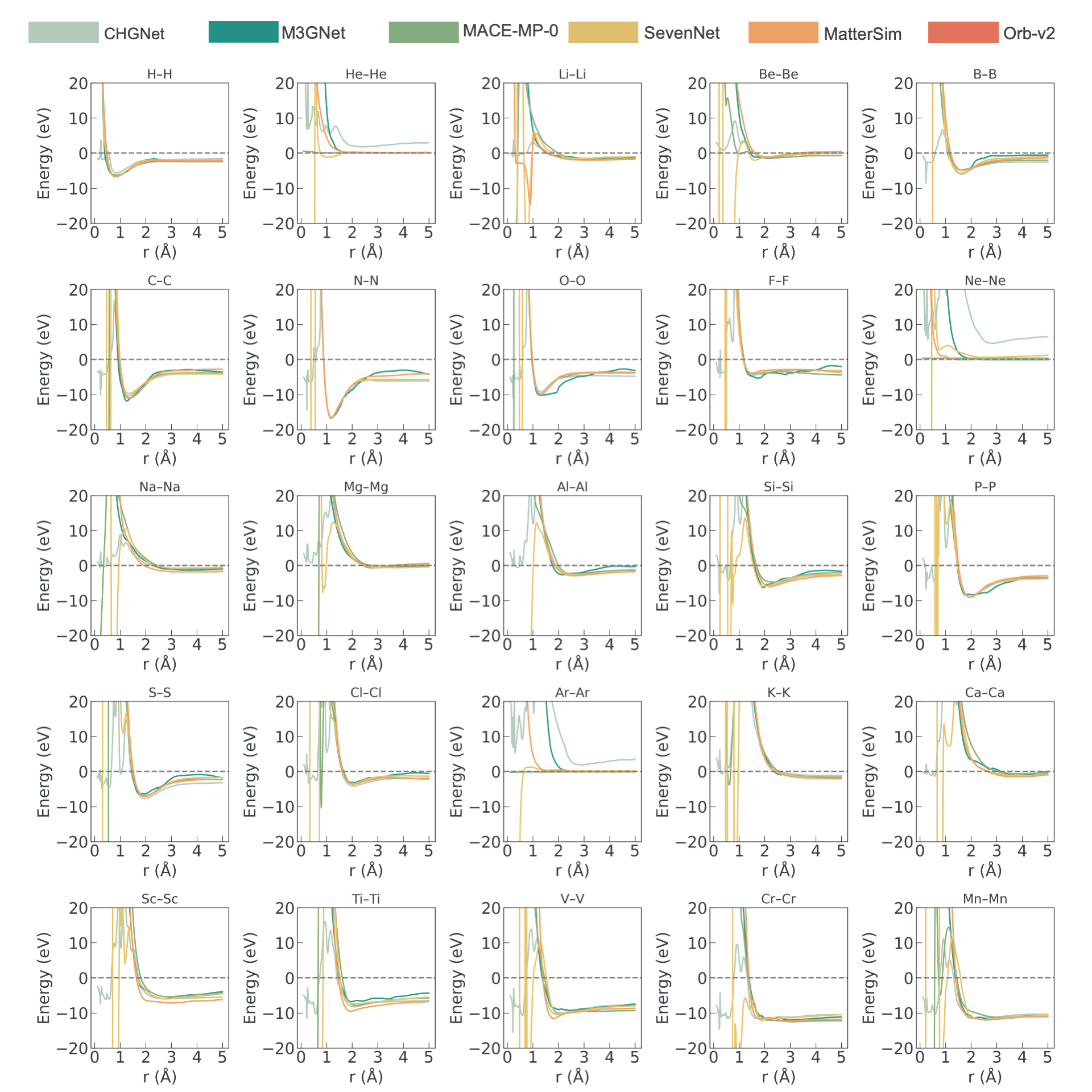}
    \caption{Pairwise energy plots}
    \label{fig:pairwiseplots1}
    \vspace{-0.2in}
\end{figure}

\begin{figure}[!ht]
    % \vspace{-0.3in}
    \centering
    \includegraphics[width=0.95\textwidth]{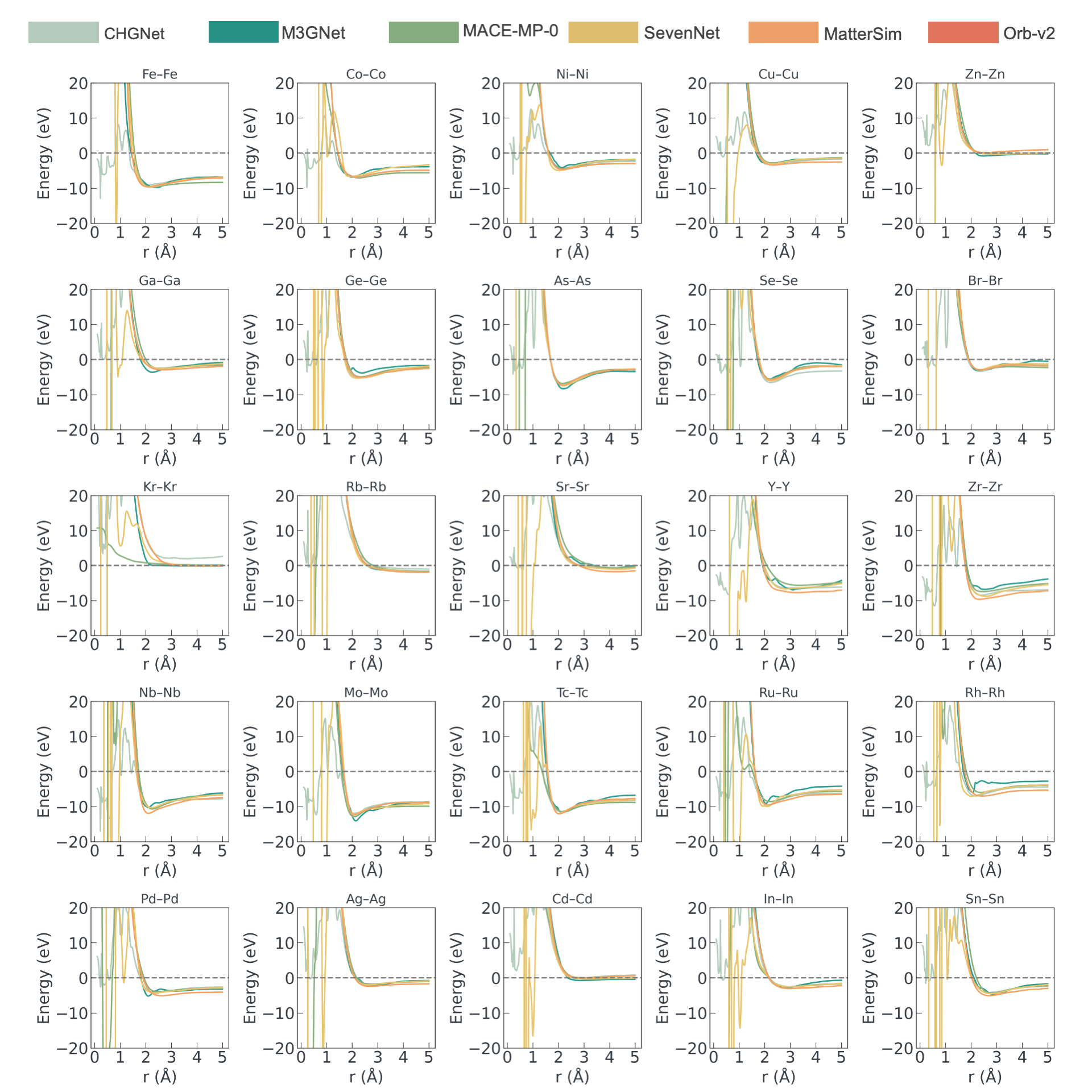}
    \caption{Pairwise energy plots}
    \label{fig:pairwiseplots2}
    \vspace{-0.2in}
\end{figure}

\begin{figure}[!ht]
    % \vspace{-0.3in}
    \centering
    \includegraphics[width=0.95\textwidth]{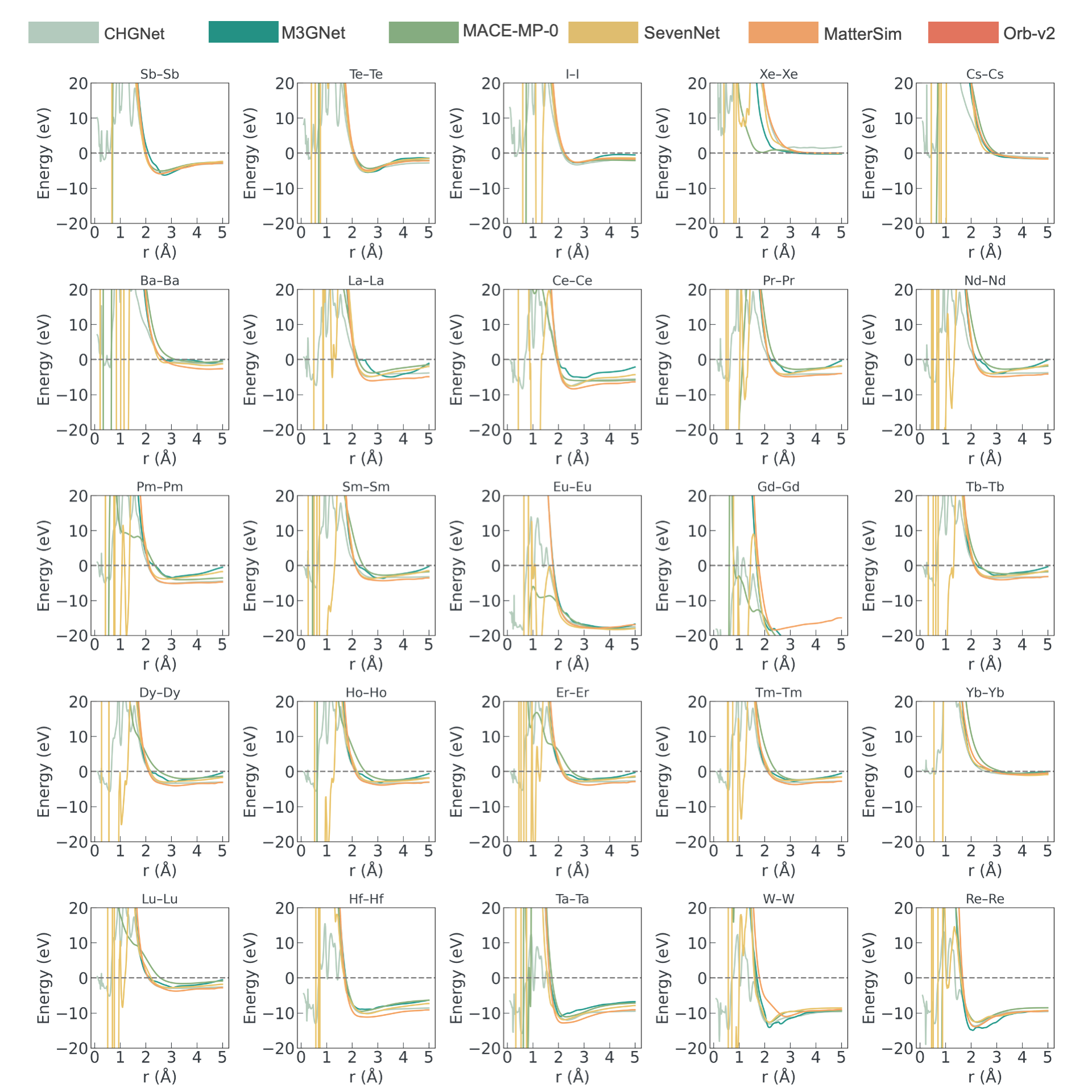}
    \caption{Pairwise energy plots}
    \label{fig:pairwiseplots3}
    \vspace{-0.2in}
\end{figure}
\begin{figure}[!ht]
    % \vspace{-0.3in}
    \centering
    \includegraphics[width=0.95\textwidth]{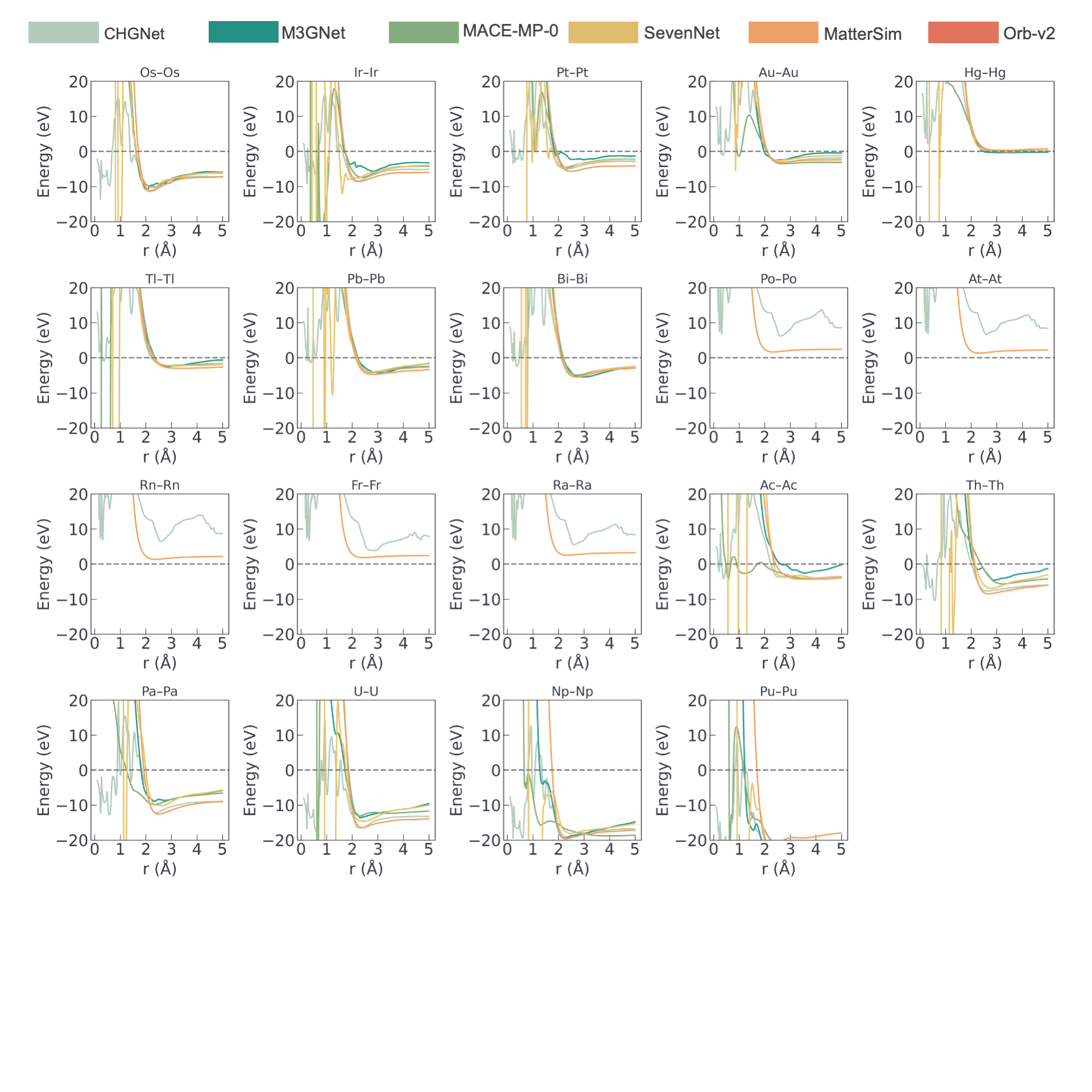}
    \caption{Pairwise Energy Plots}
    \label{fig:pairwiseplots4}
    \vspace{-0.2in}
\end{figure}

\begin{figure}[!ht]
    % \vspace{-0.3in}
    \centering
    \includegraphics[width=0.95\textwidth]{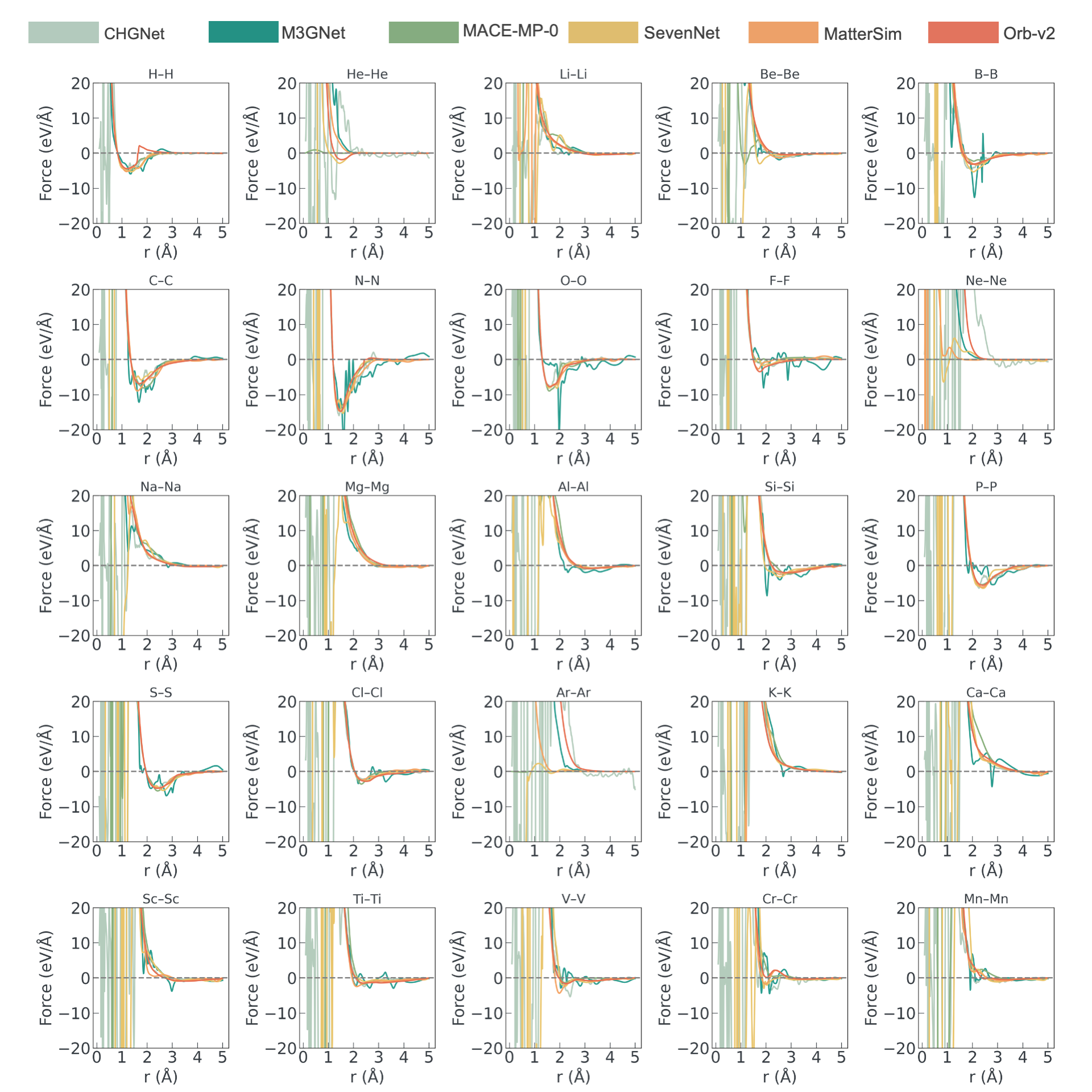}
    \caption{Pairwise Force Plots}
    \label{fig:pairwiseplots5}
    \vspace{-0.2in}
\end{figure}

\begin{figure}[!ht]
    % \vspace{-0.3in}
    \centering
    \includegraphics[width=0.95\textwidth]{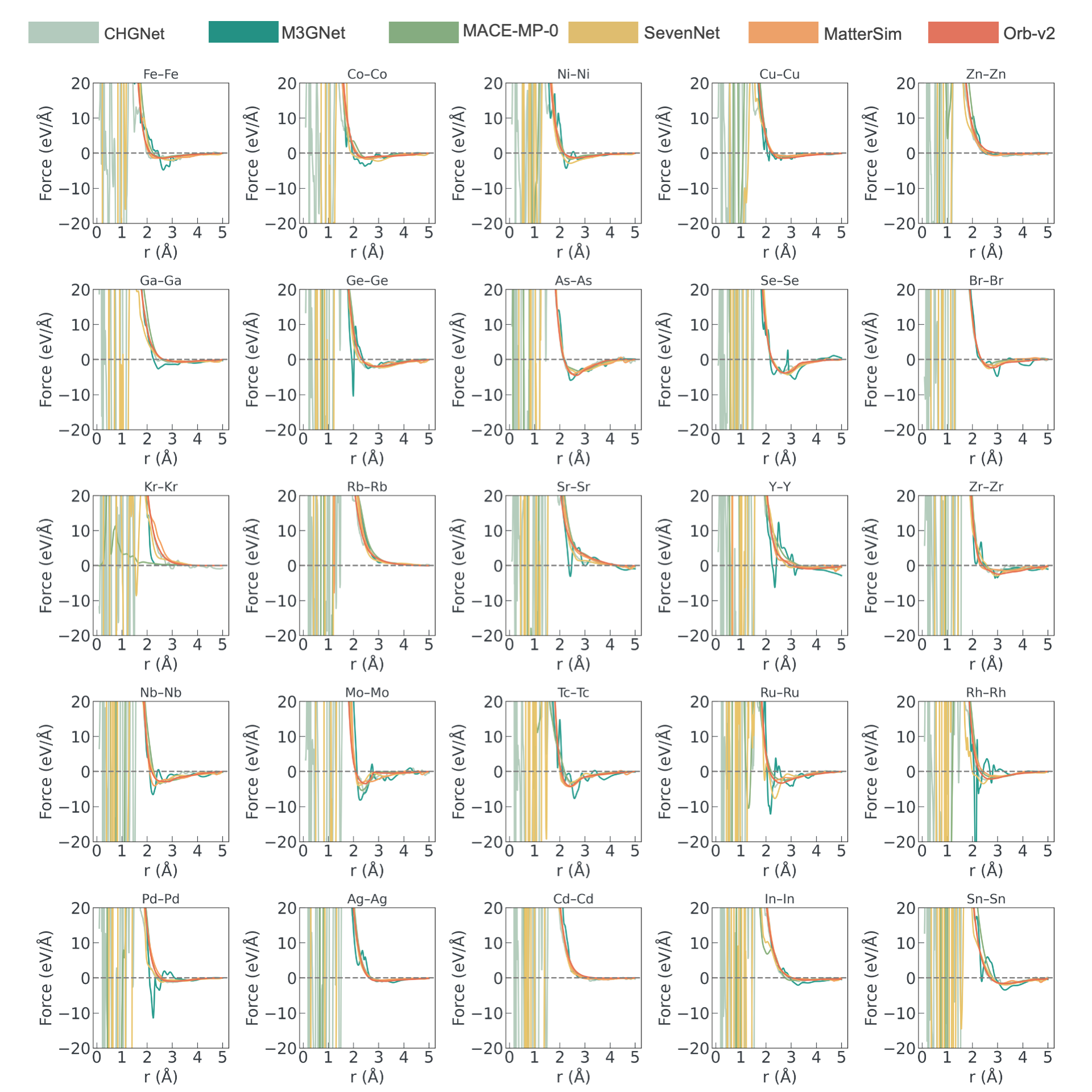}
    \caption{Pairwise Force Plots}
    \label{fig:pairwiseplots6}
    \vspace{-0.2in}
\end{figure}\begin{figure}[!ht]
    % \vspace{-0.3in}
    \centering
    \includegraphics[width=0.95\textwidth]{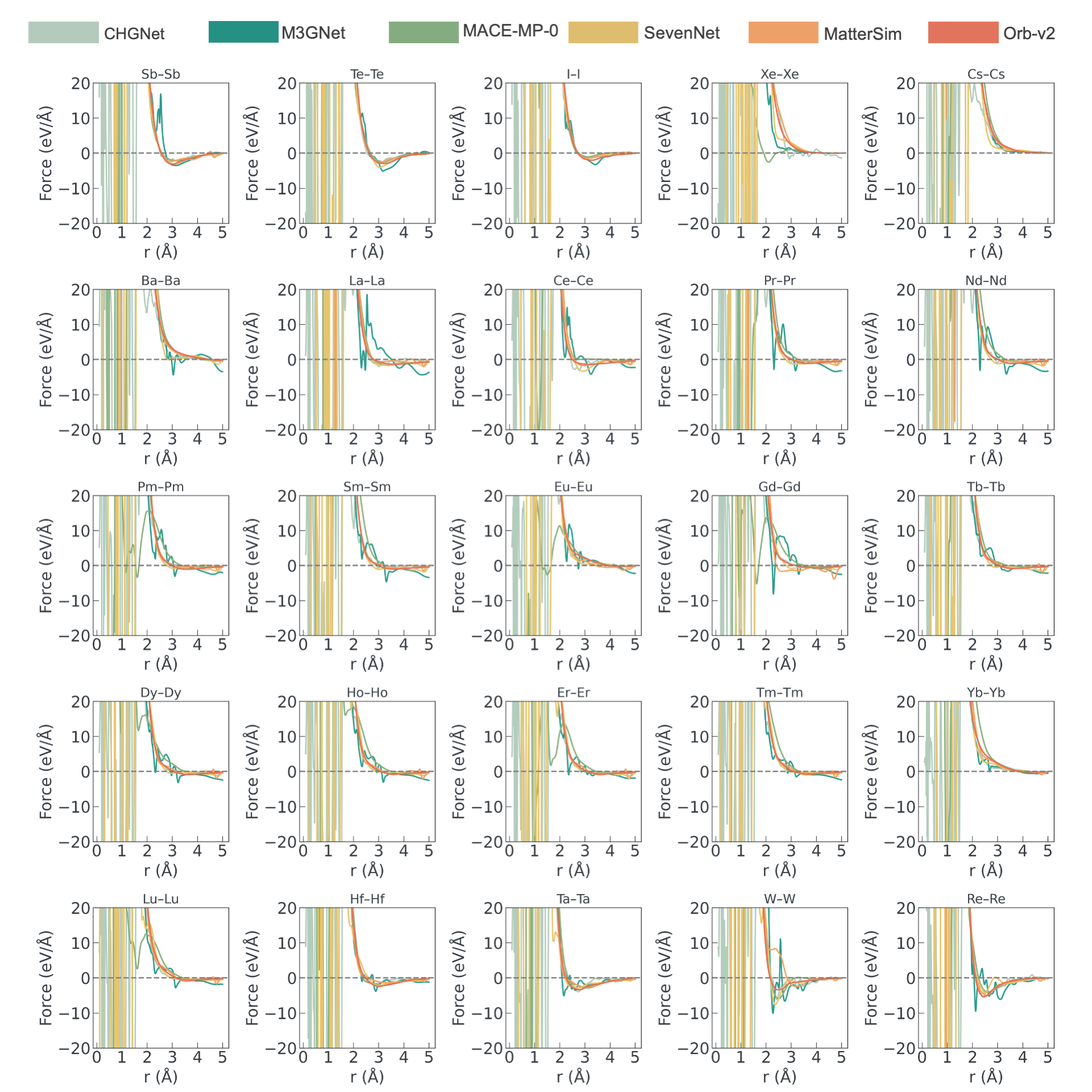}
    \caption{Pairwise Force Plots}
    \label{fig:pairwiseplots7}
    \vspace{-0.2in}

\end{figure}

\begin{figure}[!ht]
    % \vspace{-0.3in}
    \centering
    \includegraphics[width=0.95\textwidth]{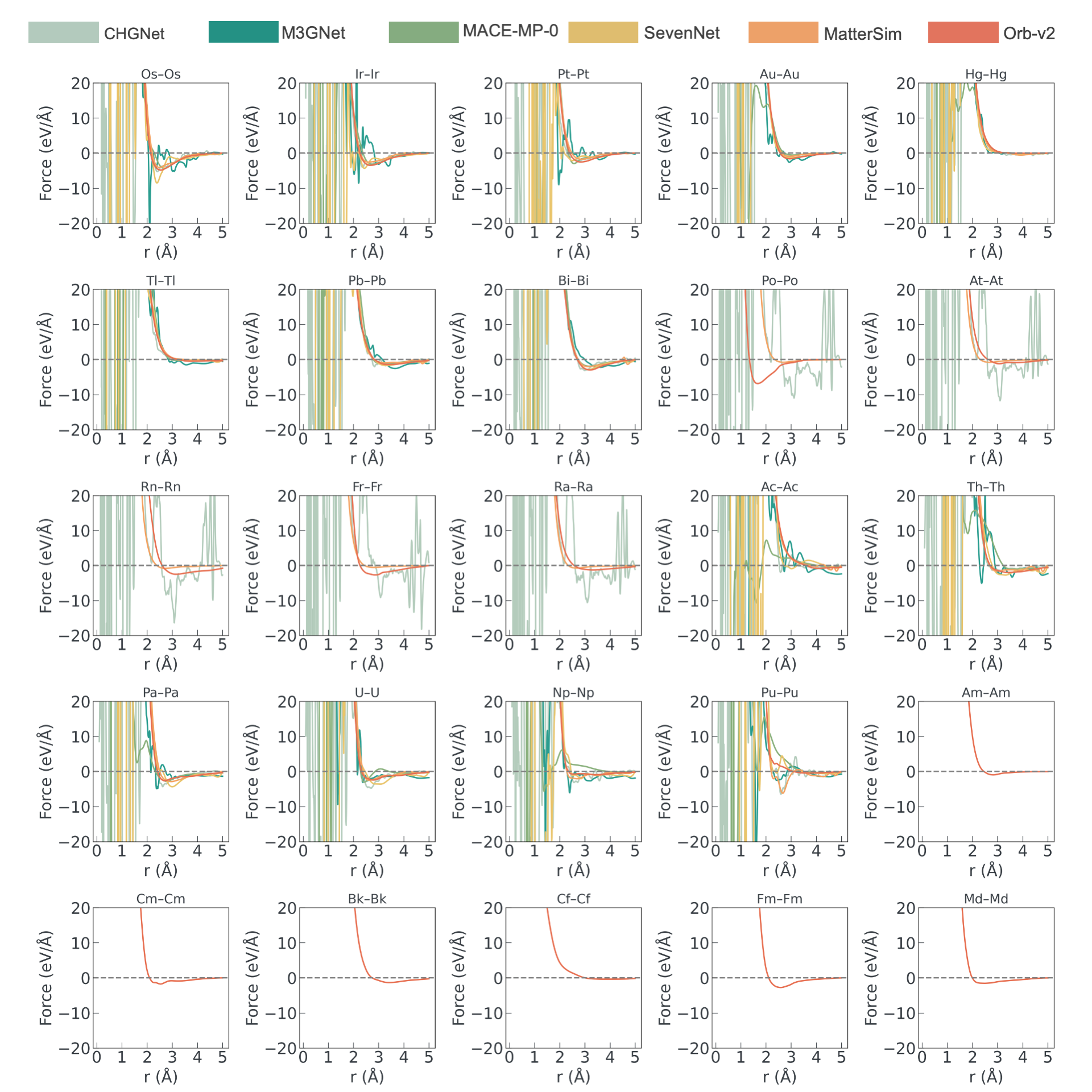}
    \caption{Pairwise Force Plots}
    \label{fig:pairwiseplots8}
    \vspace{-0.2in}
\end{figure}

\begin{figure}[!ht]
    % \vspace{-0.3in}
    \centering
    \includegraphics[width=0.95\textwidth]{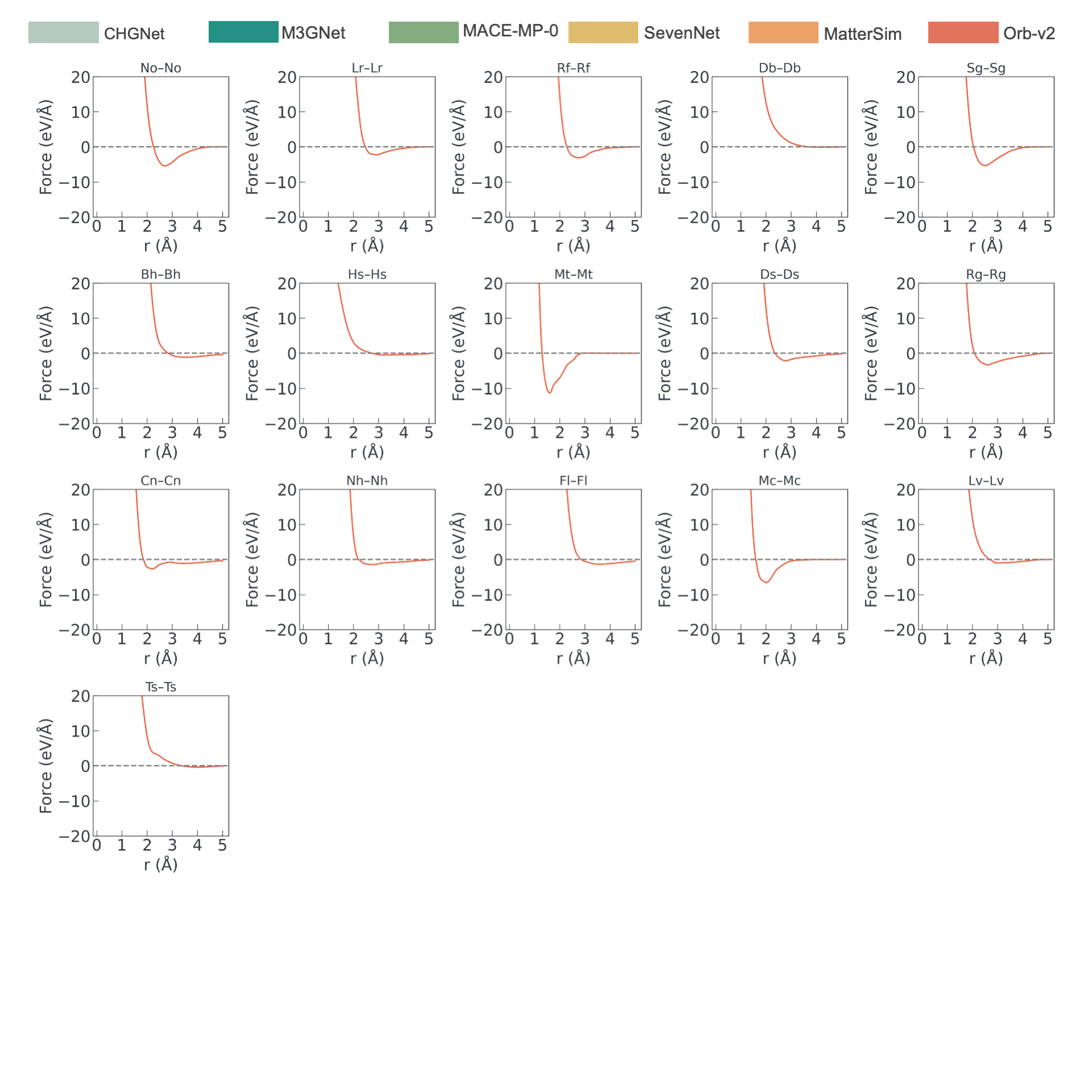}
    \caption{Pairwise Force Plots}
    \label{fig:pairwiseplots9}
    \vspace{-0.2in}
\end{figure}

\clearpage

\section{Pairwise Energy and Force: Hetero-nuclear Systems}\label{app:heteronuclear}
\begin{figure}[!ht]
    % \vspace{-0.3in}
    \centering
    \includegraphics[width=0.95\textwidth]{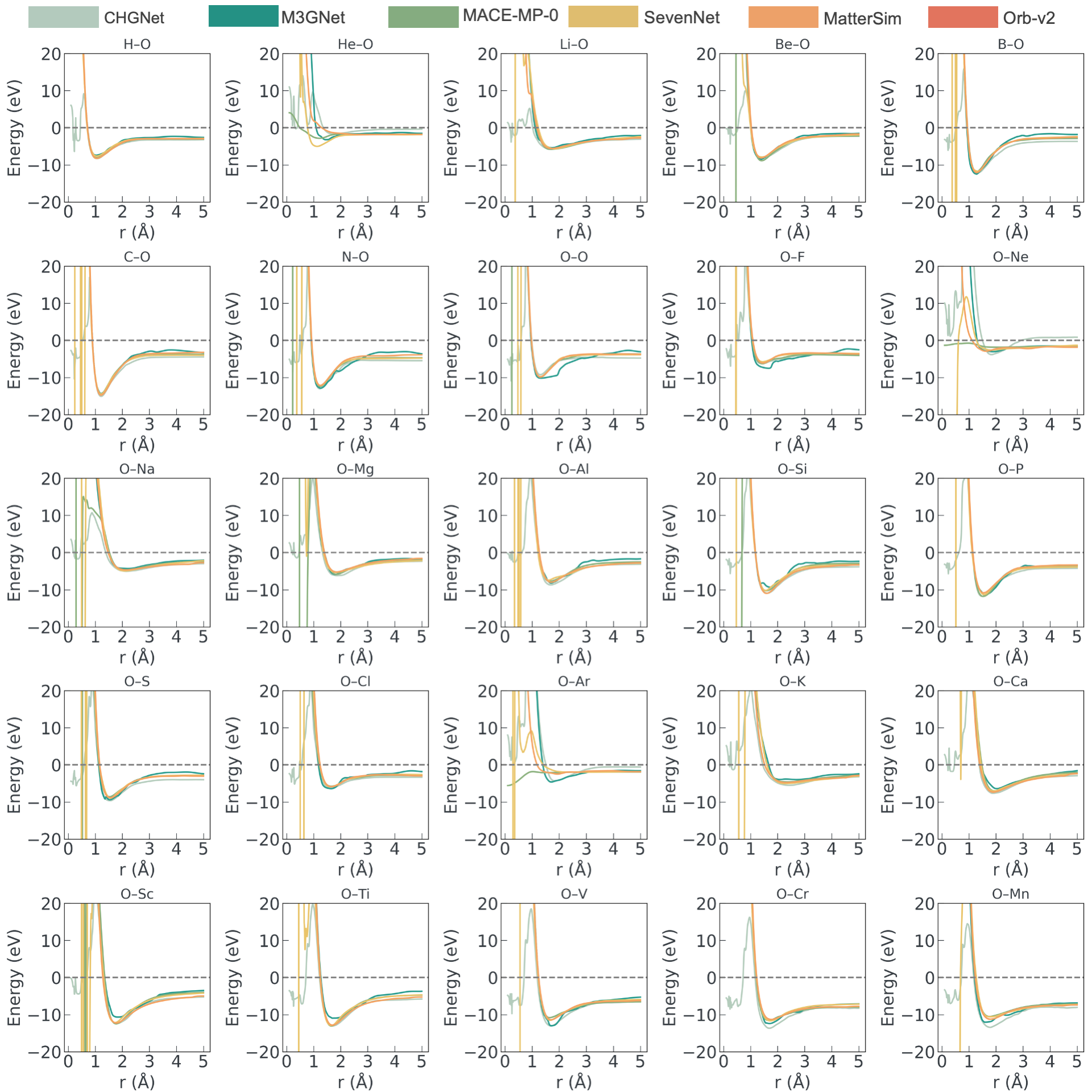}
    \caption{Pairwise Energy Plots}
    \label{fig:pairwiseplots10}
    \vspace{-0.2in}
\end{figure}

\begin{figure}[!ht]
    % \vspace{-0.3in}
    \centering
    \includegraphics[width=0.95\textwidth]{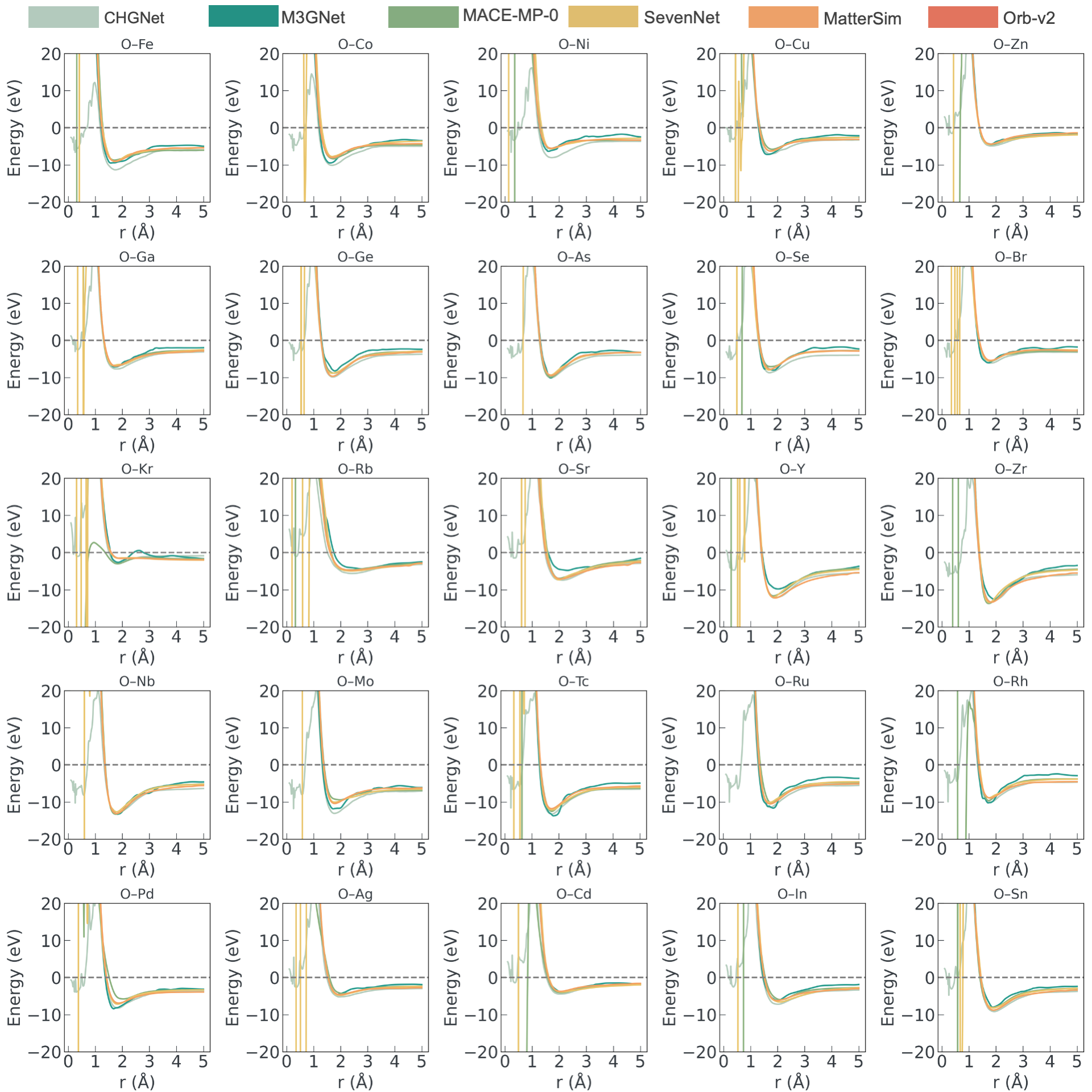}
    \caption{Pairwise Energy Plots}
    \label{fig:pairwiseplots11}
    \vspace{-0.2in}
\end{figure}

\begin{figure}[!ht]
    % \vspace{-0.3in}
    \centering
    \includegraphics[width=0.95\textwidth]{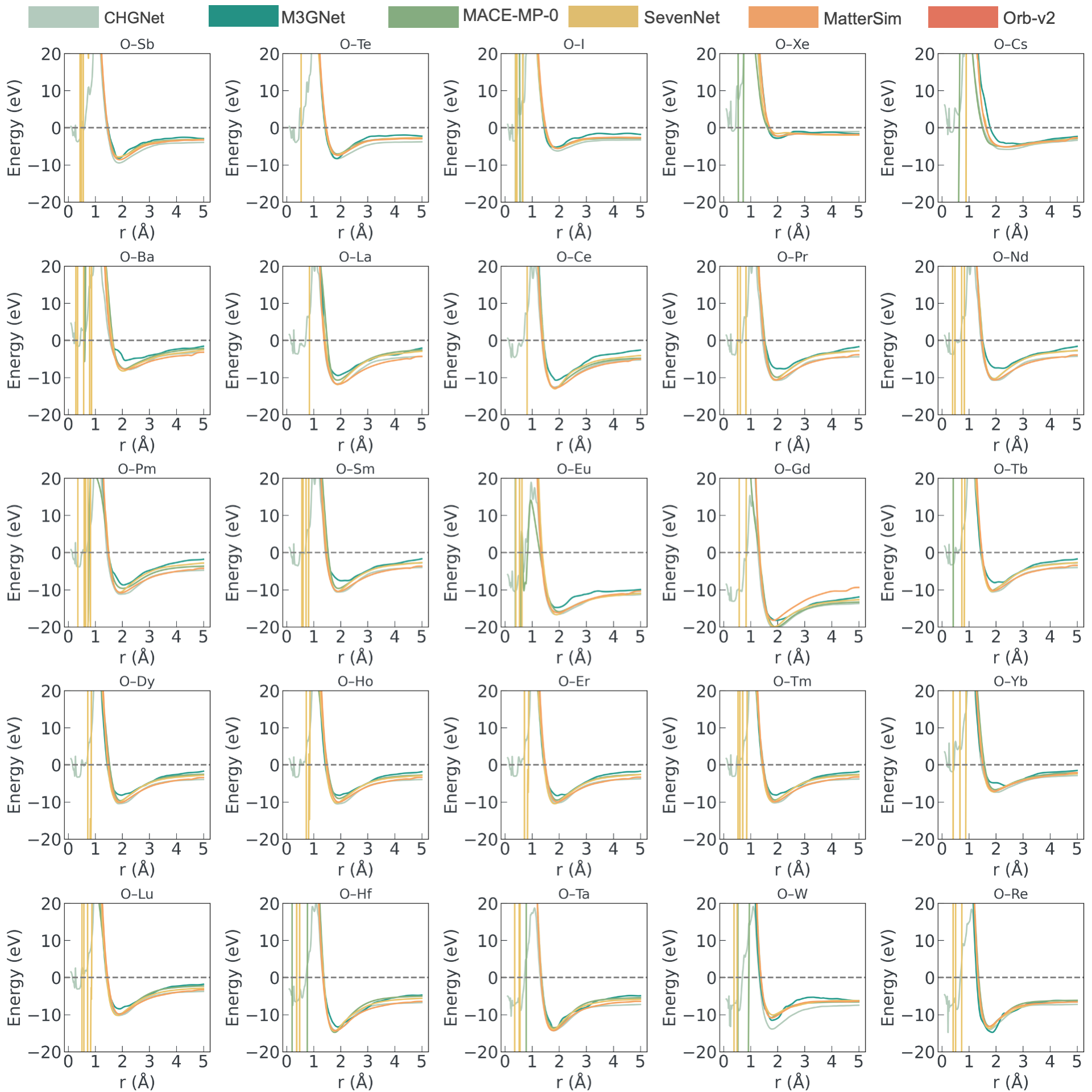}
    \caption{Pairwise Energy Plots}
    \label{fig:pairwiseplots12}
    \vspace{-0.2in}
\end{figure}

\begin{figure}[!ht]
    % \vspace{-0.3in}
    \centering
    \includegraphics[width=0.95\textwidth]{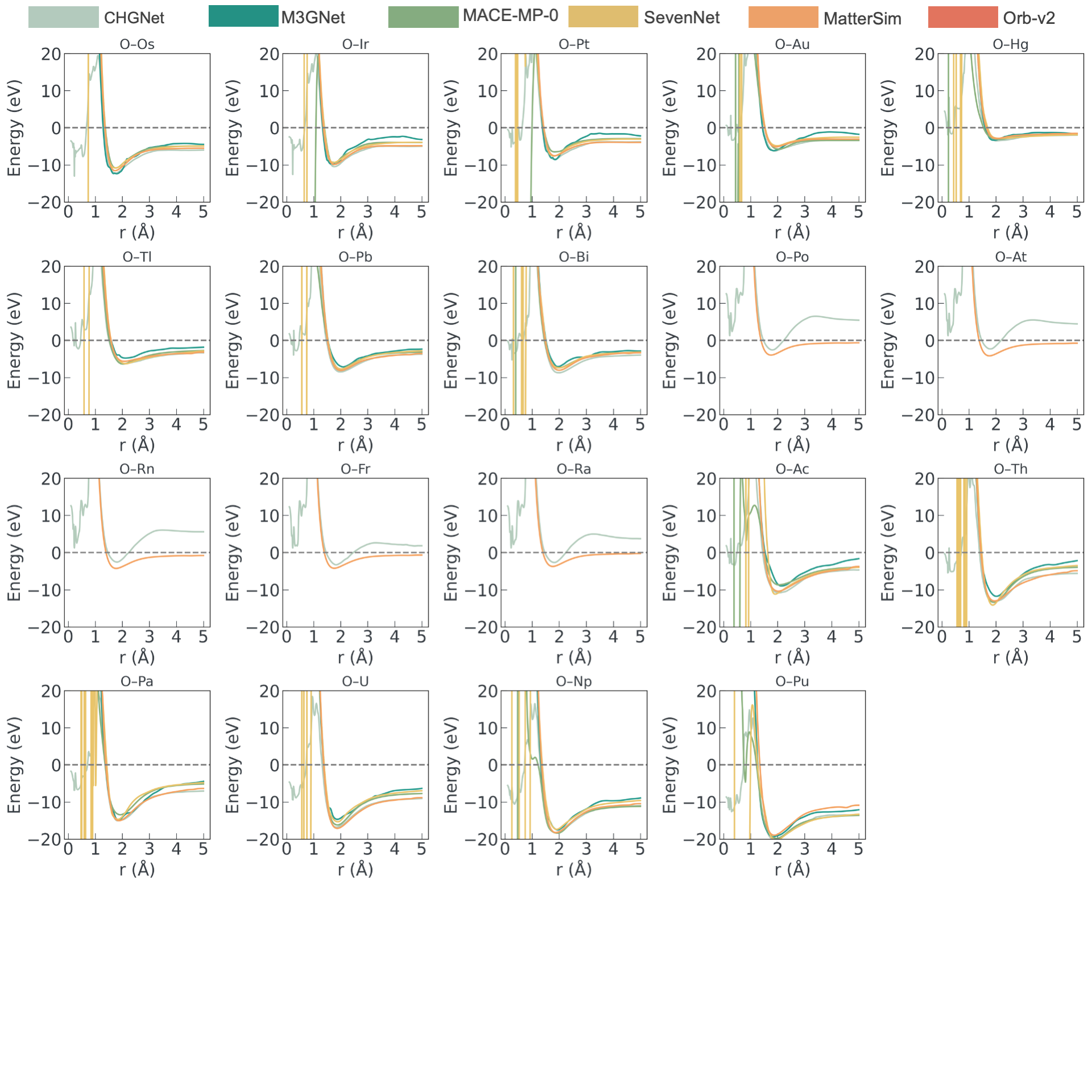}
    \caption{Pairwise Energy Plots}
    \label{fig:pairwiseplots13}
    \vspace{-0.2in}
\end{figure}

\begin{figure}[!ht]
    % \vspace{-0.3in}
    \centering
    \includegraphics[width=0.95\textwidth]{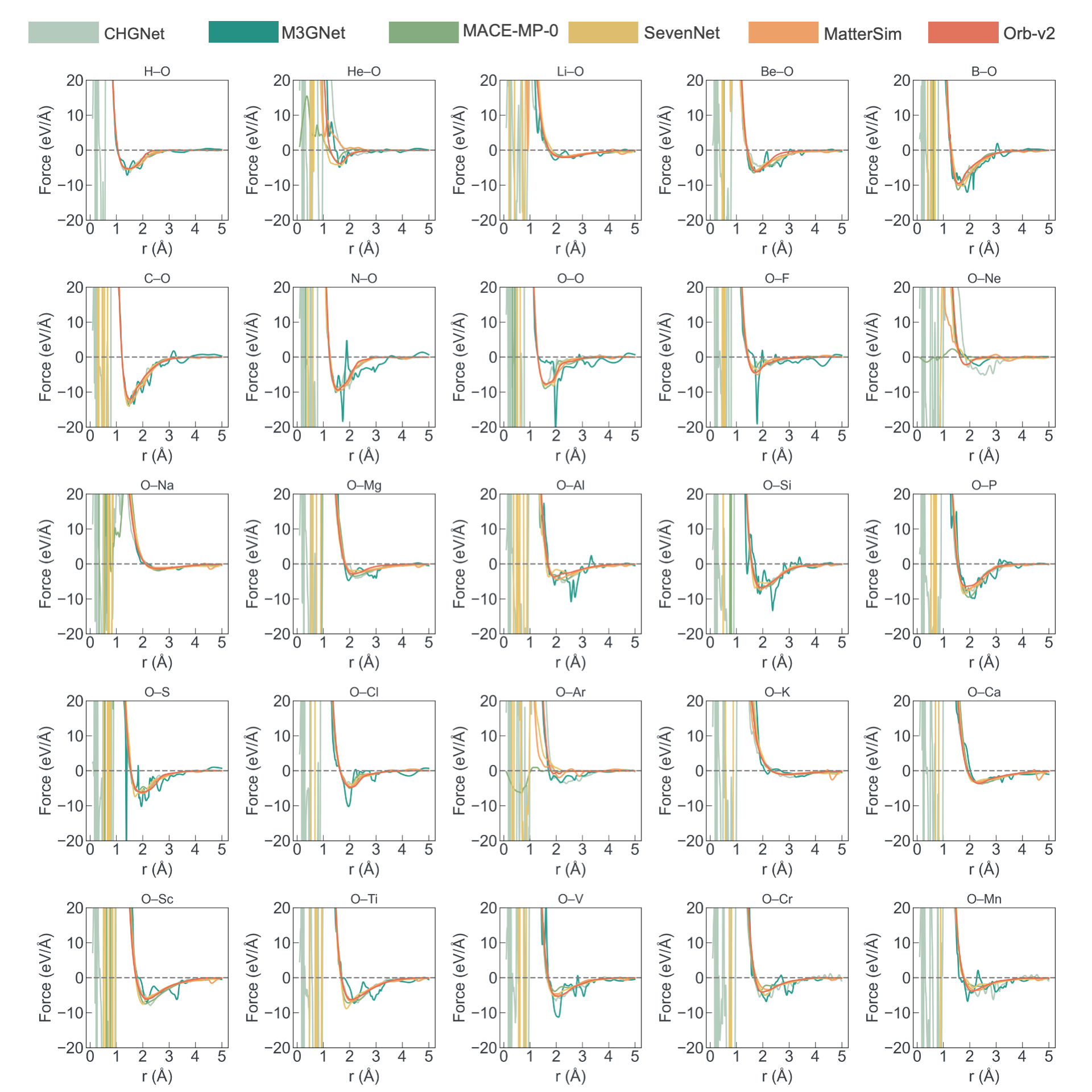}
    \caption{Pairwise Force Plots}
    \label{fig:pairwiseplots14}
    \vspace{-0.2in}
\end{figure}

\begin{figure}[!ht]
    % \vspace{-0.3in}
    \centering
    \includegraphics[width=0.95\textwidth]{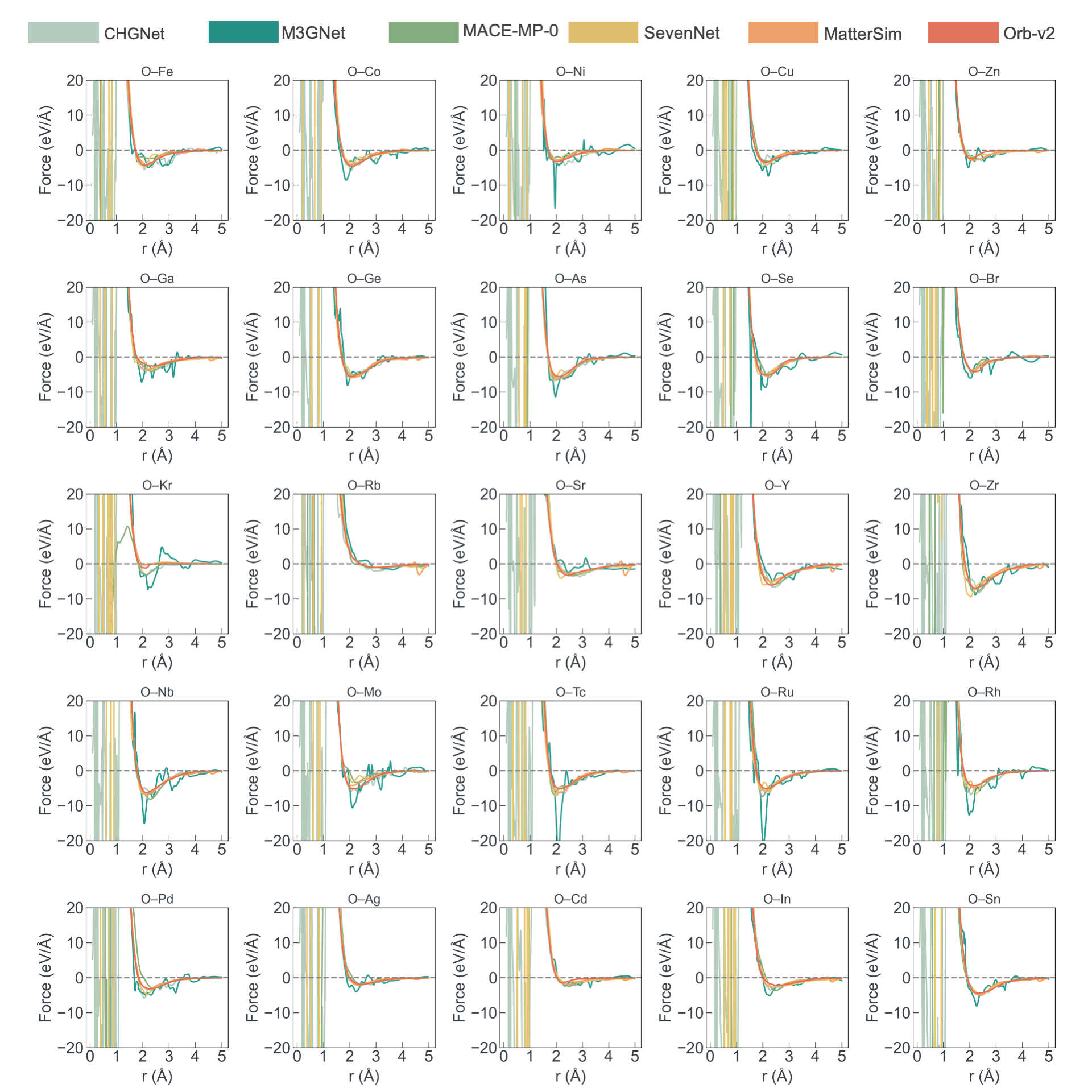}
    \caption{Pairwise Force Plots}
    \label{fig:pairwiseplots15}
    \vspace{-0.2in}
\end{figure}

\begin{figure}[!ht]
    % \vspace{-0.3in}
    \centering
    \includegraphics[width=0.95\textwidth]{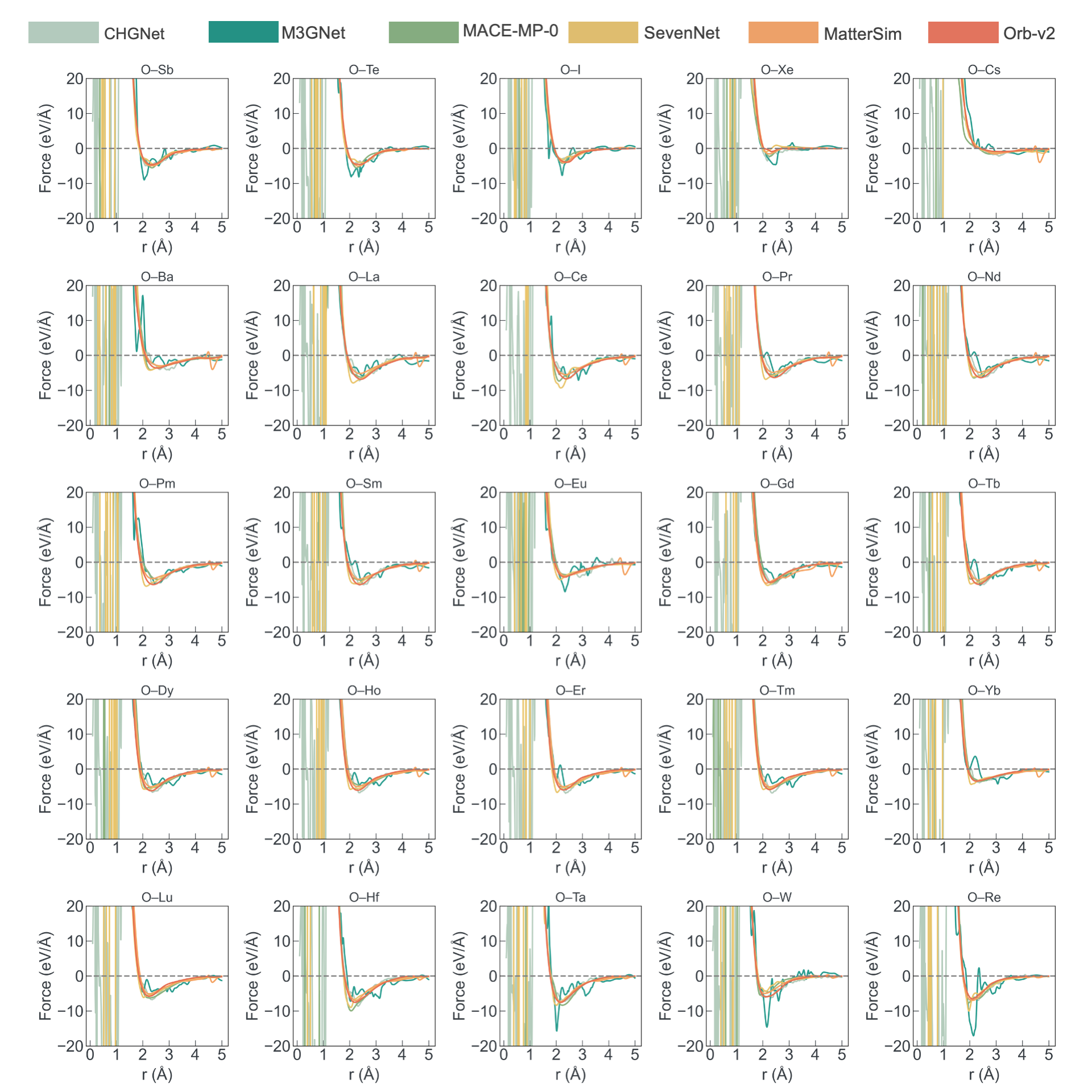}
    \caption{Pairwise Force Plots}
    \label{fig:pairwiseplots16}
    \vspace{-0.2in}
\end{figure}

\begin{figure}[!ht]
    % \vspace{-0.3in}
    \centering
    \includegraphics[width=0.95\textwidth]{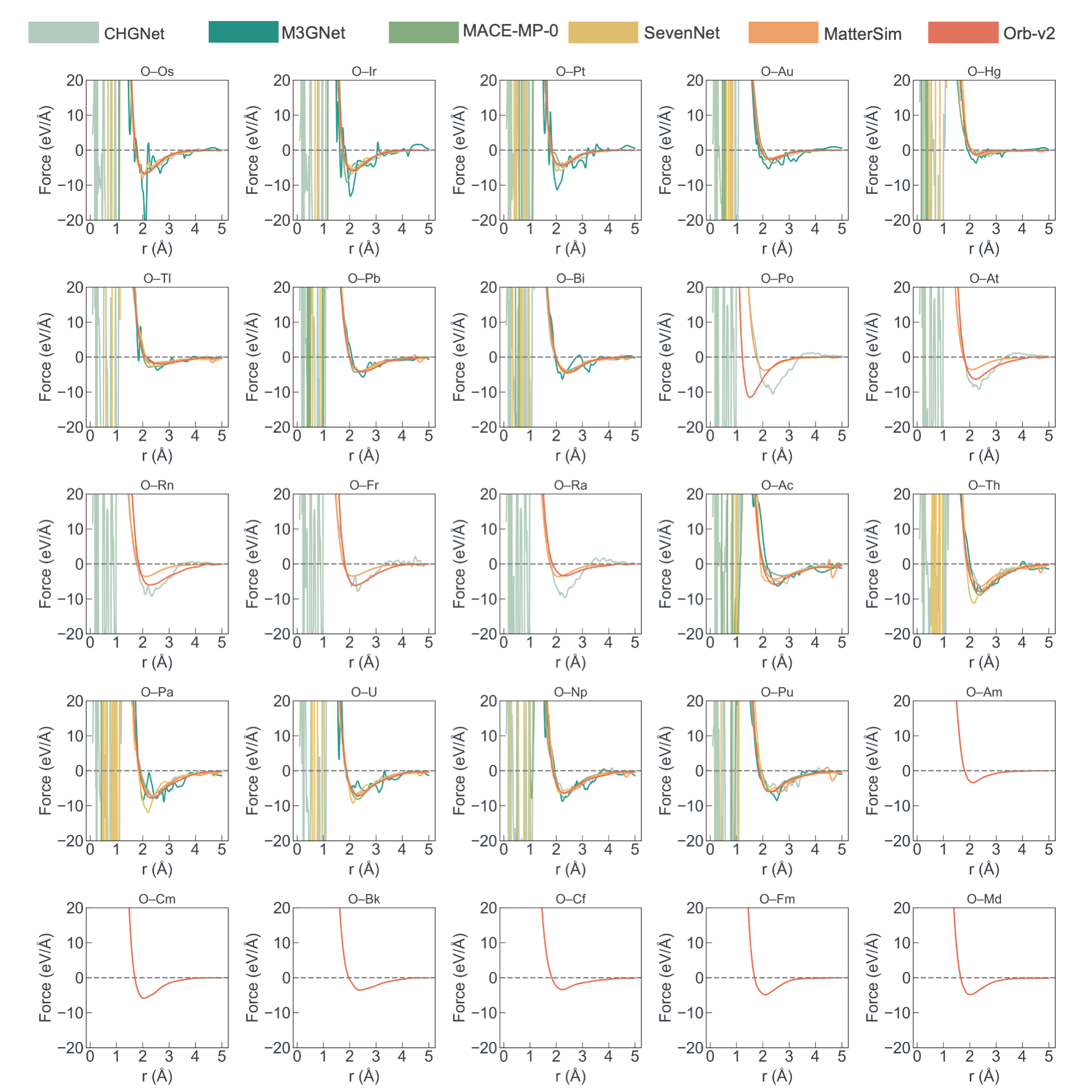}
    \caption{Pairwise Force Plots}
    \label{fig:pairwiseplots17}
    \vspace{-0.2in}
\end{figure}

In addition to analyzing the pairwise energy and force interactions within homonuclear systems, we also investigated the pairwise interactions of each element with oxygen to gain insights into hetero-nuclear bonding characteristics learned by these UMLIFFs. Supplementary~\Cref{fig:pairwiseplots10,fig:pairwiseplots11,fig:pairwiseplots12,fig:pairwiseplots13} and Supplementary~\Cref{fig:pairwiseplots14,fig:pairwiseplots15,fig:pairwiseplots16,fig:pairwiseplots17} {} shows pairwise energy and force plot for hetero-nuclear system respectively.



\begin{figure}[!ht]
    % \vspace{-0.3in}
    \centering
    \includegraphics[width=0.95\textwidth]{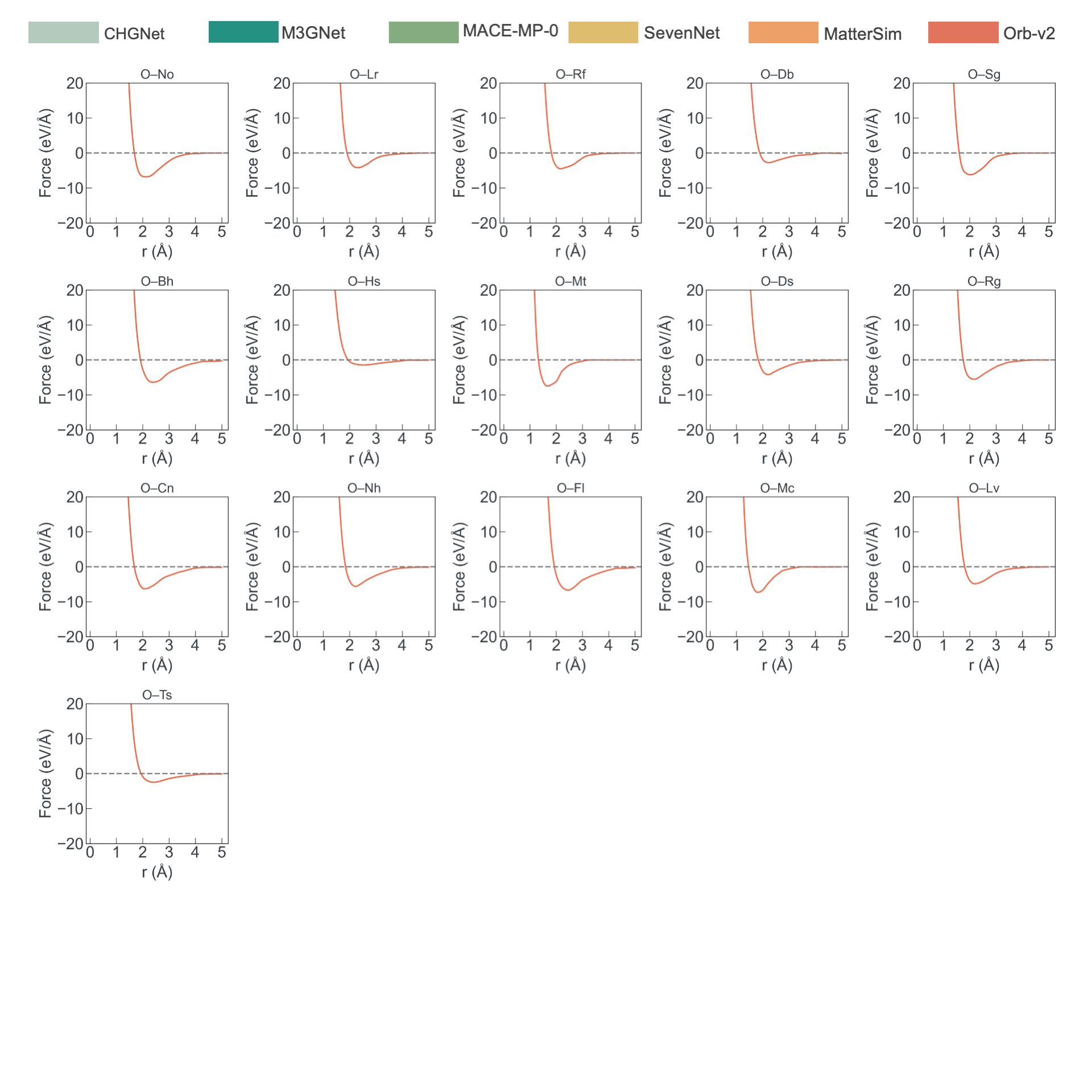}
    \caption{Pairwise Force Plots}
    \label{fig:pairwiseplots18}
    \vspace{-0.2in}
\end{figure}

\clearpage
\section{Architectural Limitations of Current UMLFFs} 
\label{app:limitation}

{The performance patterns across models reveal fundamental connections between architectural choices and physical property prediction capabilities:}

{\textbf{Invariance vs. Equivariance.} Models differ in how they encode geometric information. Invariant models like \orb{} learn representations of atomic environments based on invariant quantities, relying on large training datasets to map energy landscapes. Equivariant models like \mace{} and \sevennet{} explicitly encode directional symmetries (how properties transform under rotations), providing stronger inductive biases that typically improve data efficiency. Our results show both approaches can achieve structural stability, but neither reliably predicts mechanical properties---suggesting the limitation lies in training objectives (energy/force matching) rather than symmetry treatment alone.}

{\textbf{Energy-based vs. Force-based Prediction.} Most models learn an energy function and derive forces as gradients (F = -$\nabla$E), ensuring thermodynamic consistency. \orb{} uniquely predicts forces directly, while also employing a denoising diffusion based pretraining, enabling remarkably smooth potential energy surfaces and perfect trajectory completion. However, this design choice creates a fundamental limitation: elastic properties require second energy derivatives with respect to strain ($\frac{\partial^2E}{\partial \varepsilon^2}$), which cannot be reliably computed when forces do not derive from a consistent energy function. This explains \orb{'s} paradoxical performance---excellent structural predictions but catastrophic failure for all elastic properties. The disconnect demonstrates that simulation stability does not guarantee accurate mechanical property prediction.}

{\textbf{Dataset Scale and Property Generalization.} Increasing training data from ~1.5M structures (MPtrj) to ~500M (OMat24/OMol25) dramatically improves simulation stability (\mace{} vs. \maceomat{} being a case-in-point) by better sampling configurational space. However, property prediction accuracy improves far less: density errors decrease only ~4-5$\times$ (from 25-35\% to 5-7\%) and remain above practical thresholds ($\pm$2.5\%). This suggests that current training paradigms---focused on energy/force accuracy at 0~K---do not adequately capture the physics needed for diverse property predictions at finite temperature, including thermal expansion, anharmonic effects, and the energy-strain relationships governing elastic response.}

These architectural and training limitations highlight a central conclusion: models optimized for energy and force accuracy do not automatically predict derived properties reliably. Achieving accurate mechanical, thermodynamic, and structural property predictions may require training objectives that explicitly incorporate these properties or physics-informed constraints beyond energy/force matching.

\section{Beyond Existing Benchmarking Paradigms}
\label{app:benchmark_paradigm}

Benchmarking of UMLFFs has predominantly focused on DFT-based datasets such as MPtrj~\citep{deng2023chgnet}, OC20~\citep{chanussot2021open}, OMat24~\citep{barrosoluque2024openmaterials2024omat24}, OMol25~\citep{levine2025open}, and OC22~\citep{tran2023open}. While recent efforts have expanded evaluation scope~\citep{wood2025family, yuan2025foundation, riebesell2025framework, fu2023forcesenoughbenchmarkcritical,bihani2024egraffbench,fu2025learning}, existing benchmarks primarily assess performance on properties such as formation energies, forces, and stresses---the same quantities used as training targets. Even when evaluating derived properties like elastic moduli, validation typically relies on DFT-computed reference values generated using similar computational protocols.

% \resp{\textit{UMLFFs are trained on DFT data and achieve computational accuracies approaching DFT self-consistency for energy and force predictions. However, they systematically underperform DFT accuracy for derived properties: DFT typically predicts densities within 2.5\% of experimental values (Table~S6), while UMLFFs exhibit 5.2\% density errors despite excellent energy/force reproduction. This demonstrates fundamental limitations in how current models learn and generalize from training objectives to property predictions.}}

This creates a potential training-evaluation circularity: models learn to reproduce DFT predictions and are validated against similar DFT-computed properties, with limited assessment of whether these predictions align with measurable material behavior. UMLFFs are trained on DFT data and achieve computational accuracies approaching DFT self-consistency for energy and force predictions. However, they systematically underperform DFT accuracy for derived properties: DFT typically predicts densities within 2.5\% of experimental values (see Supplementary Table~\ref{table:baseline}), while UMLFFs exhibit 5.2\% density errors despite excellent energy/force reproduction. This demonstrates fundamental limitations in how current models learn and generalize from training objectives to property predictions. Further, to investigate this systematically, we compared formation energy $R^2$ scores from Matbench Discovery~\citep{riebesell2025framework} with experimentally measured $R^2$ scores from our benchmark across Young's modulus, shear modulus, density, and lattice parameters for the same models. The analysis reveals remarkably poor correlation between computational and experimental performance (see Supplementary Figure~\ref{fig:comp-exp-comparison}), demonstrating that models achieving excellent energy/force accuracy do not reliably predict diverse structural and elastic properties. This disconnect suggests that strong performance on training objectives does not guarantee generalization to experimentally measured properties, even when those properties should in principle be accessible from the learned potential energy surface.

Beyond the generalization gap, our benchmark evaluates model performance under conditions and for structural types that extend substantially beyond typical computational validation scenarios. Our benchmark's primary contribution is comprehensive diversity: extreme thermodynamic conditions, disordered structures, finite-temperature effects, and chemical complexity across 1,500+ systems. Specifically, \uniffbench{} uniquely assesses: (1) Extreme thermodynamic conditions---\minx-HTP evaluates stability and accuracy from 0--5000~K and 0--1000~GPa, conditions that are computationally prohibitive to validate at scale and often characterized primarily through experimental measurements; (2) Partial occupancy or compositional disorder---\minx-POcc includes minerals with crystallographic site disorder and fractional occupancies, features systematically excluded from DFT training datasets but validated primarily through experimental diffraction measurements; (3) Finite-temperature equilibrium structures---our benchmarks evaluate predictions at 298~K with realistic thermal disorder, anharmonicity, and thermal expansion, while current UMLFF training datasets consist predominantly of 0~K energy-minimized structures; (4) Diverse material properties---beyond energy and force predictions, we evaluate elastic tensors (\minx-EM), equilibrium volumes and lattice parameters (\minxeq{}), and structural stability under thermodynamic stress (\minx-HTP), revealing that increasing training dataset size improves energy/force accuracy but does not proportionally improve experimental property prediction.

While high-quality DFT could provide references for some properties (e.g., elastic tensors of ordered crystals at ambient conditions), experimental measurements offer practical validated reference data for: (1) extreme conditions where DFT is computationally prohibitive at scale, (2) disordered structures systematically excluded from DFT datasets, (3) finite-temperature equilibria requiring expensive AIMD, and (4) diverse chemistry at a scale enabling statistically robust evaluation.

We note that for elastic properties of well-ordered crystals at ambient conditions, the conclusions may remain consistent even with computational data (such as DFT) instead of experimental data. However, the combination of extreme thermodynamic conditions, disordered structures, finite-temperature effects, and diverse structures (with up to 23 elements) in \uniffbench{} creates validation tasks that are either computationally prohibitive to replicate at scale or systematically underrepresented in existing computational benchmarks. Our framework thus provides complementary validation that assesses model generalization to conditions, structural types, and property predictions that extend beyond typical training and computational evaluation scenarios, revealing systematic limitations not captured by existing benchmarks.

% New Section 4: Recommendations for Next-Generation UMLFFs
\section{Recommendations for Next-Generation Force Fields}
\label{app:reco_ff}

Based on the observations in \uniffbench{}, we provide the following recommendations for the development of next generation UMLFFs. 
\begin{itemize}
    \item \textbf{Multi-Target Training Protocols.} Future UMLFFs should incorporate experimental properties directly into training objectives to overcome current limitations. We recommend multi-task learning approaches that simultaneously optimize energy, forces, and stress tensors along with experimental properties, ensuring that higher-order derivatives of the potential energy surface are properly constrained. Specifically, training protocols incorporating elastic tensor components or phonon dispersion~\cite{gangan2025force,thaler2021learning} as direct training targets, stress-strain relationships under various deformation modes, thermodynamic constraints, such as Born stability criteria, and multi-scale consistency between local bonding and bulk mechanical response could be explored.
    \item \textbf{Architectural Features.} The success of \orb{}'s smooth pairwise interactions suggests that direct force prediction approaches may offer advantages over energy-based methods for certain applications. However, to obtain reasonable performance on material properties, such  elastic tensors, including higher-order derivatives or finetuning toward target properties would be required.
    \item \textbf{Training Data Diversification Strategies.} Addressing systematic bias requires fundamental changes in training data curation. Future datasets must achieve balanced representation across all elemental combinations, diverse local coordination environments for each atomic pair, experimental elastic property data as training targets, systematic coverage of thermodynamic conditions beyond ambient conditions, and comprehensive inclusion of materials with complex compositions exceeding ten elements and partial occupancies. Several efforts along this direction are underway~\cite{levine2025open,wood2025family}. The correlation between training data frequency and prediction accuracy demonstrates that achieving universal capability requires not just chemical diversity but environmental diversity---ensuring atomic pairs are encountered across wide ranges of local coordination environments, bonding configurations, and chemical contexts.
    \item \textbf{Evaluation Protocol Standards.} \uniffbench{} establishes essential benchmarking standards that can become community practice. Mandatory experimental validation alongside computational benchmarks ensures real-world applicability, while standardized failure reporting with simulation completion rates provides transparency about model limitations. Application-specific accuracy thresholds offer practical guidance for model deployment. Future evaluation protocols should incorporate multi-scale property assessment spanning structural, dynamic, and mechanical behaviors, temporal stability analysis through extended MD simulations, systematic failure mode characterization, chemical diversity metrics relative to training data, and computational resource transparency for practical deployment decisions.
\end{itemize}

\end{document}